\documentclass[aps, prd, superscriptaddress, nofootinbib, preprintnumbers, twocolumn, showpacs]{revtex4}
\usepackage{graphicx}
\usepackage{amsmath}
\def\fsl#1{\setbox0=\hbox{$#1$}                 
   \dimen0=\wd0                                 
   \setbox1=\hbox{/} \dimen1=\wd1               
   \ifdim\dimen0>\dimen1                        
      \rlap{\hbox to \dimen0{\hfil/\hfil}}      
      #1                                        
   \else                                        
      \rlap{\hbox to \dimen1{\hfil$#1$\hfil}}   
      /                                         
   \fi}                                         %
\newcommand{\tr}{\mbox{tr}}
\newcommand{\VEV}[1]{\langle #1 \rangle}
\newcommand{\gtrsim}{\mathop{>}\limits_{\displaystyle{\sim}}}
\newcommand{\lessim}{\mathop{<}\limits_{\displaystyle{\sim}}}

\newcommand{\NDA}{\Omega_{\rm NDA}}

\newcommand{\Ly}{\Lambda_{LY}}

\newcommand{\DxSB}{D$\chi$SB}
\newcommand{\NKKg}{N_{\rm KK}^g}
\newcommand{\NKKb}{N_{\rm KK}^{gs}}
\newcommand{\NKKf}{N_{\rm KK}^f}
\newcommand{\NKKs}{N_{\rm KK}^h}
\newcommand{\NKKi}{N_{\rm KK}^i}
\newcommand{\tMAC}{\Lambda_{\rm tM}}
\begin{document}
\title{Topped MAC with extra dimensions?}
\date{\today}

\preprint{DPNU-03-04}
\preprint{TU-683}
\preprint{hep-ph/0311165}


\author{Michio Hashimoto}
\email[E-mail: ]{michioh@eft.phys.pusan.ac.jp}
\affiliation{Department of Physics, Pusan National University, 
             Pusan 609-735, Korea}
\author{Masaharu Tanabashi}
\email[E-mail: ]{tanabash@tuhep.phys.tohoku.ac.jp}
\affiliation{Department of Physics, Tohoku University, Sendai 980-8578, Japan}
\author{Koichi Yamawaki}
\email[E-mail: ]{yamawaki@eken.phys.nagoya-u.ac.jp}
\affiliation{Department of Physics, Nagoya University, Nagoya 464-8602, Japan}

\begin{abstract}
We perform the most attractive channel (MAC) analysis in 
the top mode standard model with TeV-scale extra dimensions,
where the standard model gauge bosons and the third generation of
quarks and leptons are put in $D(=6,8,10,\cdots)$ dimensions.
In such a model, bulk gauge couplings rapidly grow in the ultraviolet 
region.
In order to make the scenario viable,
only the attractive force of the top condensate should exceed 
the critical coupling, while other channels such as the bottom and
tau condensates should not.
We then find that the top condensate can be the MAC for $D=8$,
whereas the tau condensation is favored for $D=6$.
The analysis for $D=10$ strongly depends on 
the regularization scheme.
We predict masses of the top $(m_t)$ and the Higgs $(m_H)$, 
$m_t=172-175$ GeV and $m_H=176-188$ GeV for $D=8$, 
based on the renormalization group for the top Yukawa and Higgs quartic
couplings with the compositeness conditions at the scale
where the bulk top condenses.
The Higgs boson in such a characteristic mass range will be immediately
discovered in $H \to WW^{(*)}/ZZ^{(*)}$ once the LHC starts.
\end{abstract}

\maketitle

\section{Introduction}

The origin of mass is a central mystery of the standard model (SM).
In particular, 
why are the masses of W, Z, and the top quark exceptionally large
compared with those of other particles in the SM?
The idea of the top quark condensate~\cite{MTY89, Nambu89, Marciano89}
explains naturally the large top mass of the order of
the electroweak symmetry breaking (EWSB) scale. 
In the explicit formulation of this idea~\cite{MTY89, Bardeen:1989ds}, often called the ``top mode standard model'' (TMSM),
the scalar bound state of $\bar{t}t$ plays the role of 
the Higgs boson in the SM.
(For reviews, see, e.g., 
Refs.~\cite{Miransky:vk,Yamawaki:1996vr,Hill:2002ap}.)

There are several problems in the original version of the TMSM:
We need to introduce ad hoc four-fermion interactions of the top quark
in order to trigger the EWSB\@.
The top mass $m_t$, given as a decreasing function of the composite scale
$\Lambda$, is predicted about 10\%--30\% larger than the experimental value, 
even if we take $\Lambda$
to the Planck or the GUT 
scale~\cite{MTY89,Bardeen:1989ds,Hashimoto:1998tj}. 
Such a huge $\Lambda$ also causes a serious fine-tuning 
problem.

As a possible solution to these problems, 
following the line of an earlier attempt~\cite{Dobrescu:1998dg,Cheng:1999bg} 
of the TMSM in the TeV-scale extra dimension 
scenario~\cite{Antoniadis:1990ew,Dienes:1998vh},
Arkani-Hamed, Cheng, Dobrescu and Hall (ACDH)~\cite{Arkani-Hamed:2000hv}
proposed an interesting version of such 
where the SM gauge bosons and the third generation of quarks and leptons
live in the $D(=6,8,\cdots)$-dimensional bulk, 
while the first and second generations are confined in the 3-brane 
(4-dimensional Minkowski space-time).
Gauge interactions in higher dimensions than four become strong
in a certain high-energy region.
Bulk gauge interactions are expected to trigger the top condensation 
without adding ad hoc four-fermion interactions, in contrast to
the original version of the TMSM.
 
However, the dynamics of bulk gauge theories was not concretely analyzed 
in Ref.~\cite{Arkani-Hamed:2000hv}.
In particular, as it turned out~\cite{Hashimoto:2000uk}
(see also Refs.~\cite{Agashe:2000nk,Dienes:2002bg}),
the bulk QCD coupling, which is the most relevant
interaction for the top condensation,
has an ultraviolet fixed point (UV-FP) or upper bound
within the same $\overline{\rm MS}$-scheme of the truncated
Kaluza-Klein (KK) effective theory~\cite{Dienes:1998vh}
as that Ref.~\cite{Arkani-Hamed:2000hv} was based on.
Thus, it is quite nontrivial whether 
the top condensation is actually realized or not.
In Refs.~\cite{Hashimoto:2000uk, Gusynin:2002cu}, we have studied 
the dynamical chiral symmetry breaking (D$\chi$SB) in bulk gauge theories, 
based on the ladder Schwinger-Dyson (SD)
equation.\footnote{\DxSB~in other scenario for extra dimensions 
was investigated in Ref.~\cite{Abe:2001yi}.}
Switching off the electroweak interaction in the bulk,
we then found that the bulk QCD coupling cannot become sufficiently large 
to trigger the top condensation for $D=6$, 
while the top condensation can be realized for $D=8$.

For the purpose of model building, we further need to study 
the effect of the bulk electroweak interactions:
Since the bulk $U(1)_Y$ interaction grows very quickly due to 
the power-like running behavior and reaches immediately 
its Landau pole $\Ly$,
it
may affect 
the most favored channel for condensate, i.e.,  
the most attractive channel (MAC)~\cite{Raby:1979my}.
We also need to study 
whether or not the prediction of the top mass agrees with 
the experiments.

In this paper\footnote{The preliminary report was given in Ref.~\cite{talk}.}, 
we demonstrate a possibility that 
the top condensate is actually the MAC even including 
all of the bulk SM gauge interactions.
This is quite nontrivial, because inclusion of 
the strong bulk $U(1)_Y$ interaction may favor 
the tau condensation rather than the top condensation.
In order for only the top quark to acquire the dynamical mass of 
the order of the EWSB scale, 
the binding strength should exceed the critical binding strength 
$\kappa_D^{\rm crit}$ only for the top quark (``topped MAC'' or ``tMAC'').
Namely, our scenario works only when
\begin{equation}
  \kappa_t(\Lambda) > \kappa_D^{\rm crit} > \kappa_b(\Lambda),
  \kappa_\tau(\Lambda), \cdots, \label{top-cond}
\end{equation}
where  $\kappa_t(\Lambda)$, $\kappa_b(\Lambda)$ and $\kappa_\tau(\Lambda)$ 
denote the binding strengths of the top, bottom, and 
tau condensates at the scale $\Lambda$, respectively.
We refer to the scale $\Lambda$ satisfying Eq.~(\ref{top-cond}) 
as the tMAC scale $\tMAC$.

For the MAC analysis,
we study binding strengths $\kappa_{t,b,\tau}$
by using the one-loop renormalization group equations (RGEs) of 
dimensionless bulk gauge couplings.
It is in contrast to the analysis of ACDH~\cite{Arkani-Hamed:2000hv} where
all of bulk gauge couplings are assumed equal
(and strong enough for triggering the EWSB). 
In order to check reliability of our MAC analysis, 
we also study the regularization-scheme dependence of the binding strengths.
We calculate gauge couplings in two prescriptions,
the $\overline{\rm MS}$-scheme of the truncated KK effective theory 
and the proper-time (PT) scheme~\cite{Dienes:1998vh}.

There are some 
varieties in the estimation of 
$\kappa_D^{\rm crit}$:
The naive dimensional analysis (NDA)~\cite{Manohar:1983md,Chacko:1999hg}
implies $\kappa_D^{\rm crit} \sim 1$, while the ladder SD equation
yields much smaller value 
$\kappa_D^{\rm crit} \sim 0.1$~\cite{Hashimoto:2000uk,Gusynin:2002cu}.
As the estimate of $\kappa_D^{\rm crit}$ increases in Eq.~(\ref{top-cond}), 
the region of the tMAC scale gets squeezed.
Even if we adopt the lowest possible value of $\kappa_D^{\rm crit}$
given by the ladder SD 
equation~\cite{Hashimoto:2000uk,Gusynin:2002cu},
we find that {\it the tMAC scale does not exist for 
the simplest scenario with $D=6$}.
On the other hand, {\it the tMAC scale does exist in $D=8$} 
for the value of the ladder SD equation,
$\tMAC R = 3.5$--$3.6$, where the compactification scale $R^{-1}$
is taken to be 1--100 TeV.
For $D=10$, the MAC analysis significantly depends on 
the regularization scheme.

Once we obtain the tMAC scale $\tMAC$, we can easily predict 
the top mass $m_t$ and the Higgs mass $m_H$ by using
the renormalization group equations (RGEs) 
for the top Yukawa and Higgs quartic couplings, and 
the compositeness conditions~\cite{Bardeen:1989ds} at the scale 
$\Lambda=\tMAC$.
This is in contrast to the earlier approach~\cite{Arkani-Hamed:2000hv} 
(see also Ref.~\cite{Kobakhidze:1999ce}) 
where the composite scale $\Lambda$ is treated as an adjustable free
parameter and fixed so as to reproduce the experimental value of $m_t$.
Without such an adjustable parameter,
we predict the top quark mass
\begin{equation}
  m_t = 172-175 \; \mbox{GeV}
\end{equation}
for $D=8$ and $R^{-1}=$ 1--100 TeV. 
This agrees with the experimental value,
$m_t=174.3 \pm 5.1$ GeV~\cite{PDG}.
We find that the value of $m_t$ near the compactification scale 
$R^{-1}$ is governed by the quasi infrared fixed point (IR-FP) 
for the top Yukawa coupling 
$y$~\cite{Hill:1980sq},
which is approximately obtained as 
$y = g_3 \cdot \sqrt{C_F(6+\delta)/(2^{\delta/2}N_c+3/2)}$
with the 4-dimensional QCD coupling $g_3$, 
the number of color $N_c(=3)$, the quadratic Casimir of 
the fundamental representation $C_F(=4/3)$, and $\delta \equiv D-4$.
The suppression factor $2^{-\delta/2}$ in $y$ arises from 
one-loop effects of the bulk top quark which is equivalent to 
a tower of KK modes (massive vector-like fermions).
The mechanism suppressing the top mass prediction is thus 
similar to that of the top seesaw~\cite{Dobrescu:1997nm}.
We also predict the Higgs boson mass as
\begin{equation}
  m_H=176-188 \; \mbox{GeV} .
\end{equation}
Thanks to the IR-FP property,
the prediction for $m_t$ and $m_H$ is stable.
The Higgs boson with such a characteristic mass can be distinguished
clearly from that of supersymmetric or other typical dynamical EWSB
models simply through its mass determination in experiments.
It will also be discovered immediately after the physics run of LHC\@.

The paper is organized as follows. 
In Sec.~II, we study running effects of bulk gauge couplings in 
the $\overline{\rm MS}$-, and PT-schemes.
In Sec.~III, we identify the tMAC scale. 
In Sec.~IV, we predict $m_t$ and $m_H$ for $D=8$. 
Sec.~V is devoted to summary and discussions.
In Appendix A, we present details of our procedure for the orbifold 
compactification.
We also count the total number of KK modes below the renormalization point.
Appendix B is for details of bulk gauge couplings in the PT-scheme.

\section{Running of bulk gauge couplings}

Let us consider a simple version of the TMSM with extra dimensions
where the SM gauge group and the third generation
of quarks and leptons are put in $D$-dimensional bulk, 
while the first and second generations live on the 3-brane 
(4-dimensional Minkowski space-time). 
The $D$-dimensions consist of the usual 4-dimensional Minkowski space-time 
and extra $\delta (= D-4)$ spatial dimensions compactified 
at a TeV-scale $R^{-1}$.
The number of dimensions $D$ is taken to be even, $D=6,8,10,\cdots$,
so as to introduce chiral fermions in the bulk. 
In order to obtain a 4-dimensional chiral theory and to forbid 
massless gauge scalars, 
we compactify extra dimensions on the orbifold $T^\delta/Z_2^{\delta/2}$
(see Appendix A).
We emphasize that there is no elementary field for Higgs in our model.
The chiral condensation of bulk fermions may generate dynamically
a composite Higgs field, instead.
Hence we investigate RGEs of bulk gauge couplings including 
loop effects of the composite Higgs.

\subsection{$\overline{\rm MS}$-coupling 
in the truncated KK effective theory}

\begin{figure}[tbp]
  \begin{flushleft}
    \hspace*{1cm} {\Large (a)}
  \end{flushleft}
  \vspace*{-1cm}
  \begin{center}
  \resizebox{0.48\textwidth}{!}
            {\includegraphics{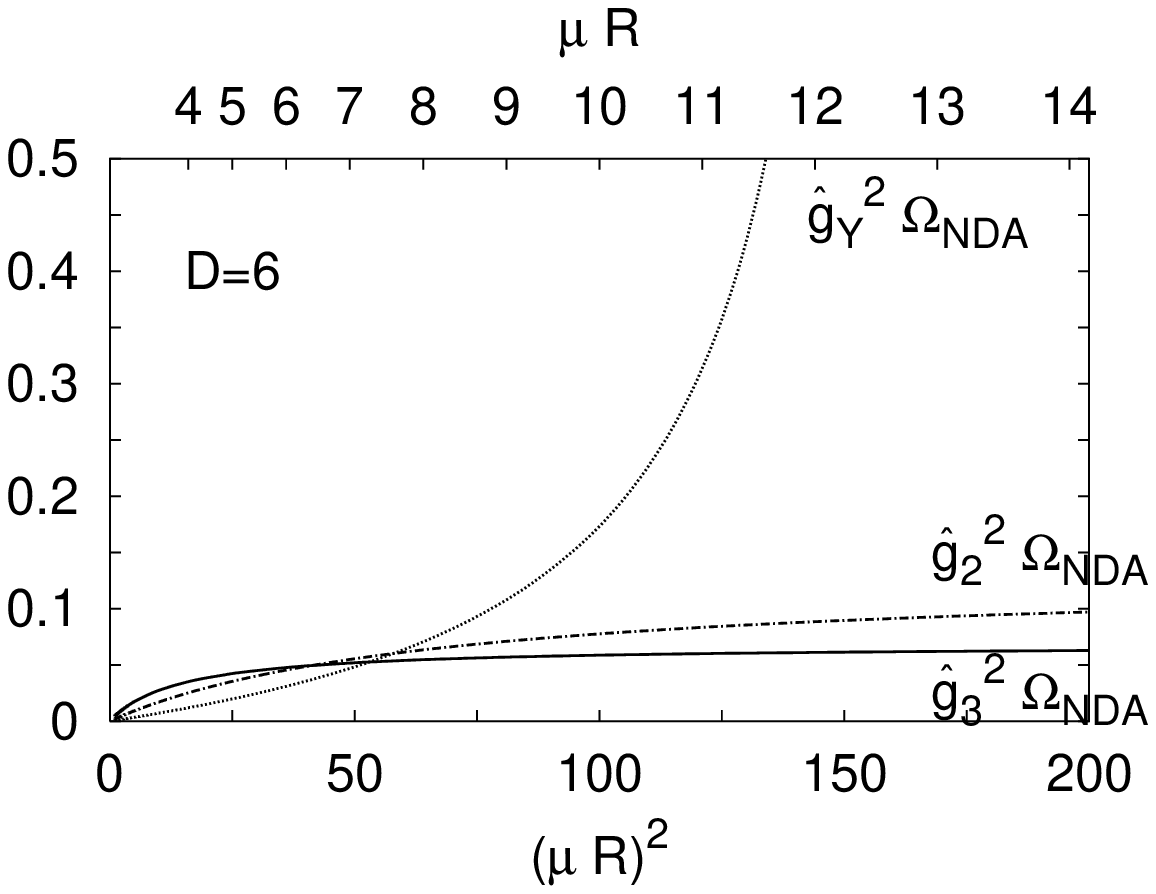}}
  \end{center}
  \begin{flushleft}
    \hspace*{1cm} {\Large (b)}
  \end{flushleft}
  \vspace*{-1cm}
  \begin{center}
  \resizebox{0.48\textwidth}{!}
            {\includegraphics{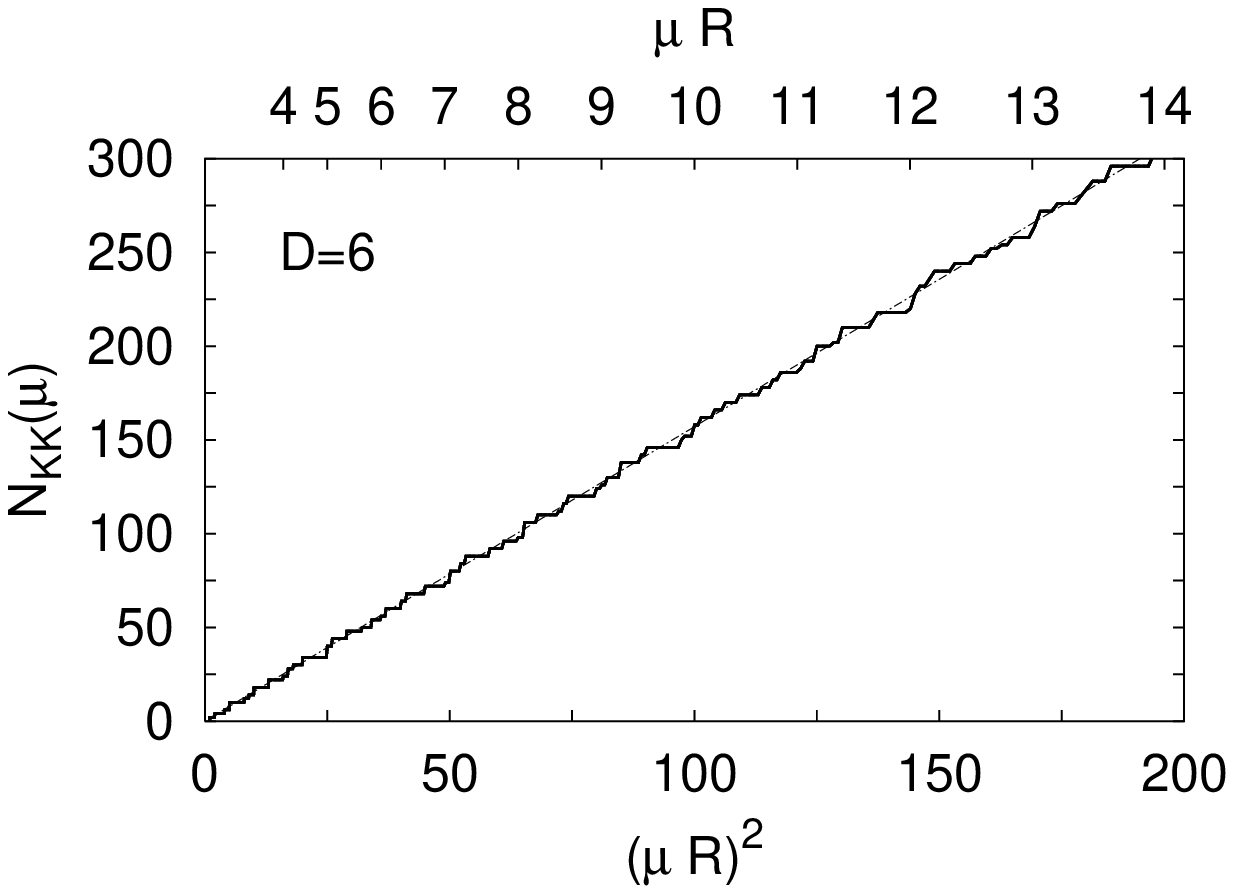}}
  \caption{Typical behaviors of (a) $\hat g_i^2 \NDA$ 
           and (b) $N_{\rm KK}^i$ for $D=6$.
           We assumed $n_g=1,R^{-1}=10$ TeV\@.
           Values of $N_{\rm KK}^i(\mu)$ are equal in $D=6$, i.e.,
           $\NKKg (\mu)=\NKKb (\mu)=\NKKs (\mu)=\NKKf (\mu)$.
           In (b), bold and dash-dotted lines 
           represent the numerical analysis of $N_{\rm KK}^i(\mu)$ and
           its approximation Eq.~(\ref{nkk_app}), respectively. 
           \label{fig-g-6d}}
  \end{center}
\end{figure}

\begin{figure}[tbp]
  \begin{flushleft}
    \hspace*{1cm} {\Large (a)}
  \end{flushleft}
  \vspace*{-1cm}
  \begin{center}
  \resizebox{0.48\textwidth}{!}
            {\includegraphics{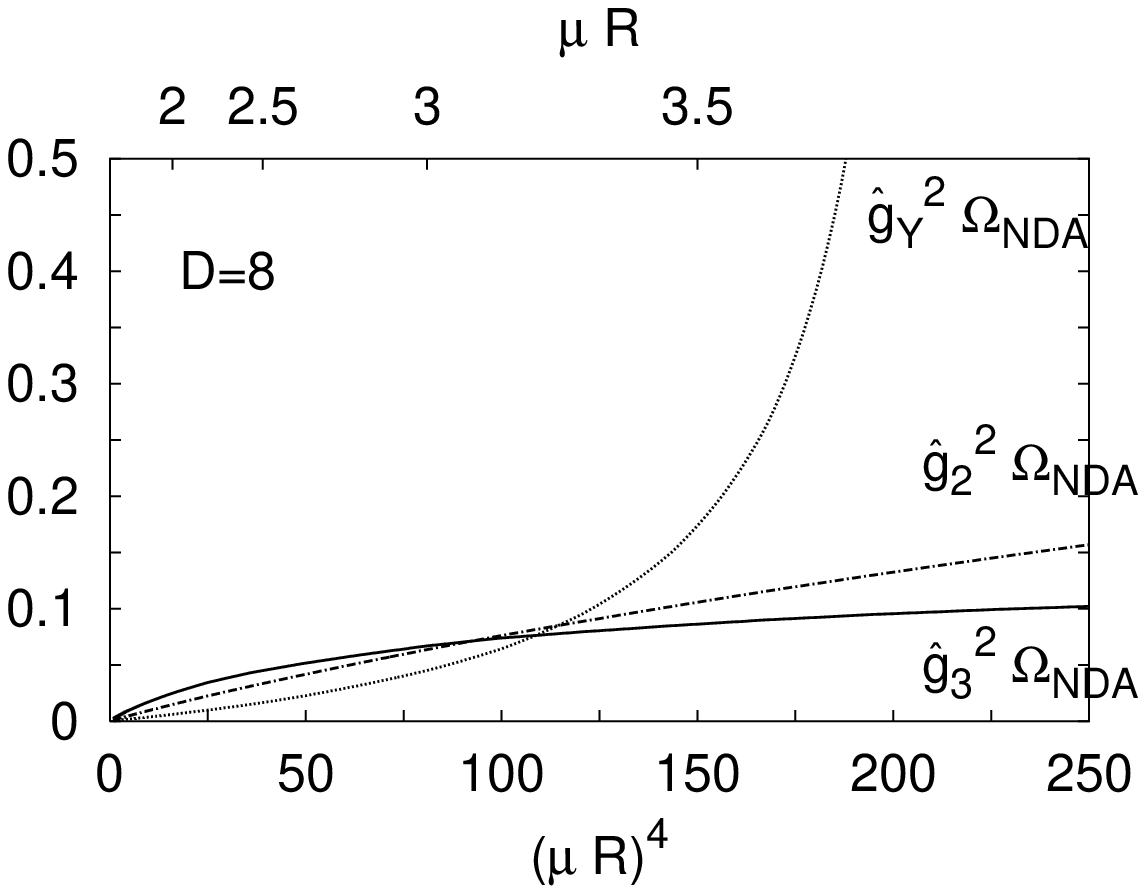}}
  \end{center}
  \begin{flushleft}
    \hspace*{1cm} {\Large (b)}
  \end{flushleft}
  \vspace*{-1cm}
  \begin{center}
  \resizebox{0.48\textwidth}{!}
            {\includegraphics{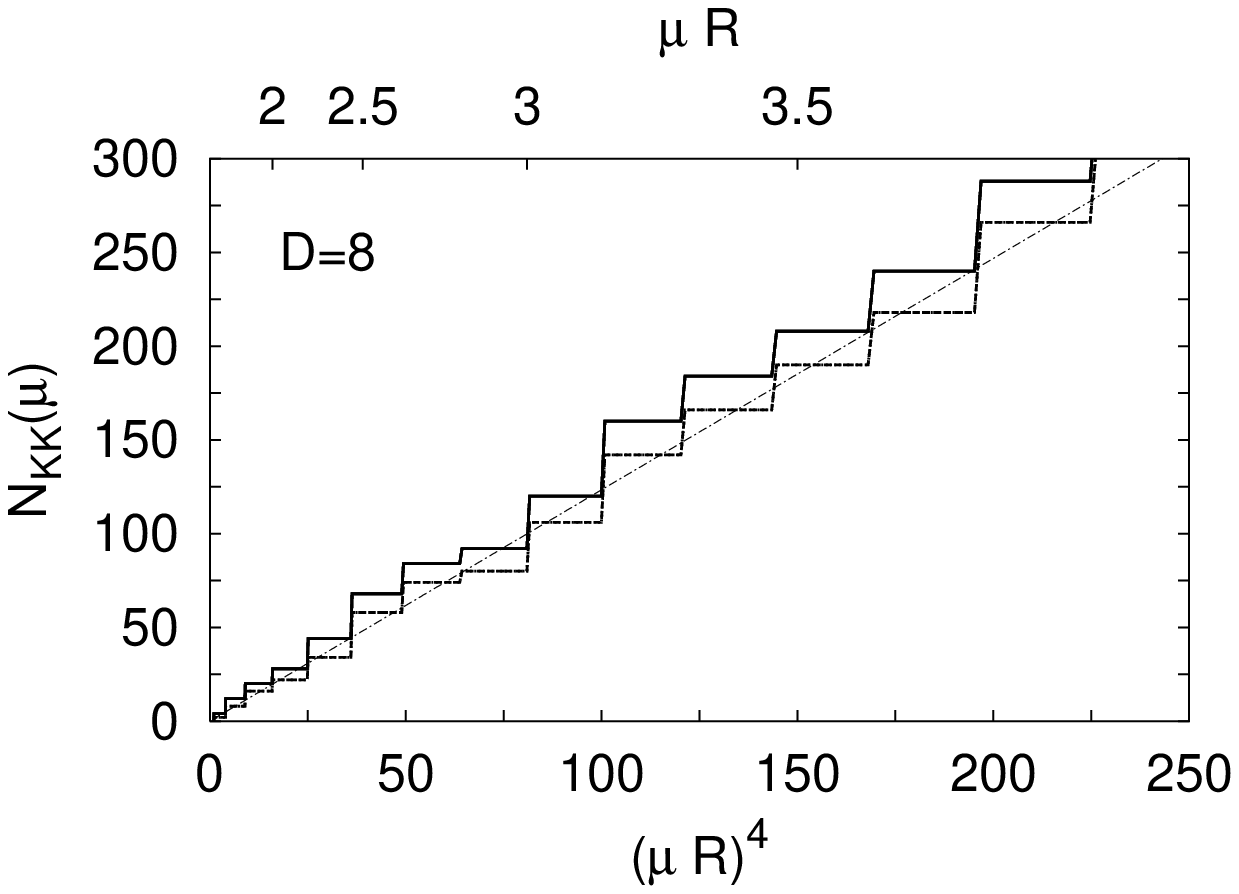}} 
  \caption{Typical behaviors of (a) $\hat g_i^2 \NDA$ 
           and (b) $N_{\rm KK}^i$ for $D=8$. 
           We assumed $n_g=1,R^{-1}=10$ TeV\@.
           In (b), bold-solid and bold-dashed lines 
           represent the numerical analysis of 
           $\NKKg (\mu)=\NKKs (\mu)$ and 
           $\NKKb (\mu)=\NKKf (\mu)/2$,
           respectively. The dash-dotted line denotes 
           the approximation Eq.~(\ref{nkk_app}). 
           \label{fig-g-8d}}
  \end{center}
\end{figure}

\begin{figure}[tbp]
  \begin{flushleft}
    \hspace*{1cm} {\Large (a)}
  \end{flushleft}
  \vspace*{-1cm}
  \begin{center}
  \resizebox{0.48\textwidth}{!}
            {\includegraphics{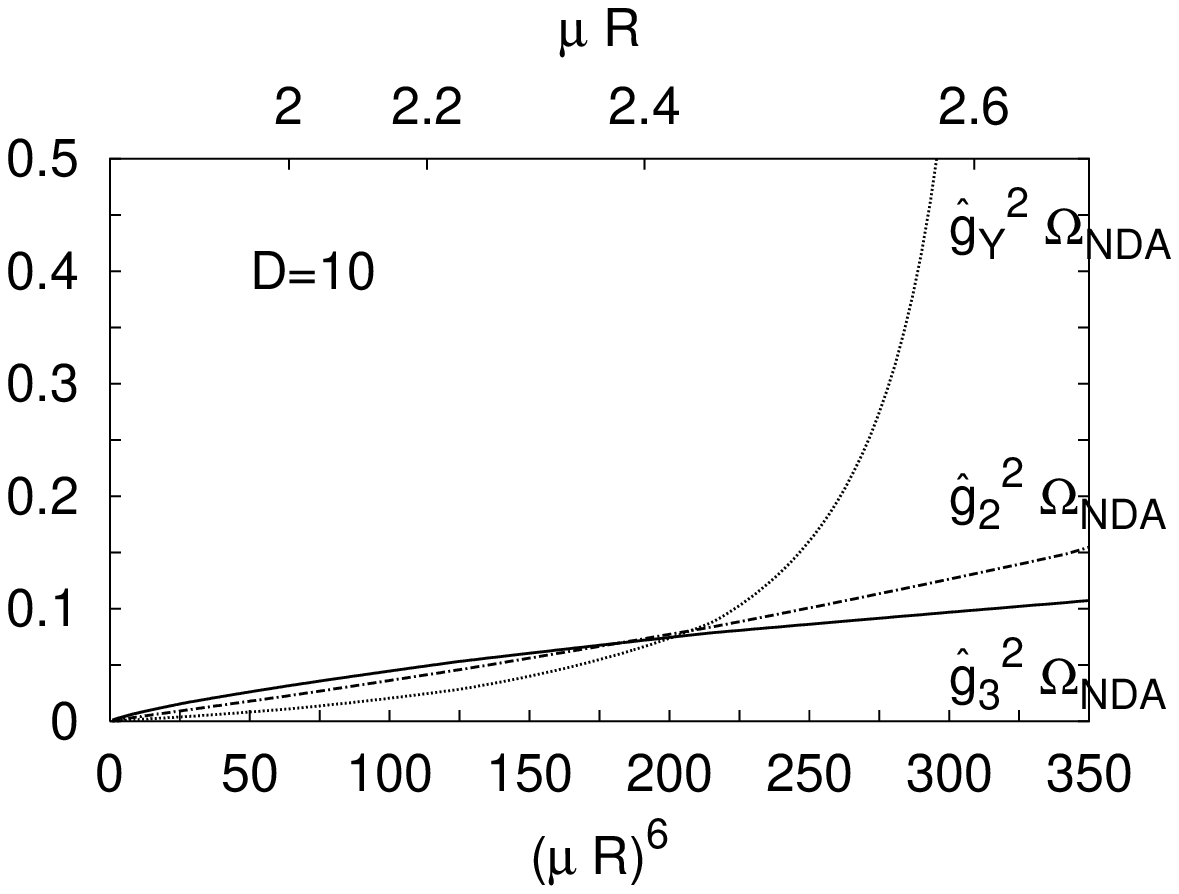}}
  \end{center}
  \begin{flushleft}
    \hspace*{1cm} {\Large (b)}
  \end{flushleft}
  \vspace*{-1cm}
  \begin{center}
  \resizebox{0.48\textwidth}{!}
            {\includegraphics{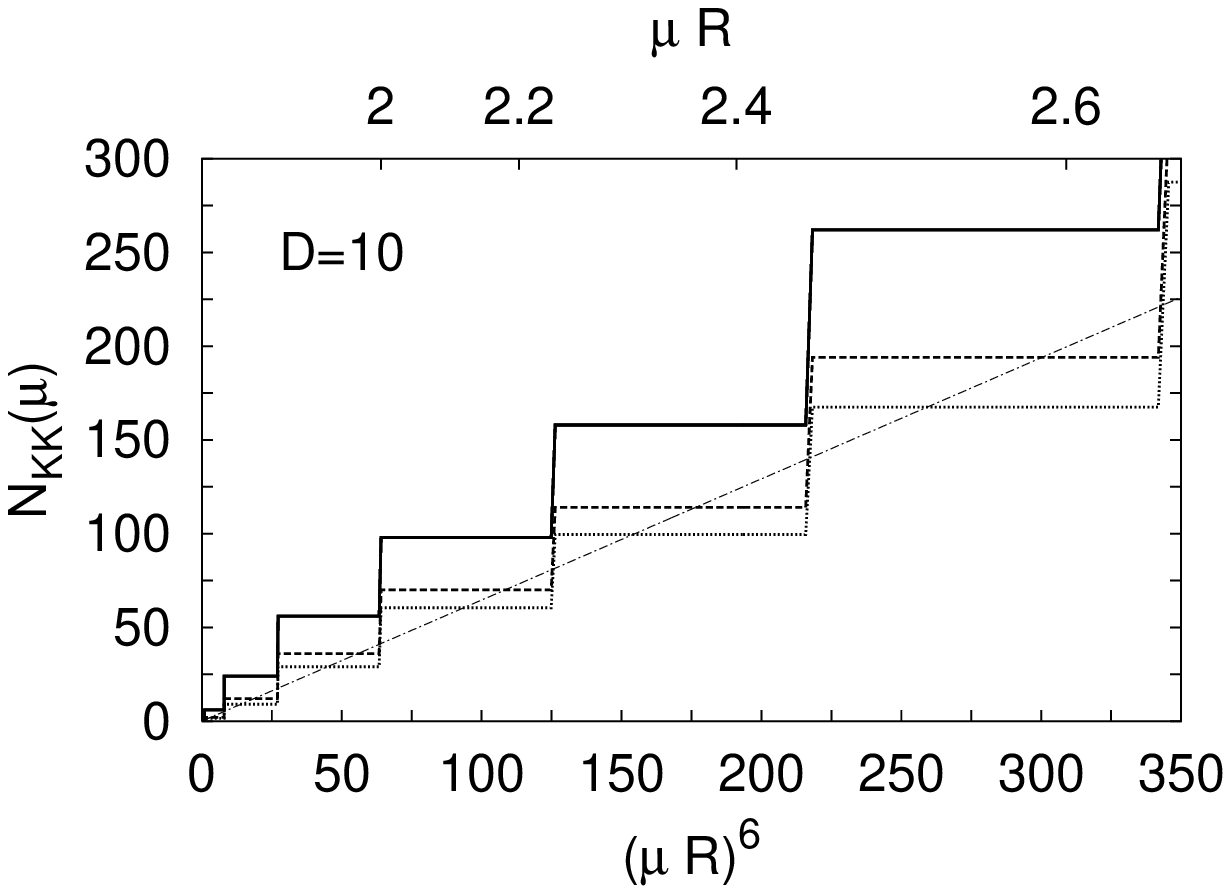}}
  \caption{Typical behaviors of (a) $\hat g_i^2 \NDA$
           and (b) $N_{\rm KK}^i$ for $D=10$. 
           We assumed $n_g=1,R^{-1}=10$ TeV\@.
           In (b), bold-solid, bold-dashed and bold-dotted lines 
           represent the numerical analysis of 
           $\NKKg (\mu)=\NKKs (\mu)$,
           $\NKKb (\mu)$, and $\NKKf (\mu)/4$,
           respectively. The dash-dotted line denotes 
           the approximation Eq.~(\ref{nkk_app}). 
           \label{fig-g-10d}}
  \end{center}
\end{figure}

We expand bulk fields into KK modes and 
construct a 4-dimensional effective theory.
In this subsection, we study running of gauge couplings in 
the ``truncated KK'' effective theory~\cite{Dienes:1998vh}
based on the $\overline{\rm MS}$-scheme.
Below the compactification scale $R^{-1}$, 
RGEs of 4-dimensional gauge couplings 
$g_i (i=3,2,Y)$ are given by those of the SM,
\begin{equation}
  (4\pi)^2 \mu \frac{d g_i}{d \mu} = b_i\,g_i^3, \quad (\mu < R^{-1})
\end{equation}
with $b_3=-7, b_2=-\frac{19}{6}$ and $b_Y=\frac{41}{6}$ for 
one (composite) Higgs doublet.
We need to take into account contributions of KK modes in $\mu \geq R^{-1}$.
Since the KK modes heavier than the renormalization scale $\mu$
are decoupled in the $\overline{\rm MS}$-RGEs,
we only need summing up the loops of the KK modes lighter than $\mu$.
Within the truncated KK effective theory, we obtain RGEs for 
gauge couplings $g_i (i=3,2,Y)$:
\begin{equation}
  (4\pi)^2 \mu \frac{d g_i}{d \mu} = b_i\,g_i^3 
   + b_i^{\rm KK}(\mu)\,g_i^3, 
  \quad (\mu \geq R^{-1}) \label{rge_ED}
\end{equation}
where RGE coefficients $b_i^{\rm KK}(\mu)$ 
are given by
\begin{eqnarray}
  b_3^{\rm KK}(\mu) &=& -11\, \NKKg (\mu)
 +\frac{\delta}{2}\, \NKKb (\mu) \nonumber \\ &&
 +\frac{8}{3}\, n_g \NKKf (\mu) \label{b3KK}
\end{eqnarray}
for $SU(3)_c$, 
\begin{eqnarray}
  b_2^{\rm KK}(\mu) &=& -\frac{22}{3} \, \NKKg (\mu)
 +\frac{\delta}{3} \, \NKKb (\mu) \nonumber \\ &&
 +\frac{8}{3} \, n_g \NKKf (\mu) \nonumber \\ &&
+ \frac{1}{6} \, n_h \NKKs (\mu) \label{b2KK}
\end{eqnarray}
for $SU(2)_W$, and
\begin{equation}
  b_Y^{\rm KK}(\mu) = 
 \frac{40}{9} \, n_g \NKKf (\mu)
 + \frac{1}{6} \, n_h \NKKs (\mu) \label{byKK}
\end{equation}
for $U(1)_Y$. Here $\delta$ is defined as
\begin{equation}
  \delta \equiv D-4  .
\end{equation}
In Eqs.~(\ref{b3KK})--(\ref{byKK}), 
$N_{\rm KK}^i (\mu), \; i=g,gs,f,h$ denote
the total number of KK modes for gauge bosons, gauge scalars, 
Dirac (4-component) fermions, and composite Higgs bosons below $\mu$, 
respectively. The number of generations in the bulk is $n_g$, 
which is unity in our model.
In the following analysis, we assume one composite Higgs doublet, $n_h=1$.
$N_{\rm KK}^i (\mu)$ for $i=g,gs,f,h$ 
are not necessarily equal,
since the orbifold boundary conditions are depending on $i$.
For details, see Appendix A.
We show numerical values for $\NKKg (\mu)=\NKKs (\mu)$, 
(the solid lines), $N_{\rm KK}^{gs} (\mu)$, (the dashed lines),
and $\NKKf (\mu)/2^{\delta/2-1}$, (the dotted lines)
in Figs.~\ref{fig-g-6d}(b), \ref{fig-g-8d}(b) and \ref{fig-g-10d}(b).

Matching the 4-dimensional action to the bulk action, we find the relation 
between the 4-dimensional gauge coupling $g$ and 
the {\it dimensionful} bulk gauge coupling $g_D$, 
$g_D^2=(2\pi R)^\delta g^2/2^{\delta/2}$. 
We define
the {\it dimensionless} bulk gauge coupling $\hat g$ as 
$\hat g^2 \equiv g_D^2 \mu^\delta$ 
and thereby obtain 
\begin{equation}
  \hat g_i^2(\mu) = \frac{(2\pi R \mu)^\delta}{2^{\delta/2}}g_i^2 (\mu).
  \label{hat-g}
\end{equation}
Combining Eq.~(\ref{hat-g}) with Eq.~(\ref{rge_ED}), 
we find RGEs for $\hat g_i$,
\begin{eqnarray}
 \mu \frac{d}{d \mu} \hat g_i &=& \frac{\delta}{2}\hat g_i
 \nonumber \\ && \hspace*{-8mm}
 + \frac{\hat g_i^3}{(4\pi)^2}\frac{2^{\delta/2}}{(2\pi R \mu)^\delta}\left[\,
   b_i + b_i^{\rm KK}(\mu) \,\right] . \label{rge_ED2}
\end{eqnarray}
In Figs.~\ref{fig-g-6d}(a), \ref{fig-g-8d}(a) and \ref{fig-g-10d}(a), 
we show typical behavior of the dimensionless bulk gauge couplings.
We have used the couplings $\alpha_i (\equiv g_i^2/(4\pi))$ 
at $\mu=M_Z(=91.1876 \;{\rm GeV})$
as inputs of RGEs:~\cite{PDG} 
\begin{equation}
  \alpha_3(M_Z)=0.1172, \label{qcd-mz}
\end{equation}
$\alpha_{\rm QED}^{-1}(M_Z)=127.922$ and 
$\sin^2 \theta_W (M_Z) = 0.23113$, i.e.,
\begin{equation}
  \alpha_2(M_Z)=0.033822, \quad \alpha_Y(M_Z)=0.010167 . \label{su2-mz}
\end{equation}
We note here that the $U(1)_Y$ has the Landau pole $\Ly$. 
The bulk gauge coupling $\hat g_Y (\mu)$ rapidly grows 
due to the power-like behavior of the running. 
As a result, the Landau pole $\Ly$ is close to the compactification
scale $R^{-1}$. 
A cutoff smaller than the Landau pole $\Ly$ needs
to be introduced in our model.

A handy approximation for $N_{\rm KK}^i (\mu \gg R^{-1})$ is widely used,
\begin{equation}
  N_{\rm KK}^i (\mu) \simeq 
  \begin{cases}
    N_{\rm KK}(\mu) & \text{for $i=g,gs,h,$} \\[3mm]
    2^{\delta/2-1} N_{\rm KK}(\mu) & \text{for $i=f,$}
  \end{cases}
  \label{nkki_app}
\end{equation}
with
\begin{equation}
N_{\rm KK}(\mu) = \frac{1}{2^{\delta/2}}
                  \frac{\pi^{\delta/2}}{\Gamma(1+\delta/2)}(\mu R)^\delta.
  \label{nkk_app}
\end{equation}
We show $N_{\rm KK}(\mu)$ in
Figs.~\ref{fig-g-6d}(b), \ref{fig-g-8d}(b) and \ref{fig-g-10d}(b) 
with the dash-dotted line.
The deviation of Eq.~(\ref{nkki_app}) from
the numerical calculations of $N_{\rm KK}^i (\mu)$ 
is negligibly small for $D=6$, whereas it is sizable for $D=8,10$ 
in $\mu \lesssim \Ly$.
We do not use the approximation of Eq.~(\ref{nkki_app}) 
for the MAC analysis\footnote{
We have used the approximation Eq.~(\ref{nkki_app}) 
in the previous report~\cite{talk}.
While the MAC analysis is somewhat affected by the approximation, 
the predictions of $m_t$ and $m_H$ are not. }. 


Here we comment on the UV-FP\footnote{
Whether or not the nontrivial UV-FP exists 
has been studied by lattice calculations
in the context of compactified extra dimensions or dimensional 
deconstruction~\cite{Ejiri:2000fc,Murata:2003uu}.}
discussed in Ref.~\cite{Hashimoto:2000uk},
which relies on the approximation of Eq.~(\ref{nkk_app}).
(See also Refs.~\cite{Agashe:2000nk,Dienes:2002bg}.)
Using Eqs.~(\ref{nkki_app}) and (\ref{nkk_app}),
Ref.~\cite{Hashimoto:2000uk} gave a RGE formula
\begin{equation}
 \mu \frac{d}{d \mu} \hat g_i = \frac{\delta}{2}\,\hat g_i
 + \left(1+\frac{\delta}{2}\right) \NDA \, b'_i\, \hat g_i^3, 
 \quad \mbox{for } \mu \gg R^{-1}, \label{rge_ED3}
\end{equation}
with $\NDA$ being the loop factor in $D$ dimensions,
\begin{equation}
  \NDA \equiv \frac{1}{(4\pi)^{D/2}\Gamma(D/2)} .
\end{equation}
The RGE coefficients $b'_i$ are
\begin{eqnarray}
  b'_3 &=& -11+\frac{\delta}{2}+\frac{4}{3}\cdot 2^{\delta/2} \, n_g ,\\
  b'_2 &=& -\frac{22}{3}+\frac{\delta}{3}+\frac{4}{3}\cdot 2^{\delta/2}\, n_g
  + \frac{1}{6}n_h , \\
  b'_Y &=& ~~\frac{20}{9}\cdot 2^{\delta/2} \, n_g + \frac{1}{6}n_h .
\end{eqnarray}
The RGE (\ref{rge_ED3}) leads to the UV-FP $g_{i*}$~\cite{Hashimoto:2000uk},
\begin{equation}
  g_{i*}^2 \NDA = \frac{1}{-(1+2/\delta)\,b'_i}, \label{UV-FP}
\end{equation}
for $b'_i < 0$.
In the case of $D=6,n_g=1,2,3$ and $D=8,n_g=1$,
$b'_3$ become negative and hence the bulk QCD has such a UV-FP\@.
On the other hand, in $D=10, n_g=1$ the bulk QCD coupling has 
a Landau pole at $4.26 R^{-1}$.

\subsection{Proper-time regularization}

There are several manners to define bulk gauge couplings.
In this subsection, we study bulk gauge couplings based on 
the proper-time (PT) regularization scheme with a matching condition 
slightly different from that in Ref.~\cite{Dienes:1998vh}.
In the PT-scheme, one-loop contributions of all KK modes are 
smoothly (exponentially) suppressed, whereas 
the contributions of KK modes heavier than the renormalization scale 
$\mu$ are sharply cut off in the truncated KK effective theory.
We here note that the Landau pole $\Ly$ is not so far from 
the compactification scale $R^{-1}$, particularly for $D=8,10$. 
Under such a situation, numerical differences between $\overline{\rm MS}$-, 
and PT-couplings may be significant.

We briefly present our definition of the PT-coupling and 
give the matching condition to the $\overline{\rm MS}$-coupling
through the effective charge. 
See Appendix B for details.

Our 4-dimensional PT-coupling $g_{\rm PT} (\Lambda)$ is 
a bare quantity defined at the cutoff $\Lambda$, in sharp contrast to 
the $\overline{\rm MS}$-coupling $g_{\overline{\rm MS}}(\mu)$, 
which is a renormalized quantity.
The effective charge $g_{\rm eff}(q_E^2)$ with Euclidean momentum 
$q_E^2 (\equiv -q^2)$ is obtained in the PT-scheme as
\begin{eqnarray}
  \left. \frac{1}{g_{\rm eff}^2(q_E^2)} \right|_{\rm PT}
 &=& \frac{1}{g_{\rm PT}^2 (\Lambda)}
    -\frac{b}{(4\pi)^2}\ln \frac{r q_E^2}{4 \Lambda^2}
    +\frac{c_0^{\rm PT}}{(4\pi)^2} \nonumber \\ &&
    -\sum_{m_{\vec n}^2 > 0}\Pi_{\rm KK}^{\rm PT}(q_E^2;m_{\vec n}^2)
 \label{eff_pt_q}
\end{eqnarray}
with $m_{\vec n}^2=|\vec n|^2/R^2$, $\vec n \equiv (n_1,n_2,\cdots,n_\delta)$
and a constant $r$ to be discussed later.
The coefficient $b$ of the logarithmic divergent term from zero modes
is given by
\begin{equation}
 b \equiv 
 -\frac{11}{3} \, C_A 
 +\frac{4 T_R}{3} \, n_f
 +\frac{T_R}{3} \, n_h ,
\end{equation}
where $C_A$ is the quadratic Casimir of the adjoint representation,
$T_R$ the trace of the product of two generator matrices, 
$\tr (T^a T^b)=T_R \delta^{ab}$, and 
$n_f$ the number of flavor of 4-component fermions. 
The constant term $c_0^{\rm PT}$ in Eq.~(\ref{eff_pt_q}) 
arises from zero modes and is given by
\begin{equation}
 c_0^{\rm PT} = -(4c_g-c_s)\,C_A 
 + ( 8 \, n_f \, c_f + n_h \,  c_s ) \, T_R ,
\end{equation}
with
\begin{equation}
  c_g \simeq 0.0365, \quad c_s \simeq 0.2343, \quad c_f \simeq -0.0501 .
\end{equation}
The KK mode summation of the vacuum polarization function 
$\Pi_{\rm KK}^{\rm PT}(q_E^2;m_{\vec n}^2)$ 
is calculated in the PT-scheme,
\begin{eqnarray}
\lefteqn{
  \sum_{|\vec n|^2>0}\Pi_{\rm KK}^{\rm PT}(q_E^2,m_{\vec n}^2) =}
  \nonumber \\ &&\hspace*{-4mm}
 \frac{C_A}{(4\pi)^2}\int_0^1dx\int_{r\Lambda^{-2}}^\infty\frac{dt}{t}
 \left[\,4-(2x-1)^2\,\right]e^{-tx(1-x)q_E^2}K_g(t) \nonumber \\ 
 &&\hspace*{-4mm}-
 \frac{\delta}{2} \, \frac{C_A}{(4\pi)^2}
 \int_0^1dx\int_{r\Lambda^{-2}}^\infty\frac{dt}{t}\,
 (2x-1)^2 e^{-tx(1-x)q_E^2} K_{gs}(t) \nonumber \\ 
 &&\hspace*{-4mm}-
  8T_R \, \frac{n_f}{(4\pi)^2}
 \int_0^1dx\int_{r\Lambda^{-2}}^\infty\frac{dt}{t}\,
 x(1-x) e^{-tx(1-x)q_E^2} K_f (t) \nonumber \\ 
 &&\hspace*{-4mm}-
  T_R \, \frac{n_h}{(4\pi)^2}
 \int_0^1dx\int_{r\Lambda^{-2}}^\infty\frac{dt}{t}\,
 (2x-1)^2 e^{-tx(1-x)q_E^2} K_h (t),  \nonumber \\ \label{pi_pt_q}
\end{eqnarray}
where $K_i$'s are defined as
\begin{equation}
  K_i(t) \equiv {\cal N}_i^{\delta,[n]_1}
  \left[\,\frac{1}{2}(\vartheta_3-1)\,\right] +
  {\cal N}_i^{\delta,[n]_2} \left[\,\frac{1}{2}(\vartheta_3-1)\,\right]^2 + 
  \cdots
\end{equation}
for $i=g,gs,f,h$, with
$\vartheta_3$ being the Jacobi $\vartheta_3$ function,
$\vartheta_3=\vartheta_3(it/(\pi R^2))$.
The definition of the factor ${\cal N}_i^{\delta,[n]_k}$ is given in
Appendix A.
(For values of ${\cal N}_i^{\delta,[n]_k}$ in $D=6,8,10$,
see Table \ref{tab1}--\ref{tab3}.)

On the other hand, the effective charge $g_{\rm eff}(q_E^2)$
is also calculated in the $\overline{\rm MS}$-scheme~\cite{Hashimoto:2000uk}, 
\begin{eqnarray}
  \left. \frac{1}{g_{\rm eff}^2(q_E^2)} \right|_{\overline{\rm MS}}
 &=&  
    \frac{1}{g^2_{\overline{\rm MS}}(\mu)}
   -\frac{b}{(4\pi)^2}\ln \frac{q_E^2}{\mu^2}
   +\frac{c_0^{\overline{\rm MS}}}{(4\pi)^2}
    \nonumber \\ &&
   -\sum_{0 < m_{\vec n}^2 < \mu^2}\Pi_{\rm KK}^{\overline{\rm MS}}
     (q_E^2=0;m_{\vec n}^2;\mu^2)
    \nonumber \\ &&
   -\sum_{m_{\vec n}^2 > 0}\overline{\Pi}_{\rm KK}^{\overline{\rm MS}}
     (q_E^2;m_{\vec n}^2), 
 \label{eff_ms_q}
\end{eqnarray}
where 
\begin{eqnarray}
\lefteqn{\hspace*{-1cm}
(4\pi)^2 \Pi_{\rm KK}^{\overline{\rm MS}}(q_E^2 = 0;m_{\vec n}^2; \mu^2)
= } \nonumber \\
&& -\frac{11}{3} C_A \ln \frac{m_{\vec n}^2}{\mu^2}
   +\frac{\delta}{6} C_A
     \ln \frac{m_{\vec n}^2}{\mu^2} \nonumber \\
&& +\frac{4 n_f T_R}{3}
      \ln \frac{m_{\vec n}^2}{\mu^2} 
   +\frac{n_h T_R}{3} \ln \frac{m_{\vec n}^2}{\mu^2},
\label{pi_ms}
\end{eqnarray}
and 
\begin{eqnarray}
\lefteqn{\hspace*{-1cm}
 (4\pi)^2 \overline{\Pi}_{\rm KK}^{\overline{\rm MS}}(q_E^2;m_{\vec n}^2)
= } \nonumber \\ &&
  - C_A \left[\,4 I_g^R(q_E^2;m_{\vec n}^2)-I_s^R(q_E^2;m_{\vec n}^2)\,\right]
    \nonumber \\ &&
  + \frac{\delta}{2} \, C_A \, I_s^R(q_E^2;m_{\vec n}^2)
    \nonumber \\ && 
  + 8 T_R \, n_f \, I_f^R(q_E^2;m_{\vec n}^2)
    \nonumber \\ && 
  + T_R \ n_h \, I_s^R(q_E^2;m_{\vec n}^2) ,
\end{eqnarray}
with 
\begin{eqnarray}
I_g^R &=& 
  \int_0^1 dx \ln \left(\,1+x(1-x)\frac{q_E^2}{m_{\vec n}^2}\,\right), \\
I_s^R &=& 
  \int_0^1 dx \, (2x-1)^2 
  \ln \left(\,1+x(1-x)\frac{q_E^2}{m_{\vec n}^2}\,\right), \\
I_f^R &=& 
  \int_0^1 dx \, x(1-x)\ln \left(\,1+x(1-x)\frac{q_E^2}{m_{\vec n}^2}\,\right).
\end{eqnarray}
The constant term from zero modes is given by
\begin{equation}
  c_0^{\overline{\rm MS}} = -\frac{67}{9} \, C_A 
  + \left(\frac{20}{9} \, n_f + \frac{8}{9} \, n_h \right)\, T_R.
\end{equation}

Now, we impose the matching condition to relate the PT- and
$\overline{\rm MS}$-couplings:
\begin{equation}
 \left. \frac{1}{g_{\rm eff}^2(q_E^2 \to 0)} \right|_{\rm PT} =
 \left. \frac{1}{g_{\rm eff}^2(q_E^2 \to 0)} \right|_{\overline{\rm MS}}
\end{equation}
i.e.,
\begin{eqnarray}
&& \frac{1}{g_{\rm PT}^2 (\Lambda)}
  -\frac{b}{(4\pi)^2}\ln \frac{r q_E^2}{4 \Lambda^2}
  +\frac{c_0^{\rm PT}}{(4\pi)^2}
 \nonumber \\ &&
  -\sum_{m_{\vec n}^2 > 0}\Pi_{\rm KK}^{\rm PT}(q_E^2 \to 0;m_{\vec n}^2)
 \nonumber \\ 
 &=& \frac{1}{g^2_{\overline{\rm MS}}(\mu)}
  -\frac{b}{(4\pi)^2}\ln \frac{q_E^2}{\mu^2}
  +\frac{c_0^{\overline{\rm MS}}}{(4\pi)^2}
 \nonumber \\ &&
  -\sum_{0 < m_{\vec n}^2 < \mu^2}\Pi_{\rm KK}^{\overline{\rm MS}}
    (q_E^2 = 0;m_{\vec n}^2;\mu^2).
 \label{match_cond}
\end{eqnarray}
Following Ref.~\cite{Dienes:1998vh},
we further impose a condition, 
$g_{\rm PT}(\Lambda) = g_{\overline{\rm MS}}(\mu=\Lambda)$
in $\Lambda \gg R^{-1}$.
The parameter $r$ of the PT-cutoff is then determined as
\begin{equation}
  r \equiv \pi X_\delta^{-2/\delta}, \quad
  X_\delta = \frac{\pi^{\delta/2}}{\Gamma(1+\delta/2)} . \label{r}
\end{equation}
The matching condition in Ref.~\cite{Dienes:1998vh} roughly 
corresponds to
\begin{equation}
  \frac{1}{g_{\rm PT}^2 (\Lambda)}-
  \sum_{m_{\vec n}^2 > 0}\Pi_{\rm KK}^{\rm PT}(q_E^2 \to 0;m_{\vec n}^2)
= \frac{1}{g_{\overline{\rm MS}}^2(\mu_0)}, \label{match_cond_DDG}
\end{equation}
with $\mu_0 \sim R^{-1}$.
There is an ambiguity associated with the matching scale $\mu_0$.
On the other hand, Eq.~(\ref{match_cond}) includes
effects of the finite parts $c_0^{\rm PT}, c_0^{\overline{\rm MS}}$ 
and hence is less ambiguous about the matching scale.

The matching condition~(\ref{match_cond}) determines
values of PT-couplings at $\Lambda = R^{-1} = 10$ TeV,
\begin{align*}
  \alpha_3^{\rm PT}(R^{-1})&=0.0731, \; \alpha_Y^{\rm PT}(R^{-1})=0.01071
  , \; (D=6), \\
  \alpha_3^{\rm PT}(R^{-1})&=0.0741, \; \alpha_Y^{\rm PT}(R^{-1})=0.01070
  , \; (D=8), \\
  \alpha_3^{\rm PT}(R^{-1})&=0.0752, \; \alpha_Y^{\rm PT}(R^{-1})=0.01013
  , \; (D=10), 
\end{align*}
while the corresponding values in the $\overline{\rm MS}$-scheme
at $\mu=R^{-1}=10$ TeV are
\begin{equation}
  \alpha_3^{\overline{\rm MS}}(R^{-1})=0.0726, \quad
  \alpha_Y^{\overline{\rm MS}}(R^{-1})=0.010724 , 
\end{equation}
where inputs are Eqs.~(\ref{qcd-mz}) and (\ref{su2-mz}).
At the compactification scale $R^{-1}$,
the scheme dependence between the PT-, $\overline{\rm MS}$-couplings 
are not significant for $D=6,8,10$.

In order to discuss the scheme dependence at
the scale beyond $R^{-1}$,
we define the dimensionless bulk gauge coupling in the PT-scheme 
$\hat g_{\rm PT}(\Lambda)$,
\begin{equation}
  \hat g_{\rm PT}^2(\Lambda) \equiv
  \frac{(2\pi R\Lambda)^\delta}{2^{\delta/2}}g_{\rm PT}^2(\Lambda) 
\end{equation}
in the same way as Eq.~(\ref{hat-g}), 
where $g_{\rm PT}(\Lambda)$ is the 4-dimensional coupling. 
In the next section, we will discuss 
the scheme dependence of the couplings, $\hat g_{\rm PT}(\Lambda)$ and 
$\hat g_{\overline{\rm MS}}(\mu=\Lambda)$: 
The scheme dependence near the Landau pole $\Ly$ is small
for $D=6,8$, while it is significant for $D=10$.

\section{Analysis of \lowercase{t}MAC scale}

In this section, we analyze the energy scale $\tMAC$ (tMAC scale)
where the top condensate is the MAC and only in the $\bar{t}t$-channel 
the binding strength exceeds the critical value, i.e.,
\begin{equation}
  \kappa_t(\tMAC) > \kappa_D^{\rm crit} > \kappa_b(\tMAC),
  \kappa_\tau(\tMAC), \cdots, \label{top-cond-sec3}
\end{equation}
where $\kappa_{t,b,\tau}$ are given in terms of
the SM gauge couplings in the bulk such as those shown in 
Figs.~\ref{fig-g-6d}(a), \ref{fig-g-8d}(a), \ref{fig-g-10d}(a)
and 
the critical binding strength $\kappa_D^{\rm crit}$ 
is determined by the ladder SD 
equation~\cite{Hashimoto:2000uk,Gusynin:2002cu}.
This is contrasted to the MAC analysis in 
Ref.~\cite{Arkani-Hamed:2000hv} where it was assumed that
$\hat g_3^2=\hat g_2^2=\hat g_1^2(=5/3 \hat g_Y^2)$
and that the value is large enough to trigger 
the \DxSB.
However, as we have shown in 
Figs.~\ref{fig-g-6d}(a), \ref{fig-g-8d}(a) and \ref{fig-g-10d}(a),
values of the couplings at the scale 
where $\hat g_3^2 \simeq \hat g_2^2 \simeq \hat g_1^2$ 
are not necessarily large, 
$\hat g_i^2 \NDA \sim 0.1$, for $D=6,8,10$.
It is thus highly non-trivial whether or not the tMAC scale $\tMAC$ exists.

As usual, we discuss the MAC based on 
the one-gauge-boson-exchange approximation~\cite{Raby:1979my}.
The binding strength
$\kappa$ of a $\bar{\psi}\chi$ channel is given by
\begin{eqnarray}
 \kappa (\mu) &\equiv& 
 -\hat g_3^2 (\mu)\NDA \mbox{\boldmath $T$}_{\bar{\psi}} \cdot
                  \mbox{\boldmath $T$}_{\chi}  \nonumber \\ &&
 -\hat g_2^2 (\mu)\NDA \mbox{\boldmath $T'$}_{\bar{\psi}} \cdot
                  \mbox{\boldmath $T'$}_{\chi}  \nonumber \\ &&
 -\hat g_Y^2 (\mu)\NDA Y_{\bar{\psi}} Y_{\chi}, 
\end{eqnarray}
where {\boldmath $T$}, {\boldmath $T'$} are 
the generators of $SU(3)_c,SU(2)_W$, and $Y$ is the hypercharge.
Noting the identity
\begin{equation}
 -\,\mbox{\boldmath $T$}_{\bar{\psi}} \cdot \mbox{\boldmath $T$}_{\chi}
 = \frac{1}{2}\left( C_2(\bar{\psi}) + C_2(\chi)
                    -C_2(\bar{\psi}\chi) \right),
\end{equation}
with $C_2(r)$ being the quadratic Casimir for the representation $r$
of the gauge group,
we calculate the binding strengths:
\begin{eqnarray}
  \kappa_t (\mu)&=& C_F \hat g_3^2 (\mu) \NDA 
                 + \frac{1}{9}\hat g_Y^2 (\mu) \NDA \label{k_t}, \\
  \kappa_b (\mu)&=& C_F \hat g_3^2 (\mu) \NDA 
                 - \frac{1}{18}\hat g_Y^2 (\mu) \NDA \label{k_b}, \\
  \kappa_\tau (\mu)&=& \phantom{C_F \hat g_3^2 (\mu) \NDA + \;\;\;}
                   \frac{1}{2} \hat g_Y^2 (\mu) \NDA \label{k_tau},
\end{eqnarray}
for the top, bottom and tau condensates, respectively, 
where $C_F (= 4/3)$ is the quadratic Casimir of 
the fundamental representation of $SU(3)_c$.
In the following analysis, we study these three channels. 

We next turn to the estimation of the critical binding strength 
$\kappa_D^{\rm crit}$.
Often applied is
the naive dimensional analysis (NDA)~\cite{Manohar:1983md,Chacko:1999hg},
$\kappa_D^{\rm crit} \sim 1$.
On the other hand, the value of $\kappa_D^{\rm crit}$ 
is fairly smaller than that of the NDA
in the approach of
the improved ladder SD equation~\cite{Hashimoto:2000uk},
\begin{equation}
 \kappa_6^{\rm crit} \simeq 0.122, \;\;
 \kappa_8^{\rm crit} \simeq 0.146, \;\;
 \kappa_{10}^{\rm crit} \simeq 0.163 . \label{kd_sd1}
\end{equation}
There are actually some issues in the estimation 
of $\kappa_D^{\rm crit}$ through 
the ladder SD equation: 
\begin{itemize}
\item Non-ladder corrections may push the value of $\kappa_D^{\rm crit}$
down as much as 1--20\%, as in the analysis of the 4-dimensional
walking technocolor~\cite{Appelquist:1988yc}.

\item 
Another momentum identification of 
the (dimensionful) running coupling constant 
requires the nonlocal gauge fixing~\cite{Georgi:1989cd}, which 
pushes up the value of $\kappa_D^{\rm crit}$ as~\cite{Gusynin:2002cu},
\begin{equation}
 \kappa_6^{\rm crit} = 0.15, \;\;
 \kappa_8^{\rm crit} \simeq 0.214, \;\;
 \kappa_{10}^{\rm crit} \simeq 0.278 . \label{kd_sd2}
\end{equation}
 
\item 
The values in Eqs.~(\ref{kd_sd1})--(\ref{kd_sd2}) were obtained 
under the assumption that the dimensionless bulk gauge 
coupling $\hat g$ is constant.
As shown in Figs.~\ref{fig-g-6d}(a), \ref{fig-g-8d}(a) and 
\ref{fig-g-10d}(a), 
all of bulk gauge couplings $\hat g_i^2(\mu)$ are monotonously 
increasing functions.
This means that the attractive force is overestimated under 
the simplification of $\hat g_i^2(\mu) =$ const., i.e.,
the momentum dependence of $\hat g$ makes
values of $\kappa_D^{\rm crit}$ larger than 
Eqs.~(\ref{kd_sd1})--(\ref{kd_sd2}).

\item Effects of the compactification were also neglected 
in the estimate of Eqs.~(\ref{kd_sd1})--(\ref{kd_sd2}).
If such effects are included,
the value of $\kappa_D^{\rm crit}$ becomes also larger.
\end{itemize}
Taking these issues into account,
we may regard the value of $\kappa_D^{\rm crit}$ in Eq.~(\ref{kd_sd1})
to be a reference value as the best compromise.

\begin{figure}[tbp]
  \begin{flushleft}
    \hspace*{1cm} {\Large (a)}
  \end{flushleft}
  \vspace*{-1cm}
  \begin{center}
  \resizebox{0.47\textwidth}{!}
            {\includegraphics{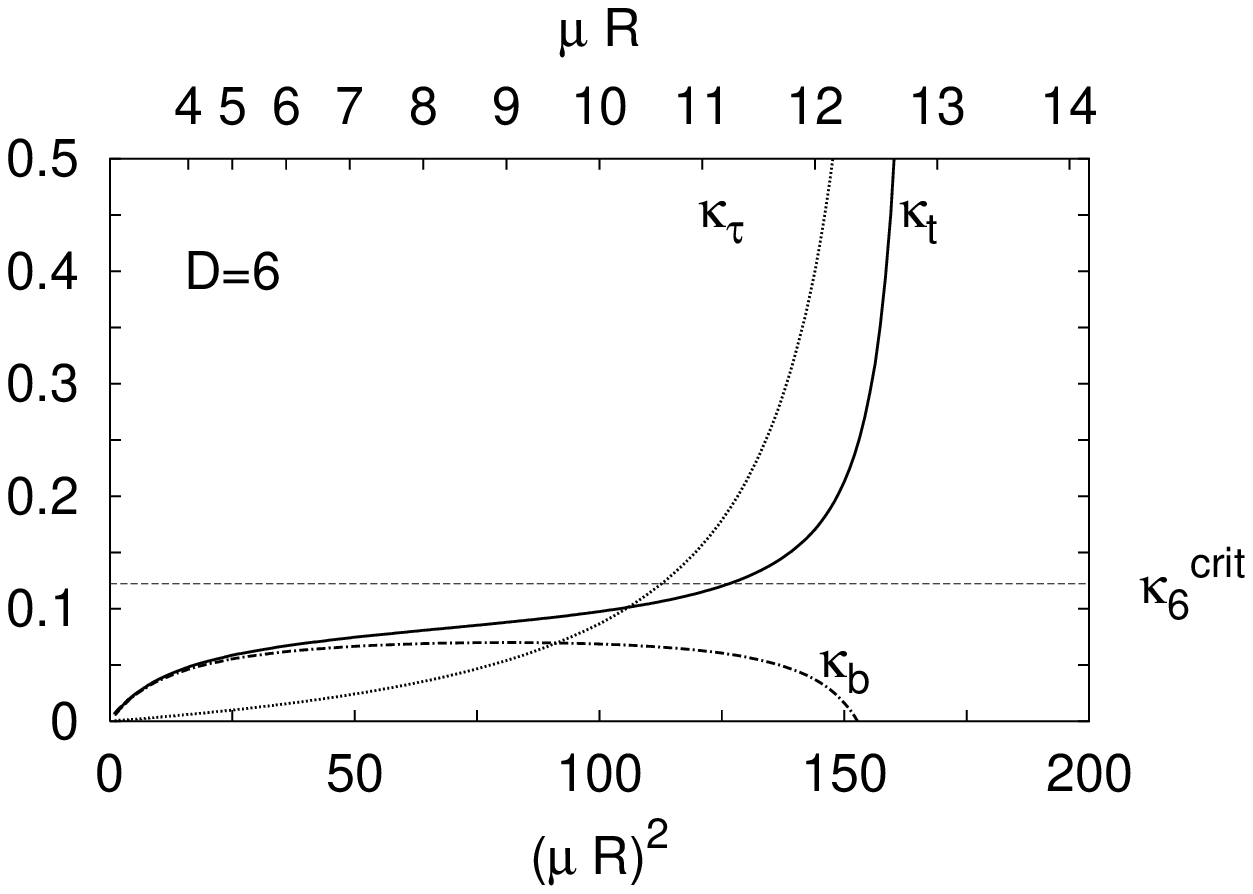}}\\[10mm]
  \end{center}
  \begin{flushleft}
    \hspace*{1cm} {\Large (b)}
  \end{flushleft}
  \vspace*{-1cm}
  \begin{center}
  \resizebox{0.47\textwidth}{!}
            {\includegraphics{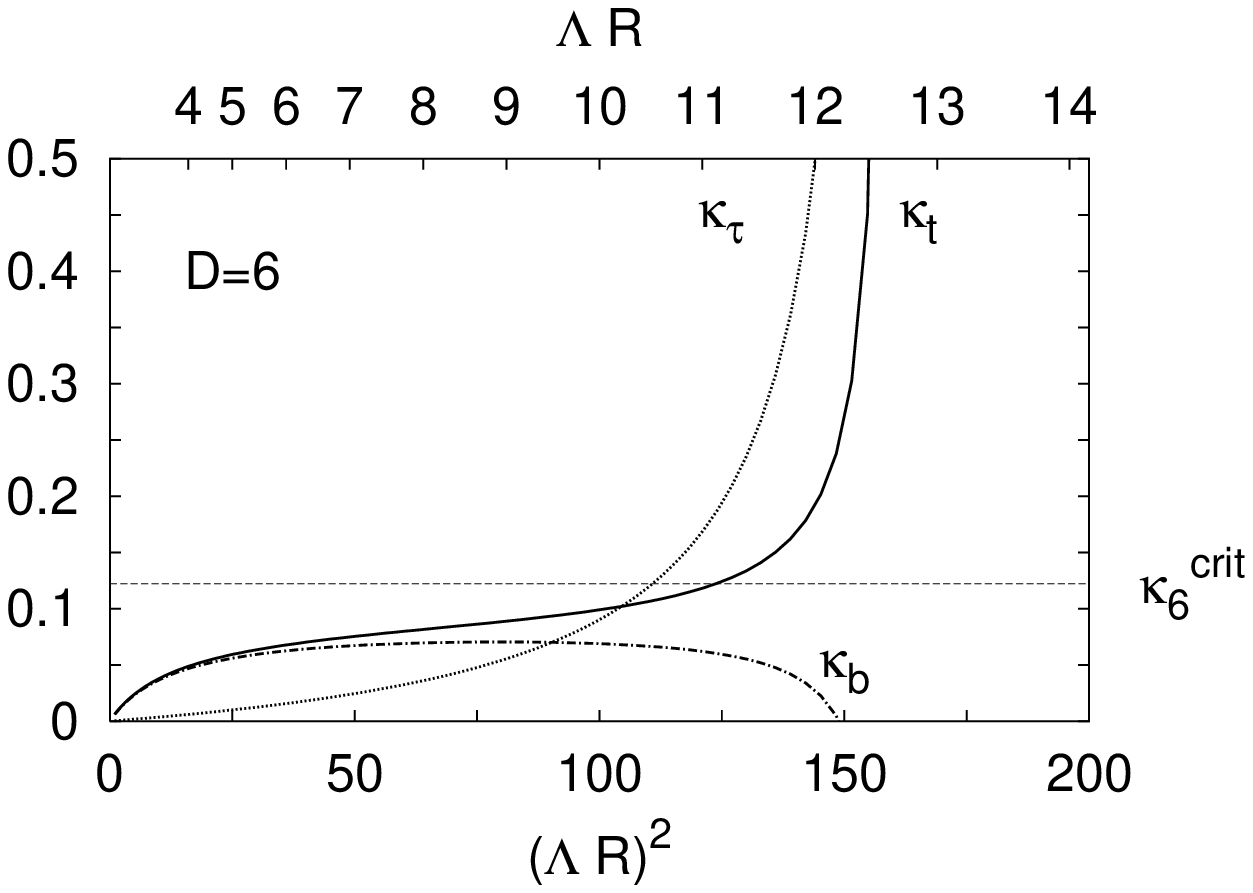}}
  \caption{Binding strengths $\kappa_{t,b,\tau}$ with 
           $D=6, n_g=1, R^{-1}=10$ TeV.
           The graphs (a) and (b) correspond to 
           the $\overline{\rm MS}$- and PT-schemes, 
           respectively. We also show $\kappa_6^{\rm crit}$ 
           of Eq.~(\ref{kd_sd1}) by the horizontal dashed line.
           \label{fig-tMAC-6D}}
  \end{center}
\end{figure}

\begin{figure}[tbp]
  \begin{flushleft}
    \hspace*{1cm} {\Large (a)}
  \end{flushleft}
  \vspace*{-1cm}
  \begin{center}
  \resizebox{0.47\textwidth}{!}
            {\includegraphics{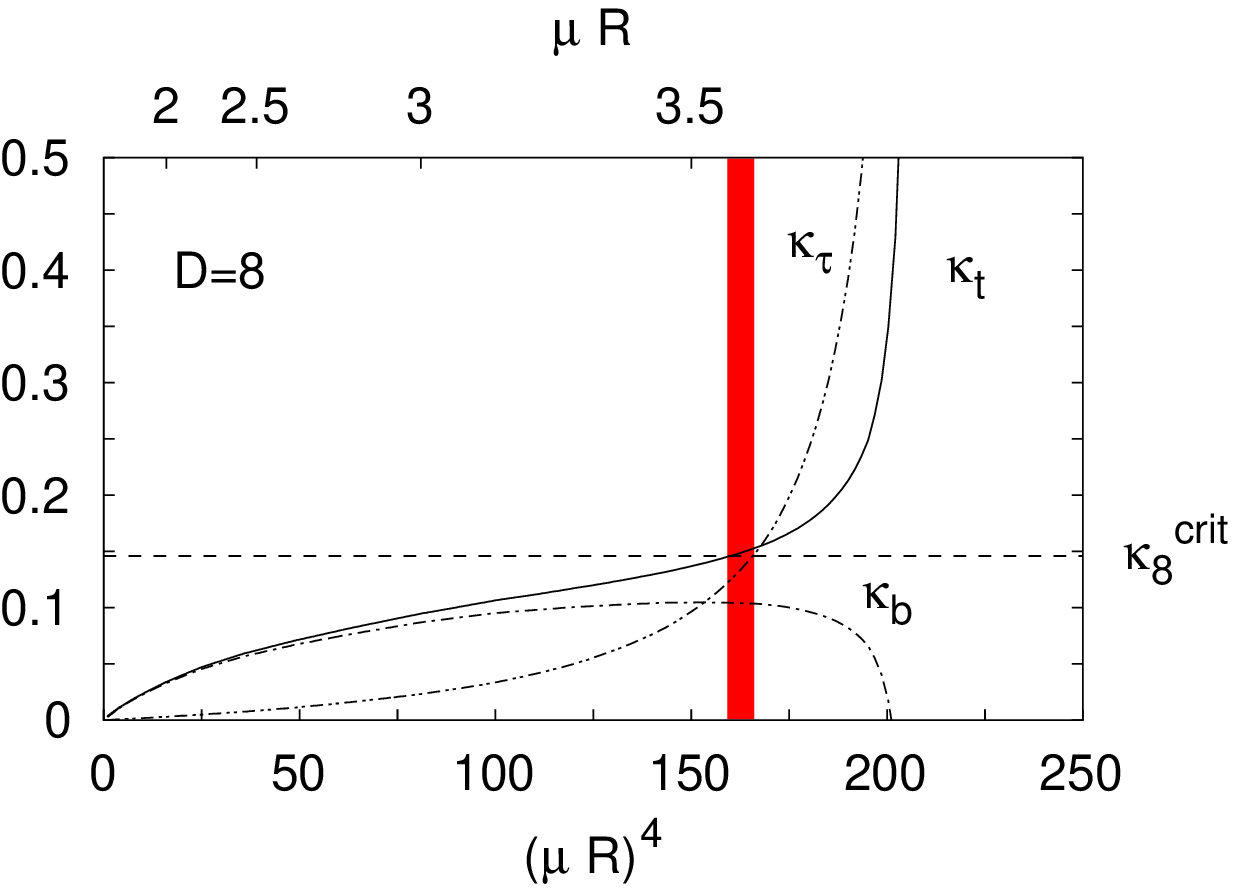}}
  \end{center}
  \begin{flushleft}
    \hspace*{1cm} {\Large (b)}
  \end{flushleft}
  \vspace*{-1cm}
  \begin{center}
  \resizebox{0.47\textwidth}{!}
            {\includegraphics{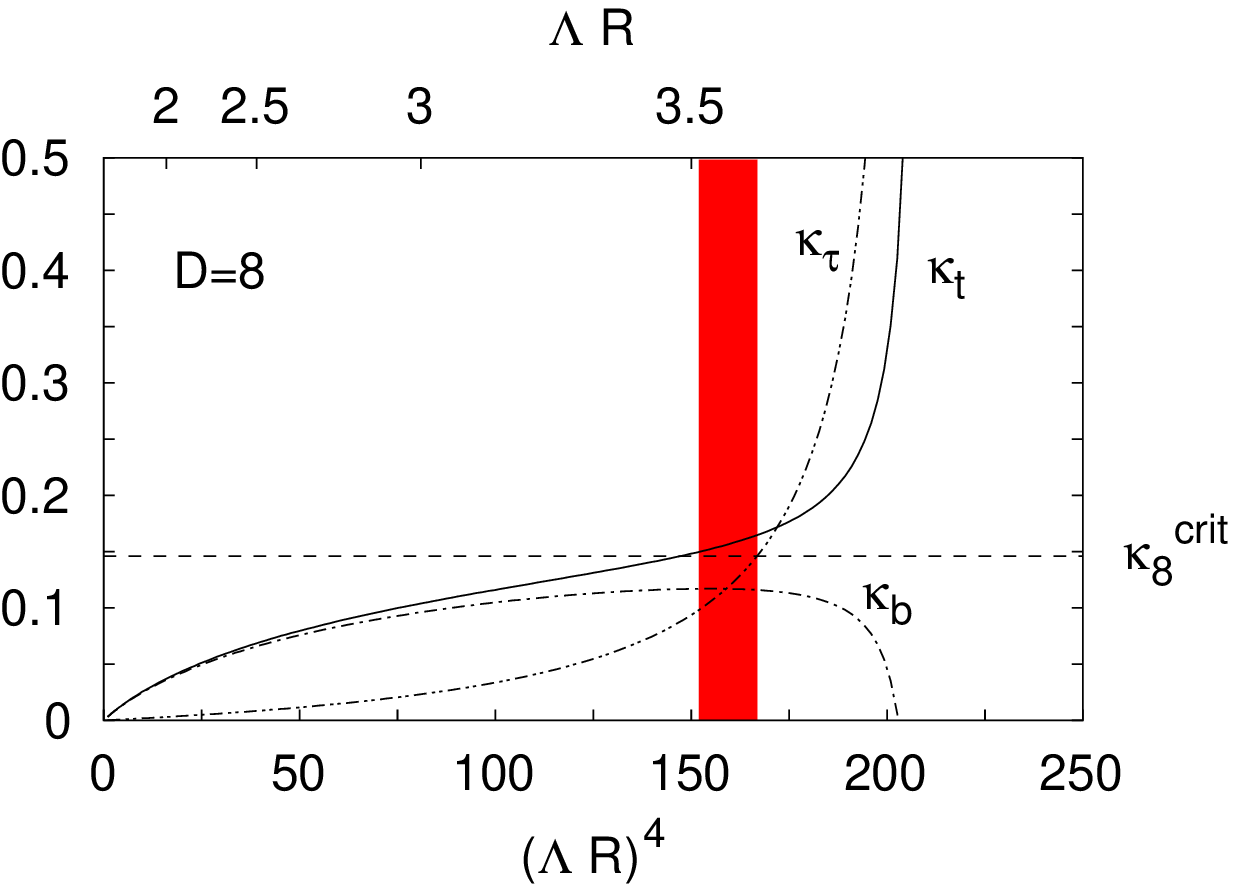}}
  \caption{Binding strengths $\kappa_{t,b,\tau}$ with 
           $D=8, n_g=1, R^{-1}=10$ TeV.
           The graphs (a) and (b) correspond with 
           the $\overline{\rm MS}$- and PT-schemes, 
           respectively. We also show $\kappa_8^{\rm crit}$ of
           Eq.~(\ref{kd_sd1}) by the horizontal dashed line.
           The shaded region is the tMAC scale $\tMAC$ satisfying 
           Eq.~(\ref{top-cond-sec3}). 
           \label{fig-tMAC-8D}}
  \end{center}
\end{figure}

\begin{figure}[tbp]
  \begin{flushleft}
    \hspace*{1cm} {\Large (a)}
  \end{flushleft}
  \vspace*{-1cm}
  \begin{center}
  \resizebox{0.47\textwidth}{!}
            {\includegraphics{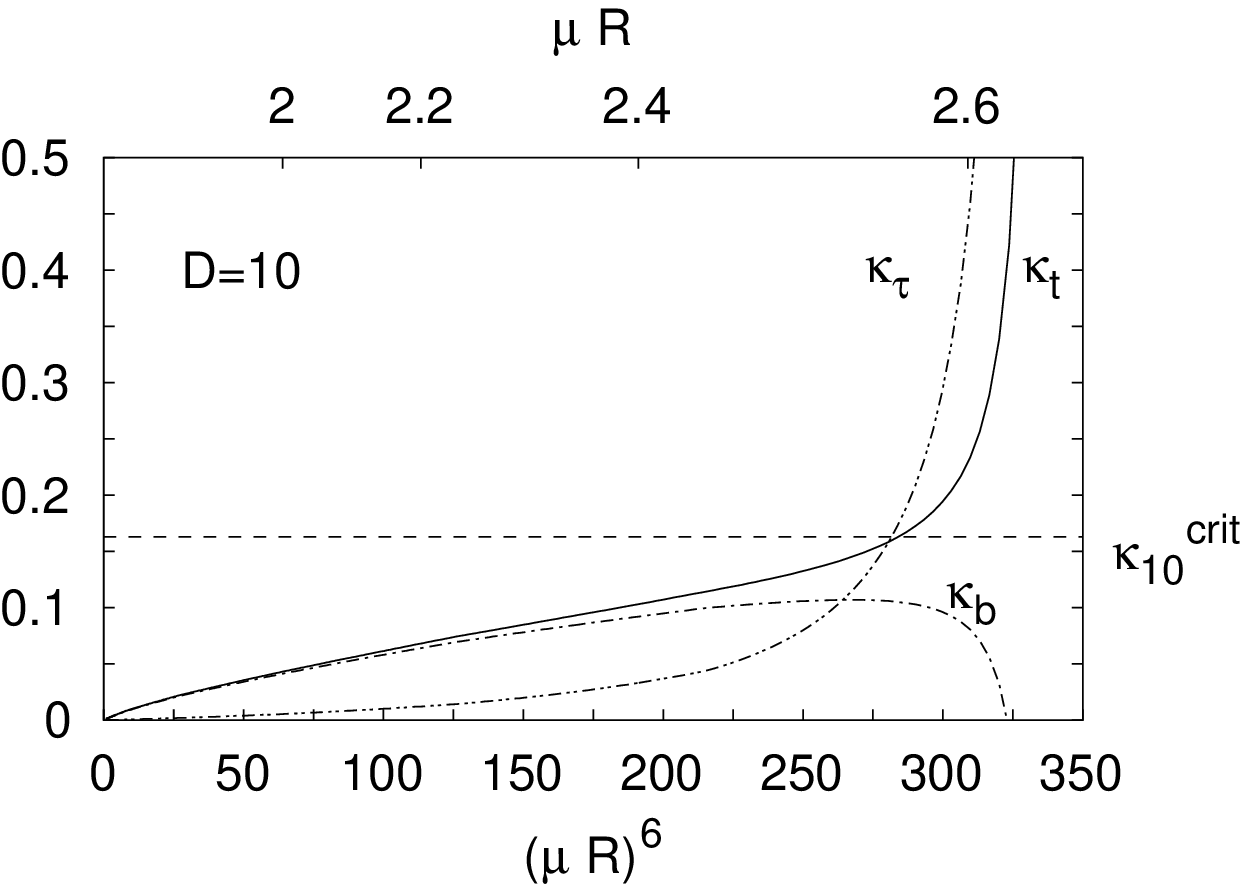}}
  \end{center}
  \begin{flushleft}
    \hspace*{1cm} {\Large (b)}
  \end{flushleft}
  \vspace*{-1cm}
  \begin{center}
  \resizebox{0.47\textwidth}{!}
            {\includegraphics{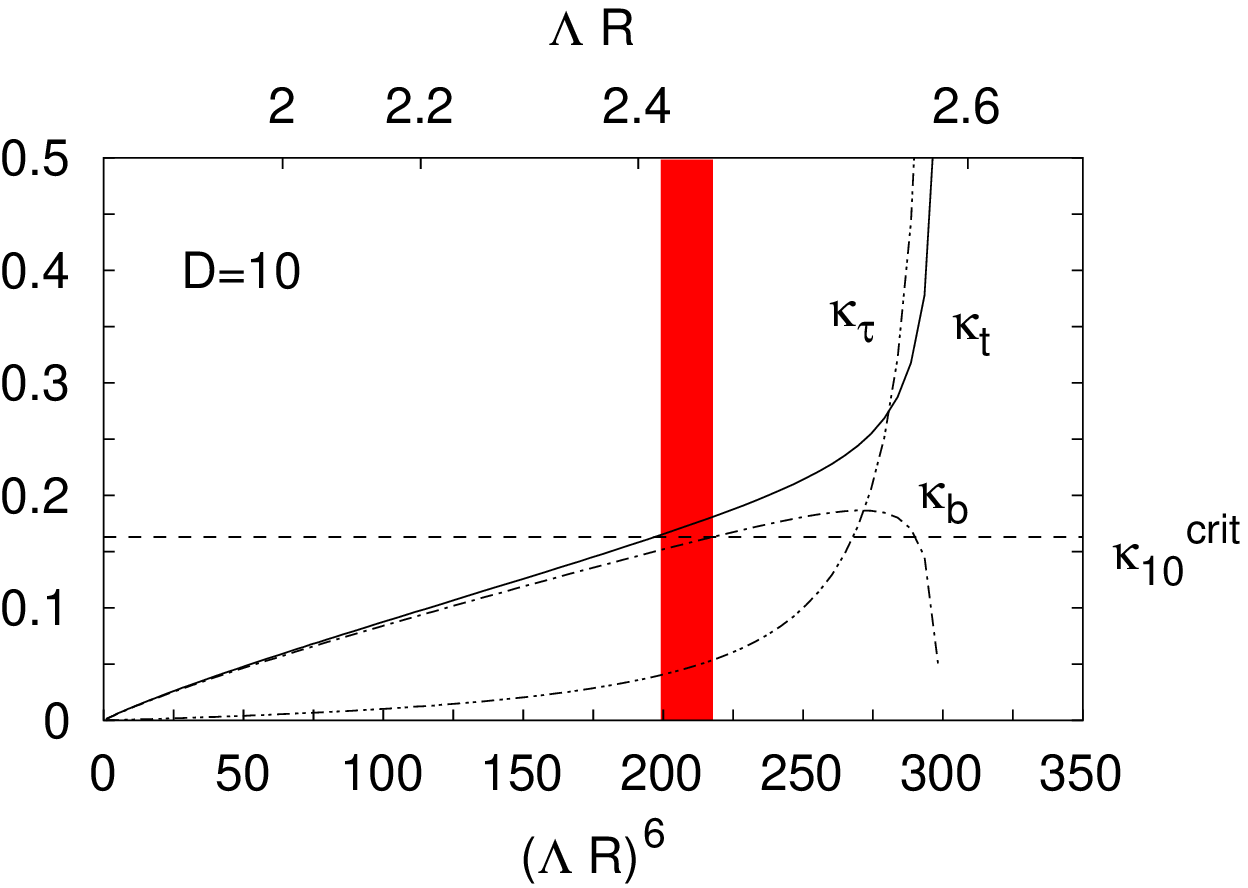}}
  \caption{Binding strengths $\kappa_{t,b,\tau}$ with 
           $D=10, n_g=1, R^{-1}=10$ TeV.
           The graphs (a) and (b) correspond with 
           the $\overline{\rm MS}$- and PT-schemes, 
           respectively. We also show $\kappa_{10}^{\rm crit}$ of
           Eq.~(\ref{kd_sd1}) by the horizontal dashed line.
           The shaded region is the tMAC scale $\tMAC$ satisfying 
           Eq.~(\ref{top-cond-sec3}). 
           \label{fig-tMAC-10D}}
  \end{center}
\end{figure}

Now, we are ready to investigate the tMAC scale.
We show behavior of the binding strengths $\kappa_{t,b,\tau}$ both in 
the $\overline{\rm MS}$- and PT-couplings in
Figs.~\ref{fig-tMAC-6D}--\ref{fig-tMAC-10D} to study the scheme dependence.

We first discuss the tMAC scale based on 
the $\overline{\rm MS}$-coupling. 
We compare attractive forces of top, bottom, and tau condensates
with 
$\kappa_D^{\rm crit}$
in 
Figs.~\ref{fig-tMAC-6D}(a), \ref{fig-tMAC-8D}(a) and 
 \ref{fig-tMAC-10D}(a) for $R^{-1}=10$ TeV\@.
As the value of $\kappa_D^{\rm crit}$ becomes larger, 
the tMAC scale gets squeezed.
 
For $D=6$, we find that {\it the tMAC scale is squeezed out}, 
if we use {\it the reference value} $\kappa_6^{\rm crit}$ 
Eq.~(\ref{kd_sd1}).
The conclusion is unchanged against varying the compactification scale
as far as $R^{-1} > 1$ TeV\@.
Actually, the tau condensation instead of the top condensation
becomes the MAC.
In order for the tMAC scale to survive,
the value of $\kappa_6^{\rm crit}$ should be lower than the 
reference value by about 20\% 
for $R^{-1}=$ 1--100 TeV, which is highly unlikely.
Thus the scenario does not work for $D=6$.

For $D=8$, on the other hand, the tMAC scale $\tMAC$
satisfying Eq.~(\ref{top-cond-sec3}) does exist, 
\begin{equation}
 \tMAC R = 3.55\,\mbox{--}\,3.59, \quad (R^{-1}=\mbox{10 TeV}),
\end{equation}
for the reference value of $\kappa_D^{\rm crit}$.
If we vary the compactification scale as $R^{-1}=1$--100 TeV, 
we find the tMAC scale,
\begin{equation}
  \tMAC R = 3.5 \, \mbox{--} \, 3.6.  \label{tmac} 
\end{equation}

For $D=10$, the tMAC scale does not exist for $R^{-1} = 10$ TeV.
If we admit $R^{-1} \leq 5$ TeV, 
we barely find the tMAC scale $\tMAC R = 2.56$--$2.57$ 
for $R^{-1} = 1$--$5$ TeV. 

We have studied the tMAC scale so far by using 
the $\overline{\rm MS}$-coupling.
In order to clarify the scheme dependence,
we also investigate the PT-coupling in the MAC analysis.
We show binding strengths $\kappa_{t,b,\tau}$ in the PT-scheme 
in Figs.~\ref{fig-tMAC-6D}(b), \ref{fig-tMAC-8D}(b) and 
\ref{fig-tMAC-10D}(b) for $R^{-1}=10$ TeV\@.  
We find that the scheme dependence is negligibly small for $D=6$. 
Although behavior of the binding strengths $\kappa_{t,b,\tau}$ 
for $D=8$ somewhat depends on the regularization scheme,
there does exist an overlapped region of the tMAC scale between 
the $\overline{\rm MS}$- and PT-schemes.
For $D=10$, however, the scheme dependence is rather large:
There is no overlap of the tMAC scale for $R^{-1}=1$--$100$ TeV.

In conclusion we have shown that the region of the tMAC scale is 
squeezed out for $D=6$, while
the tMAC scale does exist for $D=8$ without much ambiguity.
For $D=10$, we cannot obtain a reliable result because of 
the significant scheme dependence.

\section{Prediction of $m_t$ and $m_H$}

In this section, we calculate the top mass $m_t$ and 
the Higgs mass $m_H$ by using RGEs of the top Yukawa and 
Higgs quartic couplings with the compositeness 
conditions~\cite{Bardeen:1989ds} in a way used 
by ACDH~\cite{Arkani-Hamed:2000hv}.

Let us first recapitulate 
the compositeness conditions in Ref.~\cite{Arkani-Hamed:2000hv}: 
Assume that the top condensation takes place in the bulk.
Then, at the compositeness scale $\Lambda$, 
a scalar bound state $H$ (composite Higgs),
\begin{equation}
  H \sim (\bar{q}_L t_R), 
\end{equation}
is formed in the bulk, where $q_L [\equiv (t_L, b_L)^T]$ and $t_R$
are bulk fermions.
At this stage, the composite Higgs $H$ does not propagate in the bulk. 
Its kinetic term is expected to develop at the scale below $\Lambda$.
It is 
assumed that 
the effective theory 
below $\Lambda$ is described by the bulk SM\@. 
In the truncated KK effective theory of the bulk SM,
the compositeness conditions then read
\begin{equation}
 y(\mu) \to \infty , 
 \quad \frac{\lambda(\mu)}{y(\mu)^4} \to 0 , \quad
 (\mu \to \Lambda), \label{comp-cond}
\end{equation}
where $y$ and $\lambda$ denote the top Yukawa and Higgs quartic couplings,
respectively.

We shall use the above procedure to calculate $m_t$ and $m_H$.
In contrast to the ACDH analysis, however,
we have shown existence of the tMAC scale for $D=8$
without much ambiguity and hence
the extra dimension scenario of the TMSM is actually possible.
Furthermore, while the compositeness scale $\Lambda$ 
in Ref.~\cite{Arkani-Hamed:2000hv} was treated 
as a free parameter to be adjusted for reproducing
the experimental value of $m_t$, 
we identify the compositeness scale $\Lambda$ with the tMAC scale $\tMAC$,
\begin{equation}
  \Lambda = \tMAC,
\end{equation}
which is no longer an adjustable parameter but constrained as Eq.~(\ref{tmac}).
Thus we can test our model by comparing the predicted $m_t$ with 
the experimental value.

Within the truncated KK effective theory, 
RGEs for $y$ and $\lambda$
read~\footnote{
Certain of the terms in Eq.~(\ref{rge_y-ED}) are missing in the RGEs 
of Ref.~\cite{Arkani-Hamed:2000hv}.}
\begin{align}
&(4\pi)^2 \mu \frac{d y}{d \mu} = \beta_y^{\rm SM} + \beta_y^{\rm KK},  
 \label{rge_y} \\[2mm]
&(4\pi)^2 \mu \frac{d \lambda}{d \mu} =
  \beta_\lambda^{\rm SM} + \beta_\lambda^{\rm KK},  \label{rge_lam}
\end{align}
where
\begin{eqnarray}
\beta_y^{\rm SM} &=& 
  y \left[\,\left(3+\frac{3}{2}\right)y^2 
  - 8 g_3^2 - \frac{9}{4} g_2^2 
  -\frac{17}{12} g_Y^2 \,\right] , \\[2mm]
\beta_y^{\rm KK} &=& 
 \left( 6 \NKKf +\frac{3}{2} \NKKs \right)\,y^3 
 \nonumber \\ && \quad
 -\NKKg \left( 8 g_3^2 + \frac{9}{4} g_2^2 + \frac{17}{12} g_Y^2 \right)\,y
  \nonumber \\ && \quad 
 - \delta \NKKb
 \left(\frac{4}{3} g_3^2 - \frac{3}{8} g_2^2 - \frac{1}{72} g_Y^2 \right)\,y,
 \label{rge_y-ED} \\
\beta_\lambda^{\rm SM} &=&
  12 \left(\lambda y^2-y^4\right) + 12\lambda^2 \nonumber \\ && \quad 
  +\frac{3}{4}(3 g_2^4 + 2g_2^2 g_Y^2 + g_Y^4) \nonumber \\ && \quad
  -3(3g_2^2+g_Y^2)\lambda, \\
\beta_\lambda^{\rm KK} &=&
   24 \NKKf \left(\lambda y^2-y^4\right) 
 + 12 \NKKs \lambda^2  \nonumber \\ && \hspace*{-0.5cm}
 + \NKKg \left[\,
   \frac{3}{4}\left(3 g_2^4 + 2g_2^2 g_Y^2 + g_Y^4\right)
 - 3(3g_2^2+g_Y^2)\lambda\,\right] \nonumber \\ && \hspace*{-0.5cm}
 + \frac{\delta}{4} \NKKb (3 g_2^4 + 2g_2^2 g_Y^2 + g_Y^4). 
\end{eqnarray}
Note that $\beta_{y,\lambda}^{\rm SM}$ and $\beta_{y,\lambda}^{\rm KK}$
stand for contributions of zero modes and KK modes, respectively.
We calculate the RGEs by using the UV-BCs Eq.~(\ref{comp-cond})
and determine $m_t$ and $m_H$ through the conditions,
\begin{equation}
 m_t =  \frac{v}{\sqrt{2}} \, y(m_t), \quad 
 m_H = v \sqrt{\lambda(m_H)}, \label{rge-analysis}
\end{equation}
with $v=246$ GeV. 

\begin{figure}[tbp]
  \begin{center}
  \resizebox{0.47\textwidth}{!}
            {\includegraphics{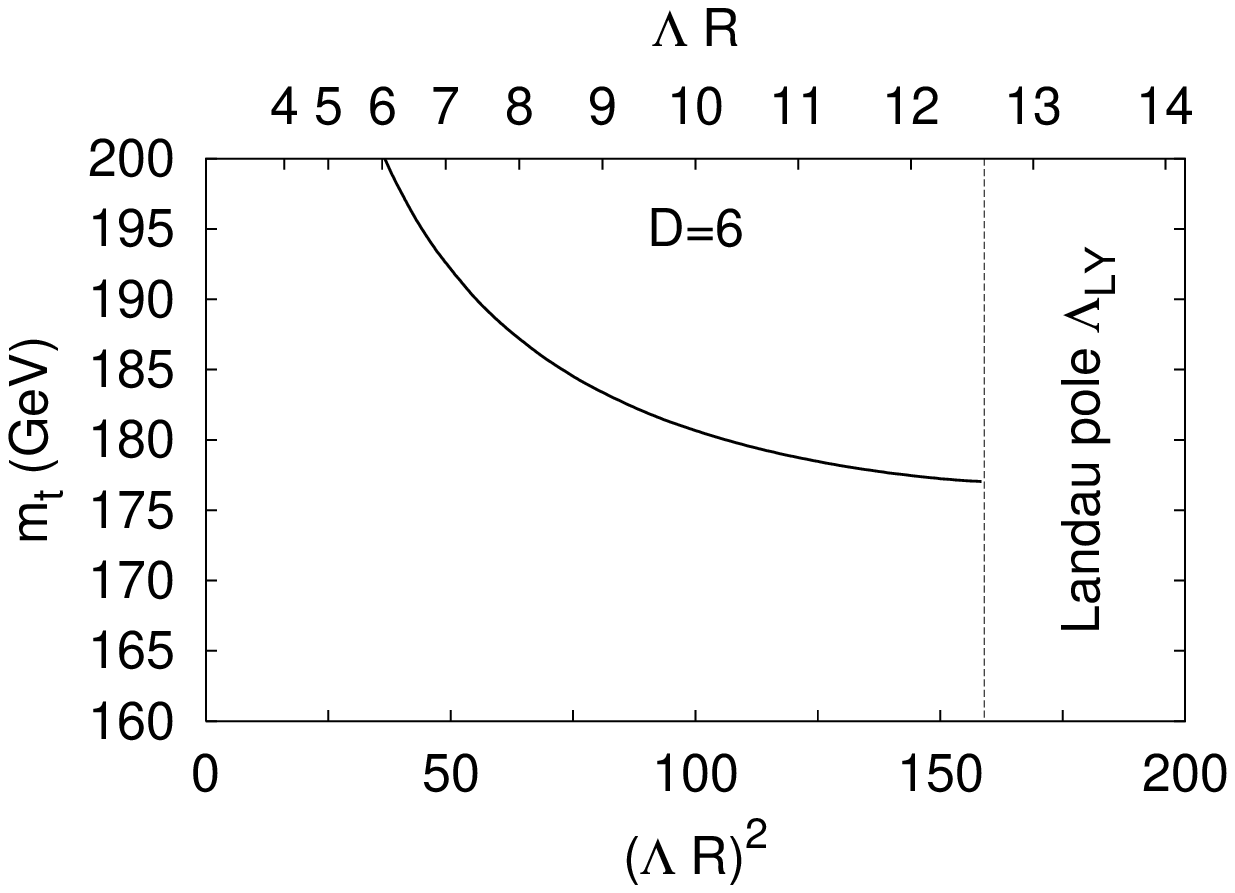}} \\[3mm]
  \resizebox{0.47\textwidth}{!}
            {\includegraphics{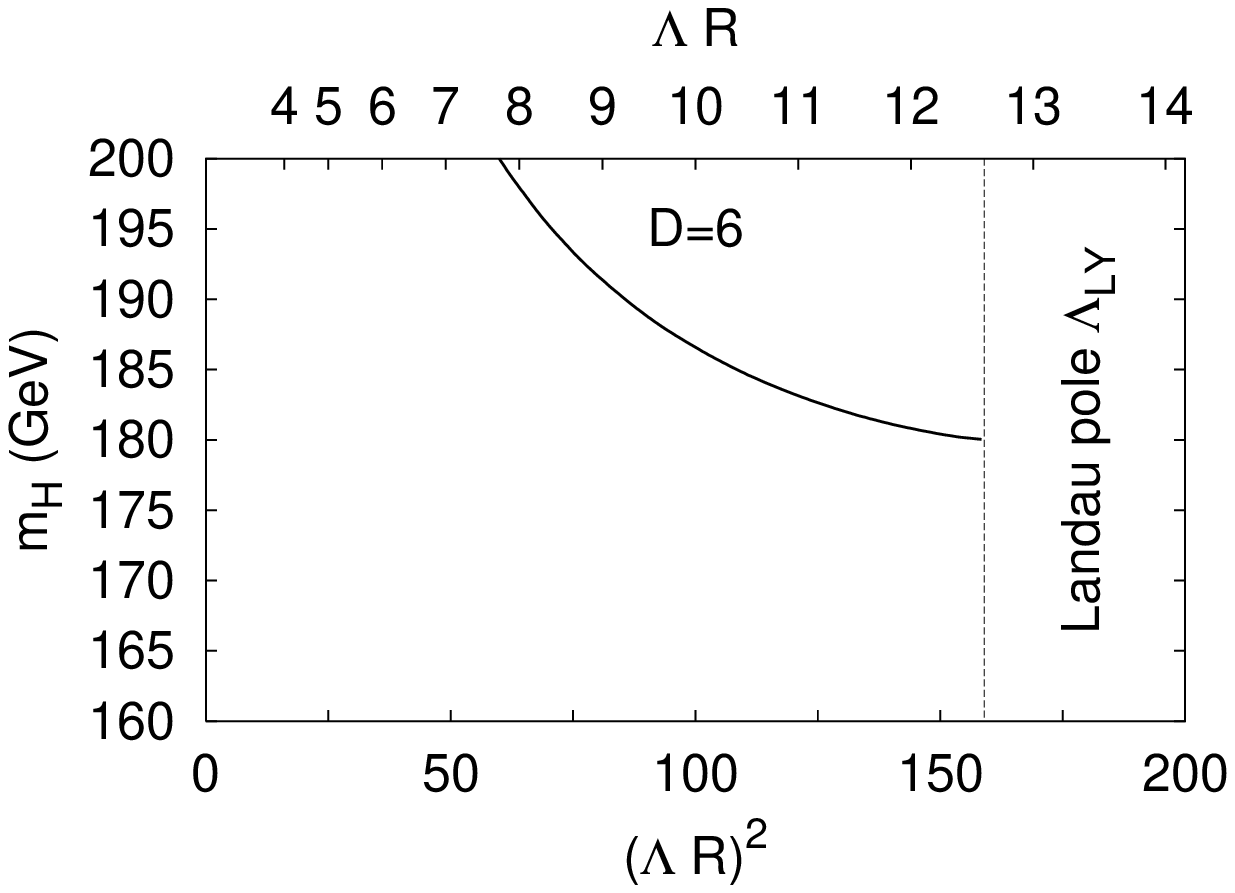}}
  \caption{Solutions $m_t$ and $m_H$ of Eq.~(\ref{rge-analysis})
           with the compositeness conditions~(\ref{comp-cond})
           for $D=6,R^{-1}=10$ TeV.
           The dashed vertical line represents the Landau pole $\Ly$.
           \label{mt-mh-6D}}
  \end{center}
\end{figure}

\begin{figure}[tbp]
  \begin{center}
  \resizebox{0.47\textwidth}{!}
            {\includegraphics{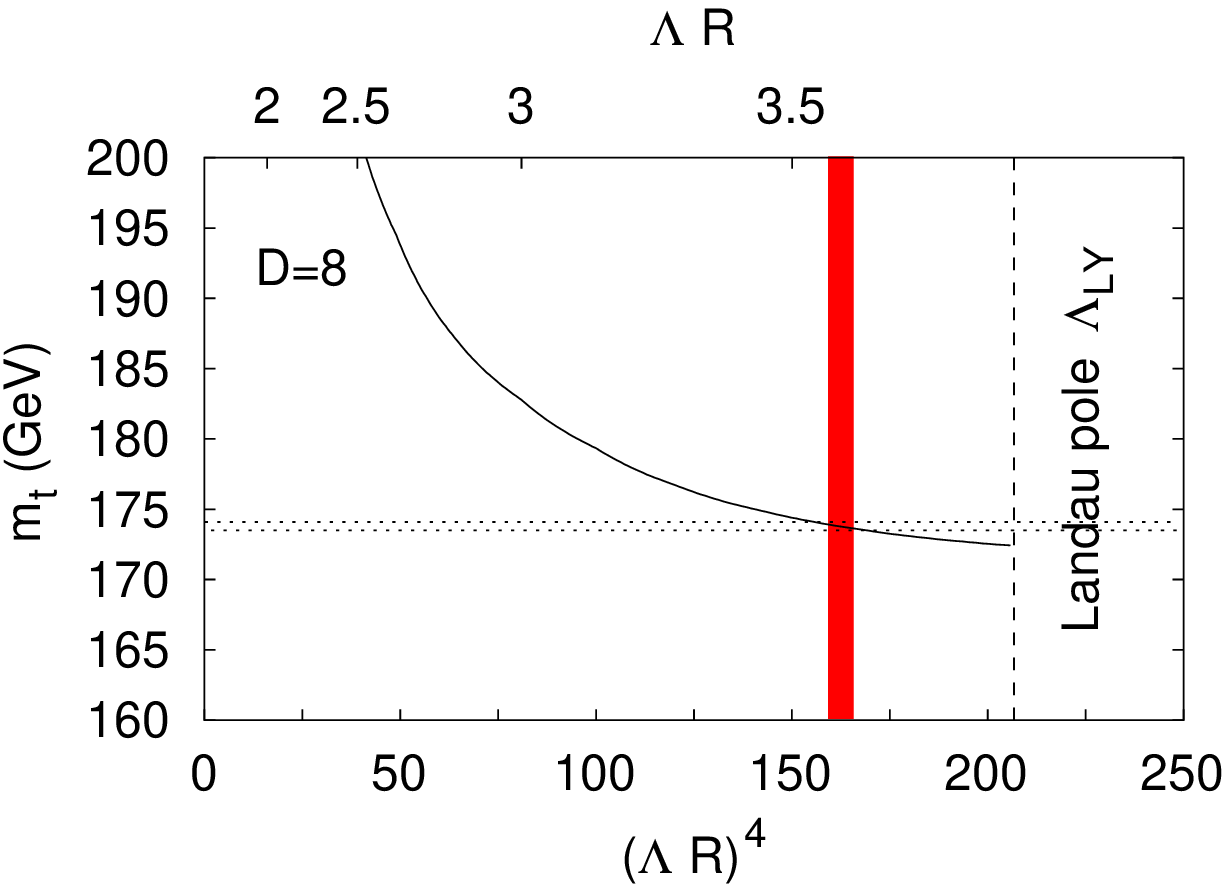}} \\[3mm]
  \resizebox{0.47\textwidth}{!}
            {\includegraphics{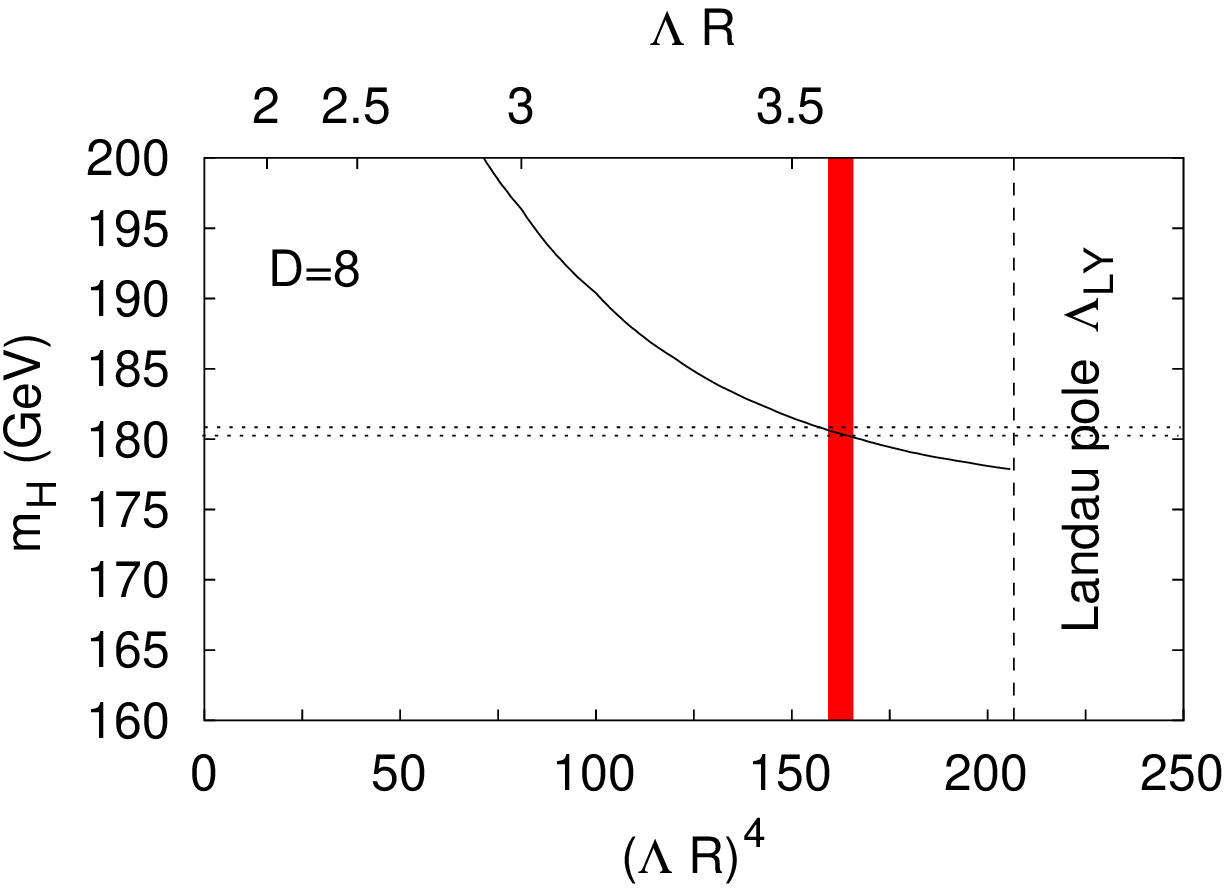}}
  \caption{Solutions $m_t$ and $m_H$ of Eq.~(\ref{rge-analysis})
           with the compositeness conditions~(\ref{comp-cond})
           for $D=8,R^{-1}=10$ TeV.
           The dashed vertical line represents the Landau pole $\Ly$.
           The shaded region is the tMAC scale $\tMAC$ satisfying 
           Eq.~(\ref{top-cond-sec3}). 
           \label{mt-mh}}
  \end{center}
\end{figure}

\begin{figure}[tbp]
  \begin{center}
  \resizebox{0.47\textwidth}{!}
            {\includegraphics{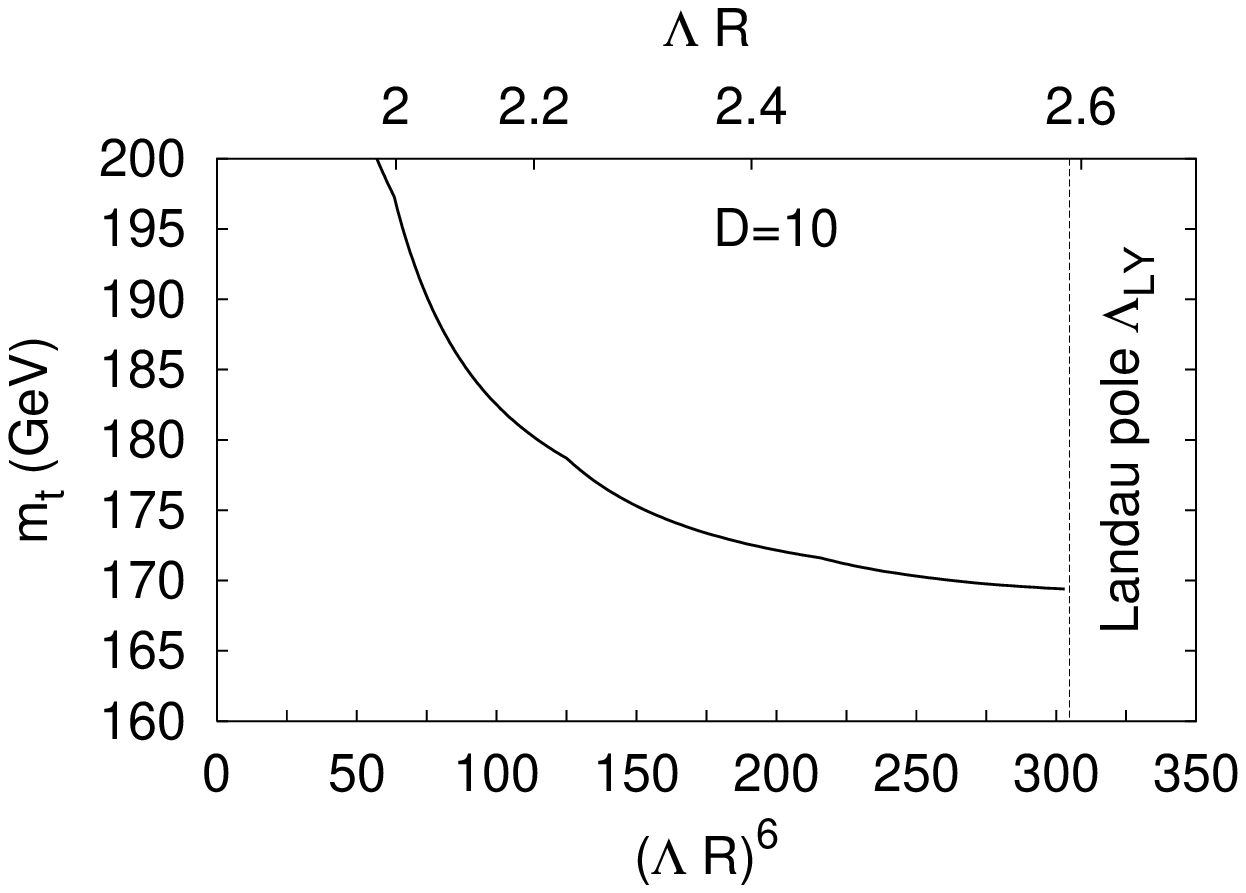}} \\[3mm]
  \resizebox{0.47\textwidth}{!}
            {\includegraphics{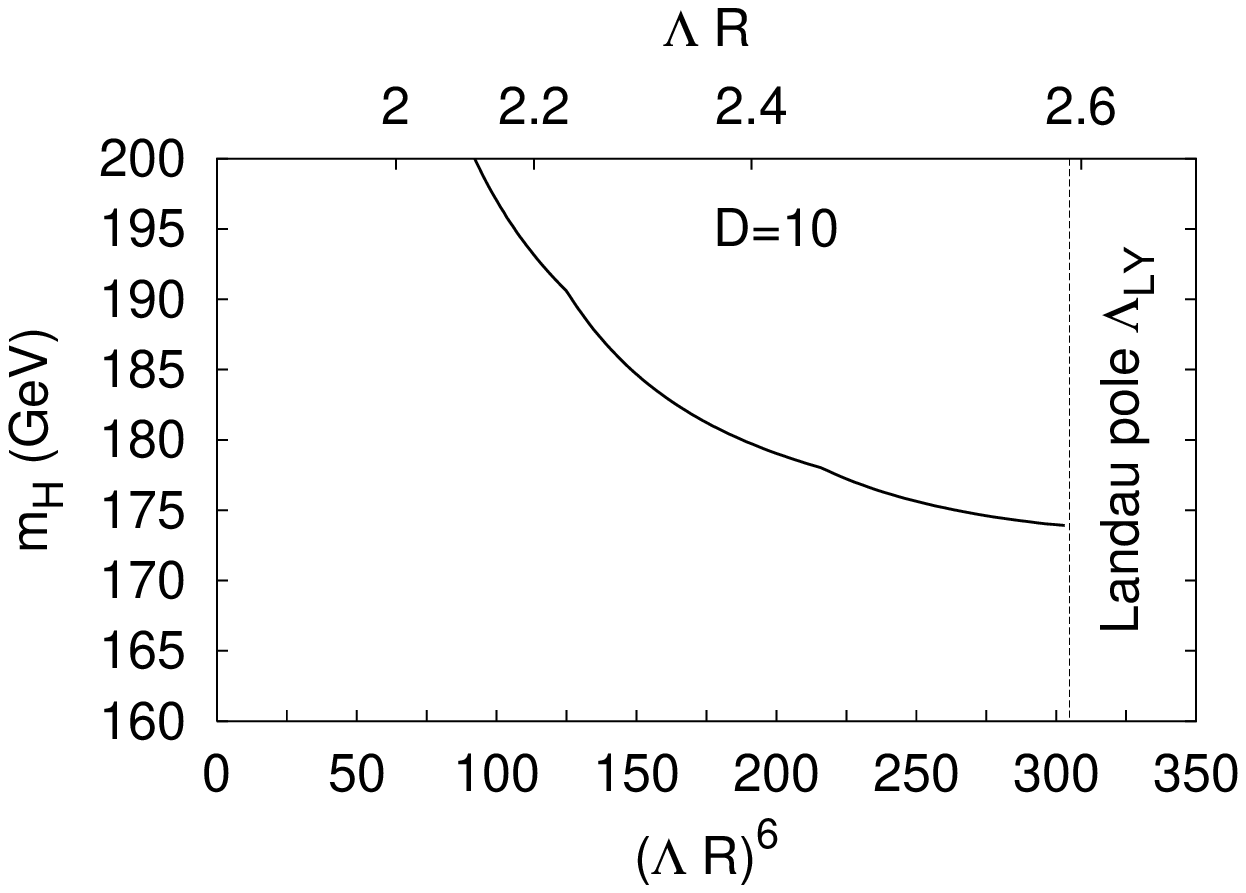}}
  \caption{Solutions $m_t$ and $m_H$ of Eq.~(\ref{rge-analysis})
           with the compositeness conditions~(\ref{comp-cond})
           for $D=10,R^{-1}=10$ TeV.
           The dashed vertical line represents the Landau pole $\Ly$.
           \label{mt-mh-10D}}
  \end{center}
\end{figure}

We show results of $m_t$ and $m_H$ 
in Figs.~\ref{mt-mh-6D}, \ref{mt-mh}, \ref{mt-mh-10D} for $D=6,8,10$, 
$R^{-1}=10$ TeV 
for various values of the compositeness scale $\Lambda$.
As we have shown in Sec.~III,
the tMAC scale does exist only for $D=8$ without much ambiguity, 
$\tMAC R = 3.5$--$3.6$.
Identifying $\Lambda$ with $\tMAC$, 
we depict the region of the tMAC scale for $D=8$ by
the shaded area in Fig.~\ref{mt-mh}. 
There is no shaded region in Figs.~\ref{mt-mh-6D} and 
\ref{mt-mh-10D}, because a sensible tMAC scale is absent for $D=6,10$.
For $D=8$ we predict
\begin{equation}
  m_t = 172-175 \; \mbox{GeV}, \label{8D-mt}
\end{equation}
and
\begin{equation}
  m_H=176-188 \; \mbox{GeV}, \label{8D-mh}
\end{equation}
for the range of the compactification scale
$R^{-1}=$ 1--100 TeV.
The uncertainties in Eqs.~(\ref{8D-mt}) and (\ref{8D-mh}) 
also include 
error of 
$\alpha_3(M_Z)=0.1172 \pm 0.0020$~\cite{PDG}.

Remarkably, the prediction Eq.~(\ref{8D-mt}) for $m_t$ 
is consistent with the reality,
the $\overline{\rm MS}$-mass $m_t=164.7 \pm 4.9$ GeV
which is
calculated from the observed value of the pole mass,
$174.3 \pm 5.1$ GeV~\cite{PDG}. 
It should be emphasized that 
the compositeness scale $\Lambda$ in Ref.~\cite{Arkani-Hamed:2000hv} 
is an arbitrary parameter and was adjusted 
to reproduce the experimental value of the top mass. 
On the contrary, our compositeness scale $\Lambda$ 
is fixed by the tMAC scale $\tMAC$
by requiring that the top quark condensation actually takes place,
while other condensations do not.
Hence the top mass as well as the Higgs mass is the prediction in 
our approach.

Compared with the prediction of Ref.~\cite{Arkani-Hamed:2000hv}, 
$m_H \sim 170-230$ GeV for $D=8$,
Eq.~(\ref{8D-mh}) is 
determined more sharply.
This is because our compositeness scale $\Lambda$ is given by
the tMAC scale $\tMAC$ which is more severely constrained 
than $\Lambda$ of Ref.~\cite{Arkani-Hamed:2000hv}, 
the value determined from the top mass within
3$\sigma$ of the experimental value.

We now discuss the reason why the value of Eq.~(\ref{8D-mt}) is 
significantly smaller than that of the original TMSM in four dimensions 
which predicted $m_t \gtrsim 200$ GeV.
Let us consider a simplified RGE for $y$ neglecting 
the electroweak gauge interactions: 
\begin{eqnarray}
 (4\pi)^2 \mu \frac{d y}{d \mu} &=& \left(
 2N_c \NKKf + \frac{3}{2} \NKKs \right)\, y^3 
 \nonumber \\ && 
 - (6 \NKKg + \delta \NKKb ) \, C_F \, y \, g_3^2 ,
 \label{y_app}
\end{eqnarray}
with $N_c=3$, $C_F=4/3$.
Using Eq.~(\ref{nkki_app}) as an approximation for $\NKKi$ and 
equating the R.H.S. of Eq.~(\ref{y_app}) to zero, 
we find the quasi IR fixed point $y_{\rm qIR}(\mu)$~\cite{Hill:1980sq}
\begin{equation}
  y_{\rm qIR}(\mu) =  g_3 (\mu) \cdot 
  \sqrt{\frac{C_F (6+\delta)}{2^{\delta/2} N_c + \frac{3}{2}}} .
  \label{qIR}
\end{equation}
The value of Eq.~(\ref{qIR}) 
decreases as $\delta (\equiv D-4)$ increases at $\mu = R^{-1}$, since
the value of $g_3(R^{-1})$ is determined 
within the SM independently of $\delta$.
In the original TMSM with $D=4$
the prediction of $m_t$ is 
governed by Eq.~(\ref{qIR}) with $\delta=0$~\cite{Bardeen:1989ds}.
As a result, 
the prediction of $m_t$ with $\delta > 0$ is substantially lower than 
that of the original TMSM with $\delta = 0$.

\begin{figure}[tbp]
  \begin{center}
  \resizebox{0.47\textwidth}{!}
            {\includegraphics{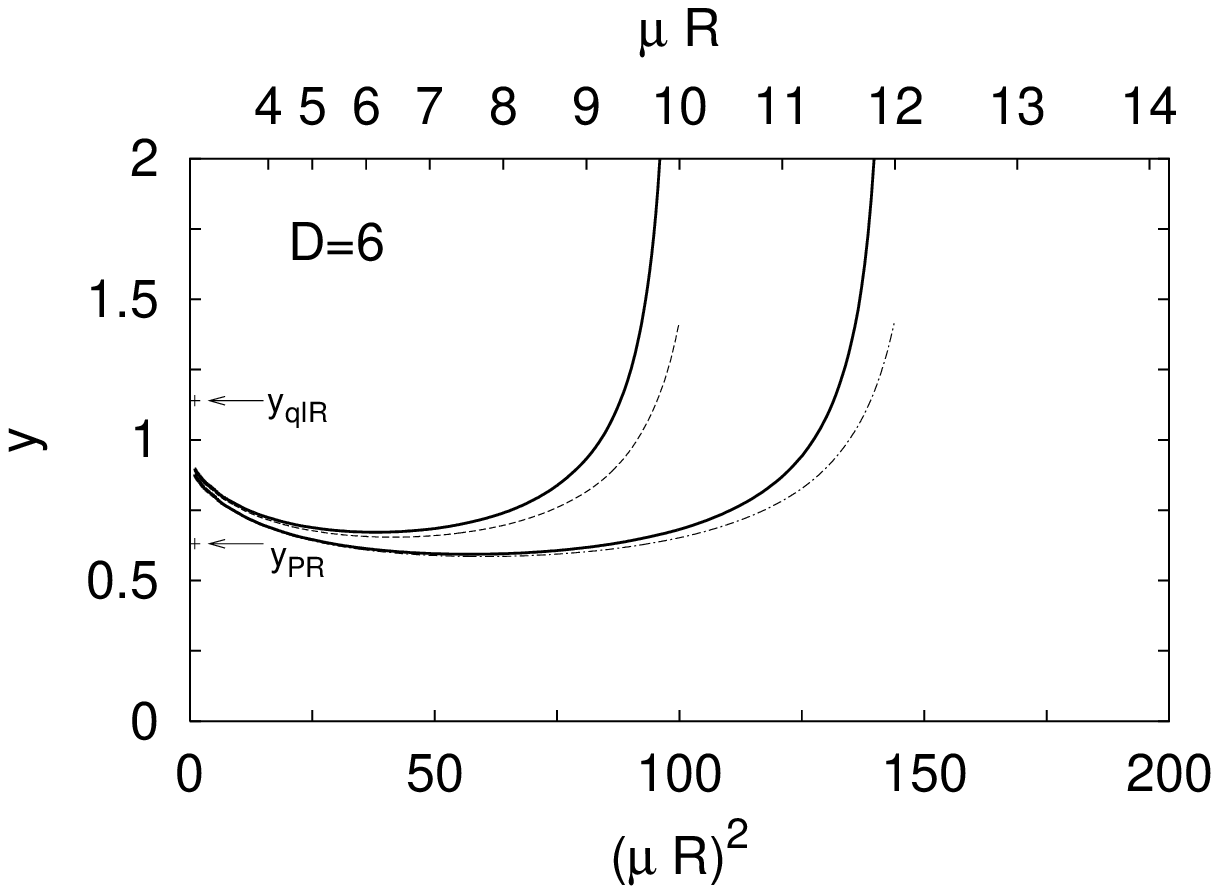}}\\[6mm]
  \resizebox{0.47\textwidth}{!}
            {\includegraphics{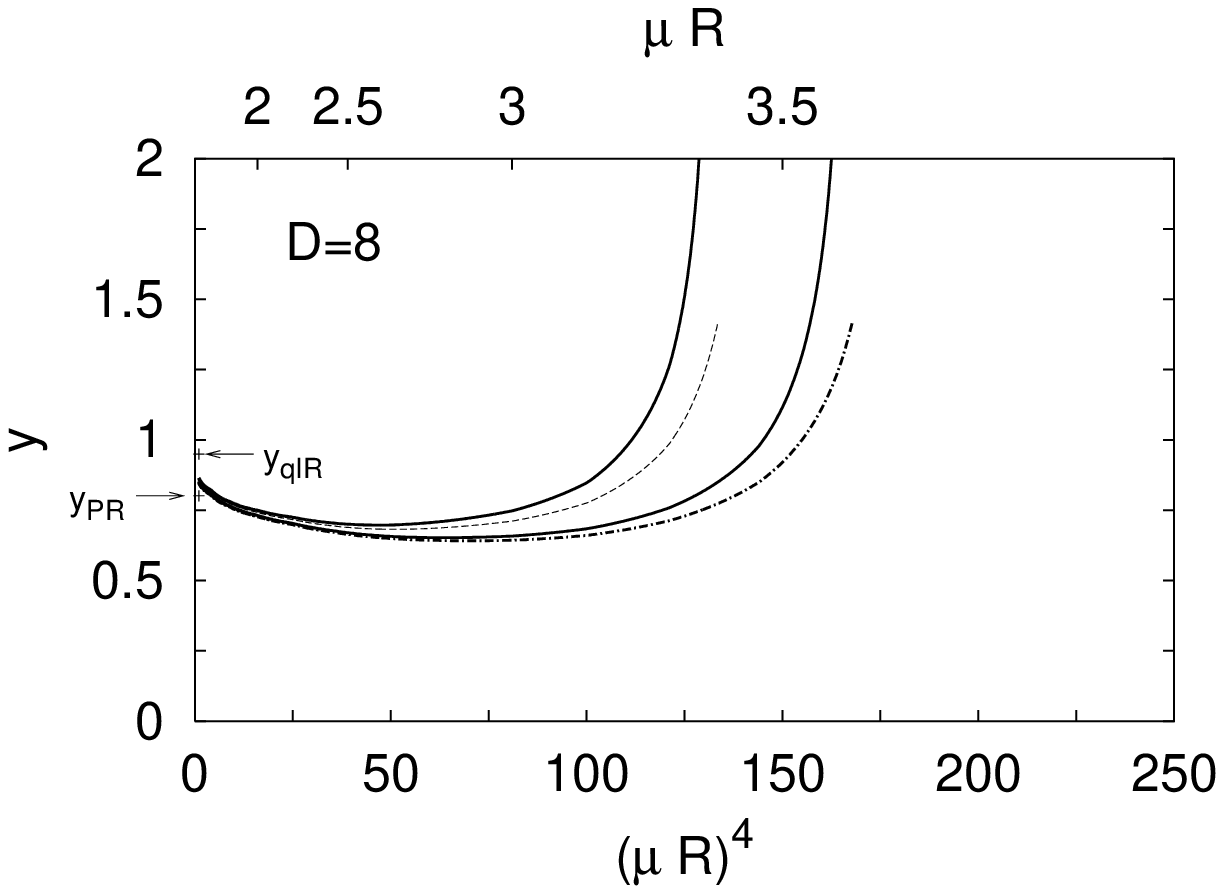}}\\[6mm]
  \resizebox{0.47\textwidth}{!}
            {\includegraphics{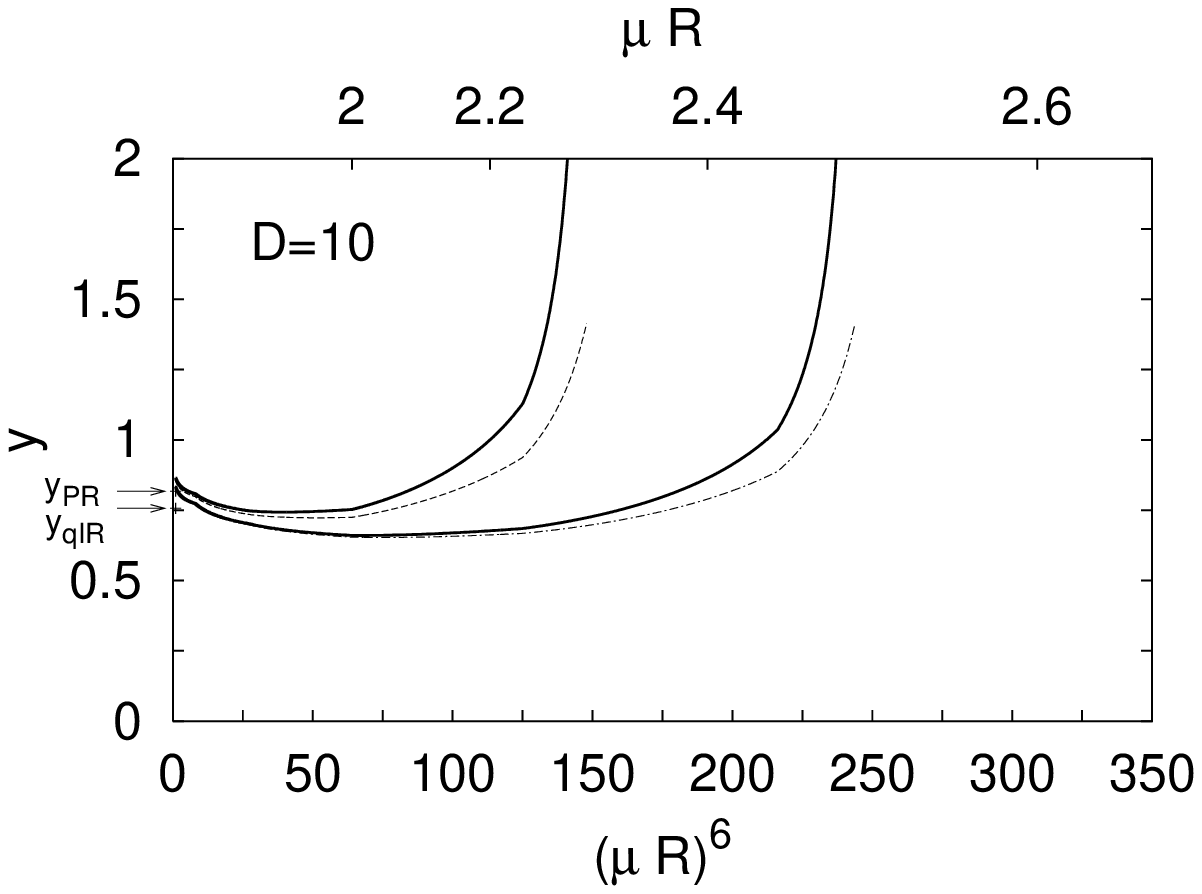}}
  \caption{RGE flows for the top Yukawa coupling $y$.
           We also show the quasi IR fixed point $y_{\rm qIR}$ 
           and the PR fixed point $y_{\rm PR}$ at $\mu=R^{-1}$.
           The graphs from top to bottom represent running of $y$
           for $D=6,8,10$, respectively.
           $R^{-1}=10$ TeV was assumed. We used 
           the full one-loop RGE~(\ref{rge_y}). 
           The UV-BCs are
           $y(\Lambda) \to \infty$ (solid lines) and 
           $y(\Lambda)=\sqrt{2}$ (dashed and dash-dotted lines) for
           two typical values of $\Lambda$.
           \label{IR-FP}}
  \end{center}
\end{figure}

The mechanism is still operative even including
the electroweak gauge interactions:
In Fig.~\ref{IR-FP}, we show the quasi IR fixed point and 
the behavior of $y$ based on 
the full one-loop RGE (\ref{rge_y}) with
various boundary conditions at $\Lambda$.
We also show the Pendleton-Ross (PR) fixed point~\cite{Pendleton:as}
determined from $\frac{d}{d \mu}\frac{y}{g_3} = 0$:
\footnote{Ref.~\cite{Arkani-Hamed:2000hv} argued 
the constraint from the PR fixed point without discussing 
the quasi IR fixed point.}
\begin{equation}
y_{\rm PR}(\mu) = g_3(\mu) \cdot \sqrt{
\frac{(6 + \delta) C_F + b'_3}{ 
        2^{\delta/2} N_c  + \frac{3}{2}}}
 \label{PR-FP}
\end{equation}
with $b'_3$ being
\begin{equation}
b'_3 = \left(-\frac{11}{3} + \frac{\delta}{6}\right)C_A  
      +\frac{8}{3} \cdot 2^{\delta/2} T_R
\end{equation}
and $C_A=N_c$, $T_R=1/2$. 
For $b'_3=0$ the PR fixed point is identical to 
the quasi IR fixed point.
As far as $b'_3 < 0$ $(D=6,8)$, 
the value of the PR fixed point is smaller than 
that of the quasi IR fixed point, 
$y_{\rm PR} < y_{\rm qIR}$,
while 
$y_{\rm PR} > y_{\rm qIR}$ for $D=10$.
The top Yukawa coupling at $R^{-1}$ for $D=6,8$ is actually between
$y_{\rm PR}$ and $y_{\rm qIR}$ for a sufficiently large top Yukawa,
$y(\Lambda) \gtrsim 1$, at high energy scale
$(\Lambda R)^\delta \gg 1$.
We note here that 
the actual prediction of $m_t$ with $D=6,8$
is even smaller than the value expected from $y_{\rm qIR}$.

We also comment that 
the predicted values of $m_t$ and $m_H$ would be stable
thanks to 
these fixed points,
even if the estimate of the tMAC scale were somewhat changed from
ours for some reason.

As was already pointed out in Ref.~\cite{Arkani-Hamed:2000hv}, 
the lower value prediction of $m_t$ than that of the original TMSM
can also be understood as follows:
Since KK modes of the top quark ($t^{(n)}$) 
as well as its zero mode ($t^{(0)}$) contribute to 
the VEV $v$, 
\begin{equation}
  v \, \propto \, 
  \VEV{\bar{t}^{(0)}t^{(0)}} + 
  \sum_{n > 0}
  \VEV{\bar{t}^{(n)}t^{(n)}},
\end{equation}
the condensate $\VEV{\bar{t}^{(0)}t^{(0)}}$ 
is suppressed compared with the original TMSM 
and so is the top mass. 

Now we discuss implication of our Higgs mass prediction Eq.~(\ref{8D-mh}).
The upper limit of $m_H$ from radiative corrections in the SM 
is $m_H < 199$ GeV at 95\% CL~\cite{PDG}.
The prediction Eq.~(\ref{8D-mh}) is still below this upper limit. 
In order to discriminate the present model from the SM 
in experiments,
we need further information of physics at the compactification scale. 
However, the Higgs boson in such a mass range, $m_H=176-188$ GeV, 
is characteristically small within the framework of the dynamical EWSB.
Yet the value is substantially 
larger than that of 
typical supersymmetric models, $m_H \lessim 130$ GeV.
Thus the present scenario is clearly distinguished from 
most of the models beyond the SM
simply through the Higgs mass observation.
The Higgs boson of this mass range
decays into weak boson pair almost 100\%.
It will be immediately discovered 
in $H \to WW^{(*)}/ZZ^{(*)}$ once the LHC starts.

\section{Summary and discussions}

We have argued 
a viable top mode standard model (TMSM) with TeV-scale extra dimensions 
where bulk $SU(3)\times SU(2) \times U(1)$ SM gauge interactions 
(without ad hoc four-fermion interactions)
trigger
condensate of only the top quark, but not of
other quarks and leptons.
In order for such a situation to be realized,
the binding strength $\kappa$ should exceed the critical binding strength
$\kappa_D^{\rm crit}$ only for the top quark (tMAC),
\begin{equation}
  \kappa_t(\tMAC) > \kappa_D^{\rm crit} > \kappa_b(\tMAC),
  \kappa_\tau(\tMAC), \cdots. \label{top-cond-fin}
\end{equation}
The binding strengths $\kappa_{t,b,\tau}$ 
were calculated by using RGEs for bulk SM gauge couplings. 
Comparing $\kappa_{t,b,\tau}$ with $\kappa_D^{\rm crit}$, 
we searched for the region of the tMAC scale $\tMAC$ satisfying
Eq.~(\ref{top-cond-fin}).
We then found that 
the region of the tMAC scale is squeezed out for $D=6$ for
the reference value of $\kappa_D^{\rm crit}$
in Ref.~\cite{Hashimoto:2000uk},
while it does exist for $D=8$, $\tMAC =$ (3.5--3.6)$R^{-1}$.
We were not able to draw a reliable conclusion for $D=10$,
since the MAC analysis for $D=10$ strongly depends on 
the regularization scheme.

For $D=8$, we predicted the top mass $m_t$ and the Higgs mass $m_H$:
\begin{equation}
  m_t=172-175 \; \mbox{GeV}  \label{mt-fin}
\end{equation}
and 
\begin{equation}
  m_H = 176-188 \; \mbox{GeV}  , \label{mh-fin}
\end{equation}
by using RGEs for the top Yukawa and Higgs quartic couplings with
the compositeness conditions at the tMAC scale $\tMAC$. 
Our predictions are governed by the quasi IR-FP and 
hence are stable against varying the composite scale.
The predicted values would not be changed so much, 
even if the region of the tMAC scale got wider than our estimate
for some reason.

Why is the value of Eq.~(\ref{mt-fin}) significantly smaller 
than that of the original TMSM in four dimensions 
which predicted $m_t \gtrsim 200$ GeV?
The value of the top Yukawa coupling at the quasi IR-FP is 
$y \simeq g_3 \cdot \sqrt{C_F(6+\delta)/(2^{\delta/2}N_c + 3/2)}$. 
Thanks to the suppression factor $2^{-\delta/2}$ in $y$,
the mass of the top quark decreases as the number of 
dimensions increases.

The predicted Higgs mass Eq.~(\ref{mh-fin}) is substantially larger
than that of typical supersymmetric models, while it is distinctively
small as a dynamical EWSB scenario.
We are thus able to distinguish the present model from typical
supersymmetric or dynamical EWSB scenarios simply through the Higgs
mass observation.
We also note that the Higgs boson in this mass range will be
immediately discovered  in $H \to WW^{(*)}/ZZ^{(*)}$ once the LHC
starts.
The upper limit of $m_H$ from radiative corrections in the SM
is $m_H < 199$ GeV at 95\% CL~\cite{PDG} and the predicted value
Eq.~(\ref{mh-fin}) is below this upper limit, however.
It is thus difficult to discriminate the Higgs in the present model
from that of the SM only from its mass.
We definitely need experimental information other than $m_H$ in order
to establish the present model.

Many issues remain to be explored:

1) Our results on the tMAC scale are sensitive to the value of 
$\kappa_D^{\rm crit}$.
Although we used the reference value of $\kappa_D^{\rm crit}$
in the approach of the ladder SD equation neglecting the effect of 
the compactification,
it would be more preferable if we can determine $\kappa_D^{\rm crit}$
more reliably. 
For such a purpose, we should take into account
effects of the compactification scale $R^{-1}$ which turned out 
not so small compared with the tMAC scale $\tMAC$ in our analysis.
We also need running of bulk gauge couplings beyond one-loop perturbation.

2) We incorporated only one composite Higgs doublet into RGEs, 
assuming other possible bound states such as vector/axial-vector bosons
are irrelevant.
In order to justify the assumption, 
we need to solve bound state problems in the bulk gauge theories.
Once such a composite scalar exists, it should be 
a tightly bound state formed by strong short distance dynamics
with large anomalous dimension.
Such a system is expected to resemble the gauged Nambu-Jona-Lasinio (GNJL) 
model where the compositeness condition is explicitly formulated.
Actually, as it happened in the 4-dimensional case~\cite{Bardeen:sm,
Kondo-Shuto-Yamawaki, Aokietal}, the pure gauge dynamics strong at short distance
in our case can also induce strong four-fermion interactions which may become
relevant operators
due to large anomalous dimensions, 
$\gamma_m \sim D/2-1$~\cite{Hashimoto:2000uk}.
We will report elsewhere the phase structure of the bulk GNJL 
model~\cite{future}.

3) There are potential constraints on our model from 
precision electroweak measurements.
The summation of KK modes below the cutoff $\Lambda$ contributes to
$\Delta\rho$ as 
$\Delta \rho \sim (\Lambda R)^{D-6} (M_W R)^2$~\cite{Nath:1999fs}.
In our case with $\Lambda=\tMAC$, $(\tMAC R)^{D-6} \sim 10$ for $D=8$,
we thus need to take $R^{-1} \gtrsim {\cal O}(\mbox{10 TeV})$, which 
may be subtle about the fine tuning.
More involved estimate will be done elsewhere.

4) Masses of other quarks and leptons have not been dealt with 
in this paper.
In the original TMSM,
these masses are 
descended  from the top condensate through ad hoc 
flavor-breaking four-fermion interactions~\cite{MTY89}.
The origin of such four-fermion interactions will be highly hoped for 
in the present scenario.

5) Our scenario crucially relies on the short distance strong dynamics 
around the composite scale.
We thus need a better-controlled theory in the UV-region.
It would be interesting to study 
a deconstructed/latticized version~\cite{Arkani-Hamed:2001ca} of our model.

\section*{Acknowledgements}
The work is supported in part by the JSPS Grant-in-Aid for the
Scientific Research (B)(2) 14340072 (K.Y. and M.T.)  
and by KRF PBRG 2002-070-C00022 (M.H.).

\appendix

\section{Counting of $N_{\rm KK}$}
\label{app:nkk}

The torus compactification of extra dimensions leaves vector-like
zero-modes in the 4-dimensional space-time, even if we start with a
chiral gauge theory in the bulk.
We thus need to compactify the extra dimensions to an orbifold
in order to obtain the conventional SM particles as zero modes.
An example of such a compactification was explicitly constructed in
Ref.\cite{Hashimoto:2000uk} based on the $T^{\delta}/Z_2^{\delta/2}$ 
orbifold for $D=4+\delta$ ($\delta=$ even) dimensions.
In this Appendix, we discuss the massive KK spectrum in the orbifold
compactification of Ref.\cite{Hashimoto:2000uk}.

\subsection{$D=4+2$}
Let us start with the 6-dimensional case ($D=4+\delta$, $\delta=2$).
The 6-dimensional space-time is decomposed into conventional and
extra dimensions,
\begin{equation}
  x^M = (x^\mu, y^m)
\end{equation}
with $M=0,1,2,3,5,6$, $\mu=0,1,2,3$ and $m=5,6$.
We assume the compactification radius $R$ and the fields at
$(x^\mu, y^5+2\pi R, y^6)$ and $(x^\mu, y^5, y^6+2\pi R)$ are
identified with the field at $(x^\mu, y^5, y^6)$.
We thus impose the periodic boundary conditions,
\begin{subequations}
\begin{eqnarray}
  A^M(x,y^5,y^6) 
    &=& A^M(x,y^5+2\pi R, y^6)
    \nonumber\\
    &=& A^M(x,y^5, y^6+2\pi R),
  \label{eq:cyclic_gauge}
\end{eqnarray}
for a gauge field and
\begin{eqnarray}
  \psi(x,y^5, y^6)
    &=& \psi(x,y^5+2\pi R, y^6)
    \nonumber\\
    &=& \psi(x,y^5, y^6+2\pi R)
  \label{eq:cyclic_fermion}
\end{eqnarray}
for a fermion.
\end{subequations}
The $Z_2$ identification of the orbifold $T^2/Z_2$ leads to the
orbifold boundary conditions (BCs),
\begin{subequations}
\begin{align}
  &A^\mu (x, y) = A^\mu (x, -y) &\; &\mbox{for $\mu=0,1,2,3$},
  \label{eq:z2_gauge_vector}
  \\
  &A^m (x, y)   = -A^m(x, -y)   &\; &\mbox{for $m=5,6$},
  \label{eq:z2_gauge_scalar}
  \\
  &\psi(x,y)    = \Gamma_{A,5} \Gamma_{A,7} \psi(x, -y) .&&
  \label{eq:z2_fermion}
\end{align}
\end{subequations}
Here $\Gamma_{A,5}$ and $\Gamma_{A,7}$ are chirality matrices in four- 
and six-dimensions, respectively\footnote{
For a convenience, we take the opposite sign for $\Gamma_{A,7}$ to 
that in Ref.~\cite{Hashimoto:2000uk}.},
\begin{equation}
  \Gamma_{A,5}\equiv i \Gamma^0 \Gamma^1 \Gamma^2 \Gamma^3, \quad
  \Gamma_{A,7}\equiv - \Gamma^0 \Gamma^1 \Gamma^2 \Gamma^3
                       \Gamma^5 \Gamma^6.
\end{equation}

It is easy to show that the gauge-vector field $A_\mu$  is decomposed
into its KK-modes,
\begin{eqnarray}
\lefteqn{
  A_\mu(x,y)
  = A_{\mu,00}(x)
} \nonumber\\
  & &
     +\sum_{n_1>0} A_{\mu, c0}^{[n_1]}
      \cos\left(\dfrac{n_1 y^5}{R}\right)
  \nonumber\\
  & &
     +\sum_{n_1>0} A_{\mu, 0c}^{[n_1]}
      \cos\left(\dfrac{n_1 y^6}{R}\right) 
  \nonumber\\
  & & + \sum_{n_1, n_2 > 0} A_{\mu,cc}^{[n_1, n_2]}
        \cos\left(\dfrac{n_1 y^5}{R}\right)
        \cos\left(\dfrac{n_2 y^6}{R}\right)
  \nonumber\\
  & &
      + \sum_{n_1, n_2 > 0} A_{\mu,ss}^{[n_1, n_2]}
        \sin\left(\dfrac{n_1 y^5}{R}\right)
        \sin\left(\dfrac{n_2 y^6}{R}\right),
\end{eqnarray}
with $n_1$, $n_2$ being positive integers, where we omitted the arguments
for the KK-modes at $x^\mu$, except for the zero mode $A_{\mu,00}$.
We note that the orbifold condition Eq.(\ref{eq:z2_gauge_vector})
leads to 
\begin{equation}
  A_{\mu, 0s}^{[n_1]} = 
  A_{\mu, s0}^{[n_1]} = 
  A_{\mu, cs}^{[n_1,n_2]} = 
  A_{\mu, sc}^{[n_1,n_2]} = 0
\end{equation}
in the KK-spectrum.
In order to investigate the KK-spectrum, we define sets of the
KK-fields,
\begin{subequations}
\begin{eqnarray}
  {\cal A}_\mu^{\delta=2, [n_1]} &\equiv& 
  \{ A_{\mu, c0}^{[n_1]}(x), \, A_{\mu, 0c}^{[n_1]}(x) \}, \\
  {\cal A}_\mu^{\delta=2, [n_1,n_2]} &\equiv& 
  \{ A_{\mu, cc}^{[n_1,n_2]}(x), \,  A_{\mu, ss}^{[n_1,n_2]}(x) \}.
\end{eqnarray}
\label{eq:gauge_vector_set}
\end{subequations}
The masses of these KK-states are given by
$m_{n_1}^2 = n_1^2 R^{-2}$ for a state contained in 
${\cal A}_\mu^{2, [n_1]}$ and
$m_{\vec n}^2 = (n_1^2 + n_2^2) R^{-2}$ for ${\cal A}_\mu^{2, [n_1,n_2]}$.
Using the number of elements of these sets,
\begin{equation}
  {\cal N}_g^{\delta=2,[n_1]} =
  {\cal N}_g^{\delta=2,[n_1,n_2]} = 2,
\end{equation}
the number of KK-modes of gauge bosons $A_\mu$ 
with $m_{\vec n}^2 \le \mu^2$ is written as
\begin{eqnarray}
\lefteqn{
  \NKKg (\mu; \delta=2) = 
} \nonumber\\
  & &  {\cal N}_g^{\delta=2,[n_1]} \cdot N_1(\mu)
     + {\cal N}_g^{\delta=2,[n_1,n_2]} \cdot N_2(\mu)
\end{eqnarray}
for $D=4+2$ dimensions, where
$N_1(\mu)$ and $N_2(\mu)$ are given by
\begin{equation}
  N_1(\mu) \equiv \sum_{n_1>0}^{n_1^2 \le \mu^2 R^2} 1 , \quad
  N_2(\mu) \equiv \sum_{n_1,n_2>0}^{n_1^2 +n_2^2 \le \mu^2 R^2} 1 ,
\label{eq:n1n2}
\end{equation}
respectively.

It is straightforward to perform similar analysis for the gauge scalar 
$A_m$.
Since the zero-mode of $A_m$ is projected out from the KK-spectrum
due to the orbifold BC (\ref{eq:z2_gauge_scalar}), 
the KK decomposition of $A_m$ is given by
\begin{eqnarray}
\lefteqn{
  A_m(x,y)
  = 
} \nonumber\\
  & &
     \sum_{n_1>0} A_{m, s0}^{[n_1]}
      \sin\left(\dfrac{n_1 y^5}{R}\right)
  \nonumber\\
  & &
     +\sum_{n_1>0} A_{m, 0s}^{[n_1]}
      \sin\left(\dfrac{n_1 y^6}{R}\right) 
  \nonumber\\
  & & + \sum_{n_1, n_2 > 0} A_{m,cs}^{[n_1, n_2]}
        \cos\left(\dfrac{n_1 y^5}{R}\right)
        \sin\left(\dfrac{n_2 y^6}{R}\right)
  \nonumber\\
  & &
      + \sum_{n_1, n_2 > 0} A_{m,sc}^{[n_1, n_2]}
        \sin\left(\dfrac{n_1 y^5}{R}\right)
        \cos\left(\dfrac{n_2 y^6}{R}\right).
\end{eqnarray}
Sets of the KK-fields are defined as 
\begin{subequations}
\begin{eqnarray}
  {\cal A}_m^{\delta=2, [n_1]} &\equiv& 
  \{ A_{m, s0}^{[n_1]}(x), \, A_{m, 0s}^{[n_1]}(x) \}, \\
  {\cal A}_m^{\delta=2, [n_1,n_2]} &\equiv& 
  \{ A_{m, cs}^{[n_1,n_2]}(x), \,  A_{m, sc}^{[n_1,n_2]}(x) \}
\end{eqnarray}
\end{subequations}
in a similar manner to Eqs.(\ref{eq:gauge_vector_set}), where
the number of elements are given by
\begin{equation}
  {\cal N}_{gs}^{\delta=2,[n_1]} =
  {\cal N}_{gs}^{\delta=2,[n_1,n_2]} = 2.
\end{equation}
We can now easily count the number of KK-modes of the gauge scalars $A_m$,
\begin{eqnarray}
\lefteqn{
  \NKKb (\mu; \delta=2) = 
} \nonumber\\
  & &  {\cal N}_{gs}^{\delta=2,[n_1]} \cdot N_1(\mu)
      +{\cal N}_{gs}^{\delta=2,[n_1,n_2]} \cdot N_2(\mu).
\end{eqnarray}

We next consider the KK spectrum of the bulk chiral fermion
$\psi_+(x,y)$,
\begin{equation}
  \Gamma_{A,7} \psi_+(x,y) = + \psi_+(x,y),
\end{equation}
in $D=4+2$ dimensions, which has obviously 4-components. 
The orbifold BC (\ref{eq:z2_fermion}) reads
\begin{equation}
  \psi_+ (x,y) = + \Gamma_{A,5} \psi_+ (x,-y) .
\end{equation}
The KK expansion of $\psi_+$ is thus given by
\begin{eqnarray}
\lefteqn{
  \psi_+(x,y) = \psi_{+,00}(x) 
} \nonumber\\
  & & + \sum_{n_1>0} \psi_{+,c0}^{[n_1]}(x)
        \cos\left(\dfrac{n_1 y^5}{R}\right)
  \nonumber\\
  & & + \sum_{n_1>0} \psi_{+,0c}^{[n_1]}(x)
        \cos\left(\dfrac{n_1 y^6}{R}\right)
  \nonumber\\
  & & + \sum_{n_1,n_2>0} \psi_{+,cc}^{[n_1,n_2]}(x)
        \cos\left(\dfrac{n_1 y^5}{R}\right)
        \cos\left(\dfrac{n_2 y^6}{R}\right)
  \nonumber\\
  & & + \sum_{n_1,n_2>0} \psi_{+,ss}^{[n_1,n_2]}(x)
        \sin\left(\dfrac{n_1 y^5}{R}\right)
        \sin\left(\dfrac{n_2 y^6}{R}\right)
  \nonumber\\
  & & + \sum_{n_1>0} \psi_{+,s0}^{[n_1]}(x)
        \sin\left(\dfrac{n_1 y^5}{R}\right)
  \nonumber\\
  & & + \sum_{n_1>0} \psi_{+,0s}^{[n_1]}(x)
        \sin\left(\dfrac{n_1 y^6}{R}\right)
  \nonumber\\
  & & + \sum_{n_1,n_2>0} \psi_{+,cs}^{[n_1,n_2]}(x)
        \cos\left(\dfrac{n_1 y^5}{R}\right)
        \sin\left(\dfrac{n_2 y^6}{R}\right)
  \nonumber\\
  & & + \sum_{n_1,n_2>0} \psi_{+,sc}^{[n_1,n_2]}(x)
        \sin\left(\dfrac{n_1 y^5}{R}\right)
        \cos\left(\dfrac{n_2 y^6}{R}\right)
  \nonumber\\
  & &
\end{eqnarray}
with 2-component fermions.
In particular, we find that the zero-mode is right-handed
under the 4-dimensional chiral rotation, 
thanks to Eq.(\ref{eq:z2_fermion}),
\begin{equation}
  \Gamma_{A,5} \psi_{+,00}(x) = + \psi_{+,00}(x). 
\end{equation}
Other KK modes include both of right, and left-handed fermions.
We thus define sets of KK-fields,
\begin{subequations}
\begin{eqnarray}
  \psi_{+,R}^{\delta=2, [n_1]} &\equiv&
  \{ \psi_{+,c0}^{[n_1]}(x), \, \psi_{+,0c}^{[n_1]}(x) \},
  \\
  \psi_{+,R}^{\delta=2, [n_1,n_2]} &\equiv&
  \{ \psi_{+,cc}^{[n_1,n_2]}(x), \, \psi_{+,ss}^{[n_1,n_2]}(x) \},
  \\
  \psi_{+,L}^{\delta=2, [n_1]} &\equiv&
  \{ \psi_{+,s0}^{[n_1]}(x), \, \psi_{+,0s}^{[n_1]}(x) \},
  \\
  \psi_{+,L}^{\delta=2, [n_1,n_2]} &\equiv&
  \{ \psi_{+,cs}^{[n_1,n_2]}(x), \, \psi_{+,sc}^{[n_1,n_2]}(x) \}.
\end{eqnarray}
\end{subequations}
The KK fields in $\psi_{+,R}^{2,[n_1]}$ and $\psi_{+,R}^{2,[n_1,n_2]}$
($\psi_{+,L}^{2,[n_1]}$ and $\psi_{+,L}^{2,[n_1,n_2]}$) are
right-handed (left-handed).
These KK-modes acquire their Dirac masses among
$\psi_{+,R}^{2,[n_1]}$ and  $\psi_{+,L}^{2,[n_1]}$
($\psi_{+,R}^{2,[n_1,n_2]}$ and  $\psi_{+,L}^{2,[n_1,n_2]}$)
in four dimensions.
The number of these KK Dirac fermions is therefore given by
\begin{eqnarray}
\lefteqn{
  \NKKf (\mu; \delta=2) = 
} \nonumber\\
  & &  {\cal N}_f^{\delta=2,[n_1]} \cdot N_1(\mu)
      +{\cal N}_f^{\delta=2,[n_1,n_2]} \cdot N_2(\mu) \label{nkk_f}
\end{eqnarray}
with
\begin{subequations}
\begin{eqnarray}
  {\cal N}_f^{\delta=2,[n_1]}
  &\equiv&  \# \psi_{+,R}^{2,[n_1]} 
  \nonumber\\
  &=& \# \psi_{+,L}^{2,[n_1]} = 2, 
  \\
  {\cal N}_f^{\delta=2,[n_1,n_2]}
  &\equiv& \# \psi_{+,R}^{2,[n_1,n_2]} 
  \nonumber\\
  &=& \# \psi_{+,L}^{2,[n_1,n_2]} = 2. 
\end{eqnarray}
\end{subequations}
We note 
\begin{equation}
  \NKKg (\mu; \delta=2) = 
  \NKKb (\mu; \delta=2) = \NKKf (\mu; \delta=2) 
\end{equation}
for $D=4+2$ dimensions.

It is straightforward to apply the above procedure to 
the bulk chiral fermion $\psi_-(x,y)$,
\begin{equation}
  \Gamma_{A,7} \psi_-(x,y) = -\psi_-(x,y).
\end{equation}
We easily find that the zero mode of $\psi_-(x,y)$ is left-handed.
We thus identify the right[left]-handed fermions in the SM to
$\psi_+(x,y)$ [$\psi_-(x,y)$].
The number of the KK Dirac fermions for 
$\psi_-$ is equal to Eq.~(\ref{nkk_f}).

For the (composite) Higgs field, we take the same $Z_2$ identification
as Eq.~(\ref{eq:z2_gauge_vector}).
This always leads to
\begin{equation}
  \NKKs (\mu;\delta) = \NKKg (\mu;\delta)
\end{equation}
for the number of KK modes of the Higgs field.
Hereafter, we abbreviate $\delta$ in the arguments of $\NKKi (\mu;\delta)$, 
$(i=g,gs,f,h)$, when the number of $\delta$ is obvious.

\subsection{$D=4+4$}
Let us next turn to the 8-dimensional case ($D=4+\delta$,
$\delta=4$),
\begin{equation}
  x^M = (x^\mu, y^m, z^{m'}), \quad
  m=5,6, \quad m'=7,8.
\end{equation}
We impose the orbifold BCs:
\begin{subequations}
\begin{eqnarray}
  A^\mu(x,y,z) &=& A^\mu(x,-y,z) \nonumber\\
               &=& A^\mu(x,y,-z), \\
  A^m(x,y,z)   &=& -A^m(x,-y,z) \nonumber\\
               &=& A^m(x,y,-z), \\
  A^{m'}(x,y,z)&=&  A^{m'}(x,-y,z) \nonumber\\
               &=& -A^{m'}(x,y,-z),
\end{eqnarray}
and
\begin{eqnarray}
  \psi(x,y,z) &=& \Gamma_{A,7'}\Gamma_{A,9} \psi(x,-y,z)\nonumber\\
              &=& \Gamma_{A,7} \Gamma_{A,9} \psi(x,y,-z),
\end{eqnarray}
\end{subequations}
with
\begin{eqnarray}
  \Gamma_{A,7'} &=& -\Gamma^0 \Gamma^1 \Gamma^2 \Gamma^3 
                     \Gamma^7 \Gamma^8 
  \nonumber\\
  \Gamma_{A,9} &=& -i\Gamma^0 \Gamma^1 \Gamma^2 \Gamma^3 
                     \Gamma^5 \Gamma^6 \Gamma^7 \Gamma^8. 
\end{eqnarray}
Noting the relations,
\begin{equation}
  \Gamma_{A,7'}\Gamma_{A,9} = \Gamma_{A,5}\Gamma_{A,7}, \quad
  \Gamma_{A,7}\Gamma_{A,9} = \Gamma_{A,5}\Gamma_{A,7'}, 
\end{equation}
we can easily see that the zero-mode of the 8-dimensional bulk
chiral fermion $\psi_+$ is chiral under the 4-dimensional chiral
symmetry. 

The bulk gauge field is decomposed into its KK-modes in a similar
manner with the $D=4+2$ case.
We obtain
\begin{subequations}
\begin{eqnarray}
\lefteqn{
  {\cal A}_\mu^{\delta=4,[n]_1} =
} \nonumber\\
  & & \{ A_{\mu, c000}^{[n]_1}, \, A_{\mu, 0c00}^{[n]_1}, \, 
         A_{\mu, 00c0}^{[n]_1}, \, A_{\mu, 000c}^{[n]_1} \},
  \\
\lefteqn{
  {\cal A}_\mu^{\delta=4,[n]_2} =
} \nonumber\\
  & & \{ A_{\mu, cc00}^{[n]_2}, \, A_{\mu, 00cc}^{[n]_2}, \, 
         A_{\mu, ss00}^{[n]_2}, \, A_{\mu, 00ss}^{[n]_2}, \,
  \nonumber\\
  & & \quad
         A_{\mu, c0c0}^{[n]_2}, \, A_{\mu, c00c}^{[n]_2}, \, 
         A_{\mu, 0cc0}^{[n]_2}, \, A_{\mu, 0c0c}^{[n]_2} \},
  \\
\lefteqn{
  {\cal A}_\mu^{\delta=4,[n]_3} =
} \nonumber\\
  & & \{ A_{\mu, ccc0}^{[n]_3}, \, A_{\mu, cc0c}^{[n]_3}, \, 
         A_{\mu, c0cc}^{[n]_3}, \, A_{\mu, 0ccc}^{[n]_3}, \,
  \nonumber\\
  & & \quad
         A_{\mu, ssc0}^{[n]_3}, \, A_{\mu, ss0c}^{[n]_3}, \, 
         A_{\mu, c0ss}^{[n]_3}, \, A_{\mu, 0css}^{[n]_3} \},
  \\
\lefteqn{
  {\cal A}_\mu^{\delta=4,[n]_4} =
} \nonumber\\
  & & \{ A_{\mu, cccc}^{[n]_4}, \, A_{\mu, ccss}^{[n]_4}, \, 
         A_{\mu, sscc}^{[n]_4}, \, A_{\mu, ssss}^{[n]_4} \},
\end{eqnarray}
\end{subequations}
for the KK-expansion of $A_\mu$.
Here $[n]_i$ stands for
\begin{equation}
  [n]_1\equiv [n_1], \quad
  [n]_2\equiv [n_1, n_2], \quad
  [n]_3\equiv [n_1, n_2, n_3], \cdots.
\end{equation}
We thus find
\begin{subequations}
\begin{eqnarray}
  {\cal N}_g^{\delta=4,[n]_1} &=& 4, \quad
  {\cal N}_g^{\delta=4,[n]_2} = 8, \\
  {\cal N}_g^{\delta=4,[n]_3} &=& 8, \quad
  {\cal N}_g^{\delta=4,[n]_4} = 4.
\end{eqnarray}
\end{subequations}
For $A_m$ ($m=5,6$) we obtain
\begin{subequations}
\begin{eqnarray}
\lefteqn{
  {\cal A}_m^{\delta=4,[n]_1} =
} \nonumber\\
  & & \{ A_{m, s000}^{[n]_1}, \, A_{m, 0s00}^{[n]_1} \},
  \\
\lefteqn{
  {\cal A}_m^{\delta=4,[n]_2} =
} \nonumber\\
  & & \{ A_{m, sc00}^{[n]_2}, \, A_{m, cs00}^{[n]_2}, \, 
         A_{m, s0c0}^{[n]_2}, \, A_{m, s00c}^{[n]_2}, \,
  \nonumber\\
  & & \quad
         A_{m, 0sc0}^{[n]_2}, \, A_{\mu, 0s0c}^{[n]_2} \},
  \\
\lefteqn{
  {\cal A}_m^{\delta=4,[n]_3} =
} \nonumber\\
  & & \{ A_{m, scc0}^{[n]_3}, \, A_{m, sc0c}^{[n]_3}, \, 
         A_{m, csc0}^{[n]_3}, \, A_{m, cs0c}^{[n]_3}, \,
  \nonumber\\
  & & \quad
         A_{m, s0cc}^{[n]_3}, \, A_{m, 0scc}^{[n]_3}, \, 
         A_{m, s0ss}^{[n]_3}, \, A_{m, 0sss}^{[n]_3} \},
  \\
\lefteqn{
  {\cal A}_m^{\delta=4,[n]_4} =
} \nonumber\\
  & & \{ A_{m, sccc}^{[n]_4}, \, A_{m, cscc}^{[n]_4}, \, 
         A_{m, scss}^{[n]_4}, \, A_{m, csss}^{[n]_4} \}.
\end{eqnarray}
\end{subequations}
It is straightforward to obtain similar expression for $A_{m'}$
($m'=7,8$).
We find
\begin{subequations}
\begin{eqnarray}
  {\cal N}_{gs}^{\delta=4,[n]_1}
  &=& 2, \quad
  {\cal N}_{gs}^{\delta=4,[n]_2} 
  = 6, \\
  {\cal N}_{gs}^{\delta=4,[n]_3} 
  &=& 8, \quad
  {\cal N}_{gs}^{\delta=4,[n]_4} 
  = 4.
\end{eqnarray}
\end{subequations}
The numbers of KK-modes of gauge fields and gauge scalars are given by
\begin{subequations}
\begin{eqnarray}
  \NKKg(\mu; \delta=4 )
  &=&       \sum_{k=1}^4 {\cal N}_g^{4,[n]_k} \cdot N_k(\mu), \\
  \NKKb(\mu; \delta=4 )
  &=&       \sum_{k=1}^4 {\cal N}_{gs}^{4,[n]_k} \cdot N_k(\mu), 
\end{eqnarray}
\end{subequations}
where $N_k(\mu)$'s are defined in Eq.(\ref{eq:n1n2}) for $k=1,2$ and
\begin{eqnarray}
  N_3(\mu) &\equiv& 
    \sum_{n_1,n_2,n_3>0}^{n_1^2 +n_2^2 + n_3^2 \le \mu^2 R^2} 1 , 
  \nonumber\\
  N_4(\mu) &\equiv& 
    \sum_{n_1,n_2,n_3,n_4>0}^{n_1^2 +n_2^2 + n_3^2 + n_4^2\le \mu^2 R^2} 1
\end{eqnarray}
for $k=3,4$.

For a bulk chiral fermion $\psi_+$ in $D=4+\delta$ ($\delta=4$)
dimensions, we find
\begin{subequations}
\begin{eqnarray}
  {\cal N}_f^{\delta=4,[n]_1} 
  &\equiv&  \# \psi_{+,R}^{4,[n]_1} 
  \nonumber\\
  &=& \# \psi_{+,L}^{4,[n]_1} = 4, 
  \\
  {\cal N}_f^{\delta=4,[n]_2} 
  &\equiv&  \# \psi_{+,R}^{4,[n]_2} 
  \nonumber\\
  &=& \# \psi_{+,L}^{4,[n]_2} = 12, 
  \\
  {\cal N}_f^{\delta=4,[n]_3} 
  &\equiv&  \# \psi_{+,R}^{4,[n]_3} 
  \nonumber\\
  &=& \# \psi_{+,L}^{4,[n]_3} = 16, 
  \\
  {\cal N}_f^{\delta=4,[n]_4} 
  &\equiv&  \# \psi_{+,R}^{4,[n]_4} 
  \nonumber\\
  &=& \# \psi_{+,L}^{4,[n]_4} = 8.
\end{eqnarray}
\end{subequations}
The number of 4-component Dirac KK-modes is then given by
\begin{equation}
  \NKKf (\mu; \delta=4)
 = \sum_{k=1}^4 {\cal N}_f^{\delta=4, [n]_k} \cdot N_k(\mu). 
\end{equation}

Here, we note 
\begin{equation}
  \NKKg(\mu; \delta=4 )
  \ne
  \NKKb(\mu; \delta=4 ), 
\end{equation}
and 
\begin{equation}
  \NKKg(\mu; \delta=4 )
  \ne
  \frac{1}{2} \NKKf(\mu; \delta=4 ), 
\end{equation}
in our orbifold compactification, because of 
\begin{equation}
    {\cal N}_g^{\delta=4,[n]_2} \ne {\cal N}_{gs}^{\delta=4,[n]_2},
    \quad 
    {\cal N}_g^{\delta=4,[n]_2} \ne \frac{1}{2}{\cal N}_f^{\delta=4,[n]_2}.
\end{equation}
The differences between them impact on the running of $\hat g_3(\mu)$ 
around $\mu R \sim {\cal O}(1)$.
The bulk QCD coupling in Eq.~(\ref{rge_ED2}) grows slowly, comparing with 
that in the approximation (\ref{rge_ED3}).
As a result, 
the region of the tMAC scale is slightly suppressed, 
if the identical value for $\hat g_3$ at the compactification scale
$R^{-1}$ is used in both RGEs.
(See Fig.~\ref{fig-tMAC-8D}(a) and Fig.~2 in Ref.~\cite{talk}.)
We also comment on the relation
\begin{equation}
  \NKKb(\mu; \delta=4 ) = \frac{1}{2} \NKKf(\mu; \delta=4 )  
\end{equation}
in our orbifold compactification.
 
\subsection{$D=4+6$}

It is now trivial task to extend the analysis to $D=4+\delta$
($\delta=6$) dimensions.
The number of the KK-modes for $\delta=6$ is given by
\begin{equation}
  \NKKi(\mu; \delta=6)
 = \sum_{k=1}^6 {\cal N}_i^{\delta=6, [n]_k} \cdot N_k(\mu)
\end{equation}
for $i=g,gs,f,h$. 

We summarize the number of elements among sets of KK-fields
in Tables~\ref{tab1},~\ref{tab2} and \ref{tab3} for $\delta=2,4,6$,
respectively. 

\begin{table}[tbp]
  \centering
  \begin{tabular}{|c||c|c|}\hline
   & ${\cal N}_i^{\delta=2,[n]_1}$ & ${\cal N}_i^{\delta=2,[n]_2}$ \\
  \hline\hline
  $i=g$ & 2 & 2 \\ \hline
  $i=gs$ & 2 & 2 \\ \hline
  $i=f$ & 2 & 2 \\ \hline
  $i=h$ & 2 & 2 \\ \hline
  \end{tabular}
  \caption{${\cal N}_i^{\delta,[n]_k}$ for $\delta=2$ ($D=6$).
           The column shows ${\cal N}_i^{\delta,[n]_k}$ 
           for $i=g$ (gauge bosons), $i=gs$ (gauge scalars),
           $i=f$ (Dirac fermions), and $i=h$ (Higgs).
           \label{tab1}}
\end{table}

\begin{table}[tbp]
  \centering
  \begin{tabular}{|c||c|c|c|c|}\hline
   & ${\cal N}_i^{\delta=4,[n]_1}$ & ${\cal N}_i^{\delta=4,[n]_2}$ 
   & ${\cal N}_i^{\delta=4,[n]_3}$ & ${\cal N}_i^{\delta=4,[n]_4}$ \\
  \hline\hline
  $i=g$ & 4 & 8 & 8 & 4 \\ \hline
  $i=gs$ & 2 & 6 & 8 & 4 \\ \hline
  $i=f$ & 4 & 12 & 16 & 8 \\ \hline
  $i=h$ & 4 & 8 & 8 & 4 \\ \hline
  \end{tabular}
  \caption{${\cal N}_i^{\delta,[n]_k}$ for $\delta=4$ ($D=8$).
           The column shows ${\cal N}_i^{\delta,[n]_k}$ 
           for $i=g$ (gauge bosons), $i=gs$ (gauge scalars),
           $i=f$ (Dirac fermions), and $i=h$ (Higgs).
           \label{tab2}}
\end{table}

\begin{table*}[tbp]
  \centering
  \begin{tabular}{|c||c|c|c|c|c|c|}\hline
   & ${\cal N}_i^{\delta=6,[n]_1}$ & ${\cal N}_i^{\delta=6,[n]_2}$ 
   & ${\cal N}_i^{\delta=6,[n]_3}$ & ${\cal N}_i^{\delta=6,[n]_4}$ 
   & ${\cal N}_i^{\delta=6,[n]_5}$ & ${\cal N}_i^{\delta=6,[n]_6}$ \\
  \hline\hline
  $i=g$ & 6 & 18 & 32 & 36 & 24 & 8 \\ \hline
  $i=gs$ & 2 & 10 & 24 & 32 & 24 & 8 \\ \hline
  $i=f$ & 6 & 30 & 80 & 120 & 96 & 32 \\ \hline
  $i=h$ & 6 & 18 & 32 & 36 & 24 & 8 \\ \hline
  \end{tabular}
  \caption{${\cal N}_i^{\delta,[n]_k}$ for $\delta=6$ ($D=10$).
           The column shows ${\cal N}_i^{\delta,[n]_k}$ 
           for $i=g$ (gauge bosons), $i=gs$ (gauge scalars),
           $i=f$ (Dirac fermions), and $i=h$ (Higgs).
           \label{tab3}}
\end{table*}

\section{Gauge couplings in extra dimensions}
\subsection{Vacuum polarization function}

We may take yet another choice for evaluating the vacuum polarization 
function, instead of the truncated KK effective 
theory~\cite{Dienes:1998vh}.
In this Appendix, we consider the proper-time (PT) scheme 
as well as the $\overline{\rm MS}$-scheme in 
the truncated KK effective theory.

The effective charge is defined by
\begin{equation}
  \frac{1}{g^2_{\rm eff}(-q^2)} \equiv \frac{1}{g^2_0}-\Pi(-q^2),
\end{equation}
where $g_0$ is the bare coupling and $\Pi$ denotes the vacuum
polarization function,
\begin{equation}
  \Pi_{\mu\nu}(-q^2) \equiv (g_{\mu\nu}q^2 - q_\mu q_\nu)\Pi(-q^2).
\end{equation}
The effective charge is closely related to physical quantities
such as scattering amplitudes. 
Through the effective coupling, we determine the relation of 
gauge couplings calculated in the two regularization schemes. 

\begin{widetext}
We consider bulk gauge theories with bulk fermions and 
bulk scalars.
In four dimensions, it is straightforward to calculate 
the contributions of fermions and scalars: 
\begin{equation}
  \Pi_f(-q^2;m^2) = n_f \cdot \frac{8T_R}{3} \cdot \frac{1}{q^2}\int_0^1 dx 
  \int \frac{d^4 \ell}{i(2\pi)^4}
  \frac{\ell^2-x(1-x)q^2-2m^2}{(\ell^2+x(1-x)q^2-m^2)^2}, \label{PiBare-f}
\end{equation}
for $n_f$ pieces of Dirac fermions of the fundamental representation, and 
\begin{equation}
  \Pi_h(-q^2;m^2) = n_h \cdot \frac{2T_R}{3} \cdot \frac{1}{q^2}\int_0^1 dx 
  \int \frac{d^4 \ell}{i(2\pi)^4}
  \frac{-2\ell^2-x(2x+1)q^2+4m^2}{(\ell^2+x(1-x)q^2-m^2)^2}, \label{PiBare-s}
\end{equation}
for $n_h$ pieces of complex scalars of the fundamental representation, 
respectively, 
where $\tr(T^a T^b)=T_R \delta^{ab}$ and 
$m$ is the mass of fermions and scalars.
For bulk fields, we perform to sum over their KK modes. 
On the other hand, it is slightly complicated to compute
the loop correction of massive gauge bosons within 
the 4-dimensional theory. 
Instead higgsing gauge theory in four dimensions,
we directly calculate the loop correction in extra dimensions.
In order to keep the gauge invariance, we use the background field 
method with the Feynman gauge.
Taking effects of compactification into account,
we replace the loop integral in $D$ dimensions to
\begin{equation}
  g_D^2 \int \frac{d^D \ell_D}{i(2\pi)^D} \to 
  g^2 \sum_{\vec n} \int \frac{d^4 \ell}{i(2\pi)^4},
\end{equation}
with the $D$-dimensional (dimensionful) coupling $g_D$ and 
the 4-dimensional one $g$. 
We here decomposed the loop momentum $\ell_D$ to the 4-dimensional one
$\ell$ and the discrete one $\vec n/R$ in extra $\delta$ dimensions, 
\begin{equation}
\ell_D = (\ell, \vec n/R), \quad \mbox{i.e.,} \quad 
\ell_D^2 = \ell^2 + m_{\vec n}^2 ,
\end{equation}
where masses of KK modes $m_{\vec n}$ are given by
\begin{equation}
m_{\vec n}^2 = |\vec n|^2/R^2, \quad \vec n \equiv (n_1,n_2,\cdots,n_\delta).
\end{equation}
Of course, the external momentum $q_D$ does not have the extra momentum,
$q_D=(q,0)$.
We thus obtain contributions of gauge bosons and gauge scalars as
\begin{eqnarray}
  \Pi_g(-q^2;m^2) &=& C_A \int_0^1 dx 
  \int \frac{d^4 \ell}{i(2\pi)^4}\left[\,
  \frac{4}{(\ell^2+x(1-x)q^2-m^2)^2} \right. \nonumber \\ && \qquad \left.
  -\frac{2}{3} \cdot \frac{1}{-q^2} \cdot
  \frac{-2\ell^2-x(2x+1)q^2+4m^2}{(\ell^2+x(1-x)q^2-m^2)^2}\,\right], 
  \label{PiBare-g}
\end{eqnarray}
and
\begin{equation}
  \Pi_{gs}(-q^2;m^2) = \frac{\delta}{3} \cdot C_A \cdot 
  \frac{1}{q^2}\int_0^1 dx 
  \int \frac{d^4 \ell}{i(2\pi)^4}
  \frac{-2\ell^2-x(2x+1)q^2+4m^2}{(\ell^2+x(1-x)q^2-m^2)^2}, \label{PiBare-gs}
\end{equation}
respectively.
We separate the vacuum polarization function to 
the contributions of zero modes $\Pi_0$, which corresponds to
the SM  corrections, and KK modes $\Pi_{\rm KK}$:
\begin{equation}
  \Pi(-q^2) = \Pi_0(-q^2)
  +\sum_{|\vec n|^2 > 0}\Pi_{\rm KK}(-q^2;m_{\vec n}^2) 
\end{equation}
with
\begin{equation}
  \Pi_0(-q^2) \equiv \Pi_g(-q^2;0)+\Pi_f(-q^2;0)+\Pi_h(-q^2;0)
\end{equation}
and
\begin{equation}
  \Pi_{\rm KK}(-q^2;m^2) \equiv 
  \Pi_g(-q^2;m^2) + \Pi_{gs}(-q^2;m^2) 
  + \Pi_f(-q^2;m^2) +\Pi_h(-q^2;m^2) .
\end{equation}

\subsection{$\overline{\rm MS}$-coupling in 
the truncated KK effective theory}

We briefly review the results in 
Refs.~\cite{Hashimoto:2000uk,Hashimoto:2002px}.
In the truncated KK effective theory, we calculate the vacuum 
polarization function by using the dimensional regularization 
(taking the loop integral to $\int d^{4-2\epsilon} \ell$ instead of
$\int d^4 \ell$):
\begin{eqnarray}
 (4\pi)^2 \Pi_g^{\overline{\rm MS}}(-q^2;m^2) &=& 
   C_A \, \left[\,4 I_g(-q^2;m^2) - I_s(-q^2;m^2)\,\right], \\
 (4\pi)^2 \Pi_{gs}^{\overline{\rm MS}}(-q^2;m^2) &=& 
  - \frac{\delta}{2} \, C_A \, I_s(-q^2;m^2), \\
 (4\pi)^2 \Pi_f^{\overline{\rm MS}}(-q^2;m^2) &=& - 8 T_R \, n_f \,
   I_f(-q^2;m^2), \\
 (4\pi)^2 \Pi_h^{\overline{\rm MS}}(-q^2;m^2) &=& - T_R \, n_h \, 
   I_s(-q^2;m^2), 
\end{eqnarray}
where we defined
\begin{eqnarray}
I_g(-q^2;m^2) &\equiv& \frac{\Gamma(\epsilon)}{(4\pi)^{-\epsilon}}
  \int_0^1 dx \left[\,m^2-x(1-x)q^2\,\right]^{-\epsilon}, \\
I_s(-q^2;m^2) &\equiv& \frac{\Gamma(\epsilon)}{(4\pi)^{-\epsilon}}
  \int_0^1 dx \, (2x-1)^2 \left[\,m^2-x(1-x)q^2\,\right]^{-\epsilon},\\
I_f(-q^2;m^2) &\equiv& \frac{\Gamma(\epsilon)}{(4\pi)^{-\epsilon}}
  \int_0^1 dx \, x(1-x) \left[\,m^2-x(1-x)q^2\,\right]^{-\epsilon} .
\end{eqnarray}
We renormalize the bare coupling in the $\overline{\rm MS}$-scheme:
\begin{equation}
  \frac{1}{g^2_{\overline{\rm MS}}(\mu)} = \frac{1}{g_0^2}
  +\frac{\Gamma(\epsilon)}{(4\pi)^{2-\epsilon}}b^{\rm SM} \mu^{-2\epsilon}
  +\frac{\Gamma(\epsilon)}{(4\pi)^{2-\epsilon}}
    b^{\rm KK}_{\overline{\rm MS}}  \mu^{-2\epsilon}
  -\sum_{\vec n}^{m_{\vec n}^2 > \mu^2}\Pi_{\rm KK}(0;m_{\vec n}^2), 
\end{equation}
with the RGE coefficient of zero mode $b^{\rm SM}$,
\begin{equation}
 b^{\rm SM} \equiv 
 -\frac{11}{3} \, C_A 
 +\frac{4 T_R}{3} \, n_f^{\rm SM}
 +\frac{T_R}{3} \, n_h^{\rm SM}
\end{equation}
and that of KK modes $b^{\rm KK}_{\overline{\rm MS}}$,
\begin{equation}
 b^{\rm KK}_{\overline{\rm MS}}(\mu) \equiv 
 -\frac{11}{3} \, C_A \, \NKKg (\mu)
 +\frac{\delta}{6} \, C_A \, \NKKb (\mu)
 +\frac{4 T_R}{3} \, n_f \, \NKKf (\mu)
 +\frac{T_R}{3} \, n_h \, \NKKs (\mu) ,
\end{equation}
where $N_{\rm KK}^i(\mu), i=g,gs,f,h$ denote the total number of KK modes
for gauge bosons, gauge scalars, fermions, and scalars 
(composite Higgs fields) below $\mu$, respectively.
We then find the effective coupling in the truncated effective theory
based on the $\overline{\rm MS}$-scheme,
\begin{equation}
  \frac{1}{g_{\rm eff}^2(-q^2)} = \frac{1}{g^2_{\overline{\rm MS}}(\mu)}
  -\frac{b^{\rm SM}}{(4\pi)^2}\ln \frac{-q^2}{\mu^2}
  -\frac{1}{(4\pi)^2}\sum_{|\vec n|^2 > 0}^{m_{\vec n}^2 \leq \mu^2} b'
   \ln \frac{m_{\vec n}^2}{\mu^2}
  -\sum_{|\vec n|^2 > 0}\overline{\Pi}_{\rm KK}^{\overline{\rm MS}}
    (-q^2;m_{\vec n}^2)
  +\frac{c_0^{\overline{\rm MS}}}{(4\pi)^2}
\end{equation}
with 
\begin{equation}
  b' =  -\frac{11}{3} \, C_A \, +\frac{\delta}{6} \, C_A \, 
        +\frac{4 T_R}{3} \, n_f \, +\frac{T_R}{3} \, n_h, \qquad 
  \sum_{|\vec n|^2 > 0}^{m_{\vec n}^2 \leq \mu^2} b' = 
  b^{\rm KK}_{\overline{\rm MS}}(\mu),
\end{equation}
where we defined
\begin{eqnarray}
 (4\pi)^2 \overline{\Pi}_{\rm KK}^{\overline{\rm MS}}(-q^2;m^2) &\equiv& 
  - C_A \left[\,4 I_g^R (-q^2;m^2) - I_s^R (-q^2;m^2)\,\right]
  + \frac{\delta}{2} \, C_A \, I_s^R(-q^2;m^2) \nonumber \\
  && \quad 
  + 8 T_R \, n_f \, I_f^R(-q^2;m^2)
  + T_R \, n_h \, I_s^R(-q^2;m^2), \\
I_g^R (-q^2;m^2) &\equiv& 
  \int_0^1 dx \ln \left(\,1-x(1-x)\frac{q^2}{m^2}\,\right), \\
I_s^R (-q^2;m^2) &\equiv& 
  \int_0^1 dx \, (2x-1)^2 
  \ln \left(\,1-x(1-x)\frac{q^2}{m^2}\,\right), \\
I_f^R (-q^2;m^2) &\equiv& 
  \int_0^1 dx \, x(1-x) \ln \left(\,1-x(1-x)\frac{q^2}{m^2}\,\right),
\end{eqnarray}
and obtained the constant term
\begin{equation}
  c_0^{\overline{\rm MS}} = -\frac{67}{9} \, C_A 
  + \frac{20 T_R}{9} \, n_f^{\rm SM} + \frac{8 T_R}{9} \, n_h^{\rm SM}.
\end{equation}

We identify the mass of the heaviest KK state to the cutoff $\Lambda$.
In the limit of $|q^2| \ll R^{-2} \ll \Lambda^2$,
we easily find 
\begin{equation}
  \sum_{|\vec n|^2 > 0}^{m_{\vec n}^2 \leq \Lambda^2}
  \overline{\Pi}_{\rm KK}^{\overline{\rm MS}}(-q^2;m_{\vec n}^2)
 = \frac{c_1}{(4\pi)^2}(-q^2 R^2) \frac{(\Lambda R)^{\delta-2}}{\delta-2}
   \frac{2\pi^{\delta/2}}{2^{\delta/2}\Gamma(\delta/2)} 
   + {\cal O}(|q^2|^2) \label{ms-q}
\end{equation}
with
\begin{equation}
 c_1 \equiv 
 -\frac{19}{30} \, C_A +\frac{\delta}{60} \, C_A 
 +\frac{4 T_R}{15} \, n_f \, 2^{\delta/2-1}
 +\frac{T_R}{30} \, n_h ,
\end{equation}
where $(\Lambda R)^{\delta-2}/(\delta-2)$ is replaced by 
$\ln \Lambda R$ for $\delta=2$. 
Since we cannot take so large value to the cutoff $\Lambda$ 
due to the Landau pole for $U(1)_Y$, 
the effects of the finite parts~(\ref{ms-q}) summing over KK modes
are negligible in a certain low energy scale, 
comparing with the finite term arising from zero modes, 
$c_0^{\overline{\rm MS}} \sim {\cal O}(10)$. 
Numerically, we find the finite sums for QCD at ${\cal O}(|q^2|)$,  
\begin{equation}
 (4\pi)^2 \sum_{|\vec n|^2 > 0}\overline{\Pi}_{\rm KK}^{\overline{\rm MS}}
 (-q^2=M_Z^2;m_{\vec n}^2)  
 = -0.103, \quad -0.656, \quad -0.786 
\end{equation}
for $D=6,8,10$, respectively, 
where we took $R^{-1}=1$ TeV, $M_Z=91.2$ GeV, $\Lambda R=13,3.7,2.6$
for $D=6,8,10$, respectively.
Of course, the effects of Eq.~(\ref{ms-q}) get smaller for larger
values of $R^{-1}$.

At the zero momentum, the effects of finite terms via KK modes
come to be completely negligible thanks to Eq.~(\ref{ms-q}).
We thus identify $g^2_{\overline{\rm MS}}(\mu=M_Z)$ to
the SM value.
We also find the $\beta$-function,
\begin{equation}
  (4\pi)^2 \frac{\partial g_{\overline{\rm MS}}}{\partial \ln \mu}
 = (b^{\rm SM}+b^{\rm KK}_{\overline{\rm MS}}(\mu))g_{\overline{\rm MS}}^3 .
\end{equation}

\subsection{Proper-time regularization}

We calculate the vacuum polarization function in the PT-scheme.
Using the trick with the PT-parameter $t$,
\begin{equation}
  \frac{1}{A^2}=\int_0^\infty dt \, t e^{-At}, 
\end{equation}
and performing the momentum integral,
\begin{equation}
  \int_0^\infty\frac{d^4 \ell_E}{(2\pi)^4}\exp(-t\ell_E^2)
  =\frac{1}{(4\pi)^2}\frac{1}{t^2}, \quad
  \int_0^\infty\frac{d^4 \ell_E}{(2\pi)^4}\,
   \ell_E^2 \cdot\exp(-t\ell_E^2)
  =\frac{2}{(4\pi)^2}\frac{1}{t^3}, \quad
\end{equation}
we can rewrite vacuum polarization functions in 
Eqs.~(\ref{PiBare-f}), (\ref{PiBare-s}), (\ref{PiBare-g}), and 
(\ref{PiBare-gs}) as follows:
\begin{eqnarray}
  (4\pi)^2 \Pi_g^{\rm PT}(q_E^2;m^2) &=& C_A \, 
   \left[\,4 J_g(q_E^2;m^2) - J_s(q_E^2;m^2)\,\right], \\
  (4\pi)^2 \Pi_{gs}^{\rm PT}(q_E^2;m^2) &=& 
  - \frac{\delta}{2} \, C_A \, J_s(-q^2;m^2), \\
  (4\pi)^2 \Pi_f^{\rm PT}(q_E^2;m^2) &=& - 8 T_R \, n_f \, 
   J_f(q_E^2;m^2), \\
  (4\pi)^2 \Pi_h^{\rm PT}(q_E^2;m^2) &=& - T_R \, n_h \, 
   J_s(q_E^2;m^2), 
\end{eqnarray}
with
\begin{align}
  J_g(q_E^2;m^2)&= \int_0^1 dx \int_0^\infty
  \frac{dt}{t} \exp[-t(x(1-x)q_E^2+m^2)], \\
  J_s(q_E^2;m^2)&= \int_0^1 dx \int_0^\infty
  \frac{dt}{t} (2x-1)^2 \exp[-t(x(1-x)q_E^2+m^2)], \\
  J_f(q_E^2;m^2)&= \int_0^1 dx \int_0^\infty
  \frac{dt}{t} x(1-x) \exp[-t(x(1-x)q_E^2+m^2)],
\end{align}
where the suffix $E$ denotes the Euclidean momentum.
Since the PT-integral diverges near $t \sim 0$,
we regularize the integral with the cutoff $\Lambda$:
\begin{equation}
  \int_0^\infty \frac{dt}{t} \to  \int_{r\Lambda^{-2}}^\infty \frac{dt}{t} .
\end{equation}

We first calculate $\Pi_0$.
Performing the Feynman parameter integration,
we obtain
\begin{eqnarray}
  J_g(q_E^2;0) &=& \int_{\frac{r q_E^2}{4\Lambda^2}}^\infty \frac{dt}{t}
  {}_1F_1(1,3/2;-t), \\
  J_s(q_E^2;0) &=& \frac{1}{3} \int_{\frac{r q_E^2}{4\Lambda^2}}^\infty 
   \frac{dt}{t}{}_1F_1(1,5/2;-t), \\
  J_f(q_E^2;0) &=& \frac{1}{12} \int_{\frac{r q_E^2}{4\Lambda^2}}^\infty
   \frac{dt}{t}\left[\,3 {}_1F_1(1,3/2;-t)-{}_1F_1(1,5/2;-t)\,\right],
\end{eqnarray}
where we used the confluent hypergeometric function ${}_1F_1(a,c;z)$ 
and the relation between the confluent hypergeometric function and 
the error function,
\begin{equation}
  {\rm Erfi} (x) \equiv \int_0^x e^{u^2} du = 
  x \cdot {}_1F_1(1/2,3/2;x^2) .
\end{equation}
Here, we note the Kummer's transformation,
\begin{equation}
{}_1F_1(a,c;z)=e^z{}_1F_1(c-a,c;-z), \quad c \ne \mbox{negative integer},
\end{equation}
and the recurrence formula,
\begin{equation}
  \frac{1}{t}\left[\,{}_1F_1(1,c;-t)-1\,\right] = 
 -\frac{1}{c}{}_1F_1(1,c+1;-t) .
\end{equation}
Since the behaviors of ${}_1F_1(a,c;z)$ around $z \sim 0$ and 
$z \to -\infty$ are given by
\begin{equation}
  {}_1F_1(a,c;z)=1+\frac{a}{c}z+{\cal O}(z^2) , 
  \quad 
  (z \sim 0),
\end{equation}
and 
\begin{equation}
  {}_1F_1(a,c;z)=\frac{\Gamma(c)}{\Gamma(c-a)}(-z)^{-a}+{\cal O}(z^{-a-1}),
  \quad 
  (z \to -\infty),
\end{equation}
respectively, 
we separate the PT-integration to the singular part and finite ones:
\begin{eqnarray}
  \int_{\frac{r q_E^2}{4\Lambda^2}}^\infty \frac{dt}{t}{}_1F_1(1,c;-t)
 &=& \int_{\frac{r q_E^2}{4\Lambda^2}}^1 \frac{dt}{t} + 
   \int_0^1 \frac{dt}{t} 
   \left[\,{}_1F_1(1,c;-t)-1\,\right] \nonumber \\ && -
   \int_0^{\frac{r q_E^2}{4\Lambda^2}} \frac{dt}{t} 
   \left[\,{}_1F_1(1,c;-t)-1\,\right] +
   \int_1^\infty \frac{dt}{t}{}_1F_1(1,c;-t) , \\
 &=& -\ln \frac{r q_E^2}{4\Lambda^2} + \frac{1}{c}
   \int_0^{\frac{r q_E^2}{4\Lambda^2}} dt\,
   {}_1F_1(1,1+c;-t) \nonumber \\ && -\frac{1}{c}
   \int_0^1 dt\, {}_1F_1(1,1+c;-t) + 
   \int_1^\infty \frac{dt}{t}{}_1F_1(1,c;-t) .
\end{eqnarray}
Noting the integration of the confluent hypergeometric function,
\begin{equation}
  \int_0^z dt {}_1 F_1(1,1+c;-t)
 = z \cdot {}_2F_2(1,1;2,1+c;-z)
 = z - \frac{z^2}{2(1+c)}+\cdots, \quad   
\end{equation}
we find
\begin{eqnarray}
  J_g(q_E^2;0) &=& -\ln \frac{r q_E^2}{4\Lambda^2} + c_g
  + \frac{2}{3}\frac{r q_E^2}{4\Lambda^2} + {\cal O}((q_E^2)^2), \\
  J_s(q_E^2;0) &=& -\frac{1}{3} \ln \frac{r q_E^2}{4\Lambda^2}
  + c_s + \frac{2}{15}\frac{r q_E^2}{4\Lambda^2} + {\cal O}((q_E^2)^2),\\
  J_f(q_E^2;0) &=& -\frac{1}{6} \ln \frac{r q_E^2}{4\Lambda^2}
  + c_f + \frac{2}{15}\frac{r q_E^2}{4\Lambda^2} + {\cal O}((q_E^2)^2),
\end{eqnarray}
with
\begin{eqnarray}
  c_g &\equiv&  -\frac{2}{3} \int_0^1 dt {}_1F_1(1,5/2;-t) + 
   \int_1^\infty \frac{dt}{t}{}_1F_1(1,3/2;-t) \simeq 0.0365, \\
  c_s &\equiv&  -\frac{2}{15} \int_0^1 dt {}_1F_1(1,7/2;-t) + 
  \frac{1}{3} \int_1^\infty \frac{dt}{t}{}_1F_1(1,5/2;-t) \simeq 0.234, \\
  c_f &\equiv&  -\frac{1}{6} \int_0^1 dt {}_1F_1(1,5/2;-t) + 
  \frac{1}{30} \int_0^1 dt {}_1F_1(1,7/2;-t) \nonumber \\ && + 
  \frac{1}{4} \int_1^\infty \frac{dt}{t}{}_1F_1(1,3/2;-t) -
  \frac{1}{12} \int_1^\infty \frac{dt}{t}{}_1F_1(1,5/2;-t)
  \simeq -0.0501 .
\end{eqnarray}
We can read $\Pi_0^{\rm PT}$ from the above results:
\begin{eqnarray}
  \Pi_0^{\rm PT}(q_E^2) &\equiv& 
   \Pi_g^{\rm PT}(q_E^2;0)+\Pi_f^{\rm PT}(q_E^2;0)
 + \Pi_h^{\rm PT}(q_E^2;0), \\
 &=& \frac{b^{\rm SM}}{(4\pi)^2} \ln \frac{r q_E^2}{4\Lambda^2}-
 \frac{c_0^{\rm PT}}{(4\pi)^2} + {\cal O}\left(\frac{q_E^2}{\Lambda^2}\right)
 \label{pt0}
\end{eqnarray}
with
\begin{equation}
 c_0^{\rm PT} = -C_A \, (4c_g-c_s)
 +8T_R \, n_f^{\rm SM} \, c_f + T_R \, n_h^{\rm SM} \, c_s .
\end{equation}
Here, we note $c_0^{\rm PT} \ll c_0^{\overline{\rm MS}}$
thanks to $c_g,c_s,c_f \ll 1$.

Now, we study the contributions of the KK modes.
We can perform the summation over all KK states 
by using the Jacobi $\vartheta_3$ function,
\begin{equation}
  \vartheta_3(\tau) \equiv \sum_{n=-\infty}^{\infty}\exp(i\pi\tau n^2) ,
\end{equation}
which has the remarkable property,
\begin{equation}
  \vartheta_3(i\tau) = \frac{1}{\sqrt{\tau}}\vartheta_3(i/\tau) , \quad
  \mbox{i.e.,} \quad \vartheta_3(i\tau) \to \frac{1}{\sqrt{\tau}} 
  \quad (\tau \to 0_+).
\end{equation}
We then obtain the summation of the vacuum polarization via 
KK modes:
\begin{eqnarray}
\lefteqn{
  \sum_{|\vec n|^2>0}\Pi_{\rm KK}^{\rm PT}(q_E^2,m_{\vec n}^2) =}
  \nonumber \\ &&
 \frac{C_A}{(4\pi)^2}\int_0^1dx\int_{r\Lambda^{-2}}^\infty\frac{dt}{t}
 \left[\,4-(2x-1)^2\,\right]\exp[-tx(1-x)q_E^2]K_g(t) \nonumber \\ &&-
 \frac{\delta}{2} \cdot \frac{C_A}{(4\pi)^2}
 \int_0^1dx\int_{r\Lambda^{-2}}^\infty\frac{dt}{t}\,
 (2x-1)^2 \exp[-tx(1-x)q_E^2]K_{gs}(t) \nonumber \\ &&-
  8T_R \cdot \frac{n_f}{(4\pi)^2}
 \int_0^1dx\int_{r\Lambda^{-2}}^\infty\frac{dt}{t}\,
 x(1-x) \exp[-tx(1-x)q_E^2]K_f (t) \nonumber \\ &&-
  T_R \cdot \frac{n_h}{(4\pi)^2}
 \int_0^1dx\int_{r\Lambda^{-2}}^\infty\frac{dt}{t}\,
 (2x-1)^2 \exp[-tx(1-x)q_E^2]K_h (t) ,
\end{eqnarray}
where we defined 
\begin{equation}
  K_i(t) \equiv {\cal N}_i^{\delta,[n]_1}
  \left[\,\frac{1}{2}(\vartheta_3-1)\,\right] + 
  {\cal N}_i^{\delta,[n]_2} \left[\,\frac{1}{2}(\vartheta_3-1)\,\right]^2 + 
  {\cal N}_i^{\delta,[n]_3} \left[\,\frac{1}{2}(\vartheta_3-1)\,\right]^3 + 
  \cdots
\end{equation}
with $i=g,gs,f,h$ and $\vartheta_3=\vartheta_3(it/(\pi R^2))$.
The values of ${\cal N}_i^{\delta,[n]_k}$ are given 
in Table~\ref{tab1}--\ref{tab3}.
Combining with Eq.~(\ref{pt0}),
we obtain the effective coupling as
\begin{eqnarray}
  \frac{1}{g_{\rm eff}^2(q_E^2)} &=& \frac{1}{g^2_0}
  -\frac{b^{\rm SM}}{(4\pi)^2}\ln \frac{q_E^2}{\Lambda^2}
  -\sum_{|\vec n|^2 > 0}\Pi_{\rm KK}^{\rm PT}(q_E^2=0;m_{\vec n}^2)
 \nonumber \\ &&
  -\sum_{|\vec n|^2 > 0}
   \overline{\Pi}_{\rm KK}^{\rm PT}(q_E^2;m_{\vec n}^2)
  +\frac{c_0^{\rm PT}}{(4\pi)^2}-\frac{b^{\rm SM}}{(4\pi)^2}\ln \frac{r}{4}
\end{eqnarray}
with
\begin{equation}
 \overline{\Pi}_{\rm KK}^{\rm PT}(q_E^2;m_{\vec n}^2) \equiv 
 \Pi_{\rm KK}^{\rm PT}(q_E^2;m_{\vec n}^2)-
 \Pi_{\rm KK}^{\rm PT}(q_E^2=0;m_{\vec n}^2), 
\end{equation}
where we neglected the corrections of the order of
${\cal O}(q_E^2/\Lambda^2)$.
In the limit of $(q_E^2 R^2)(\Lambda R)^{\delta-2} \ll 1$,
we find 
\begin{equation}
 \sum_{|\vec n|^2 > 0}
  \overline{\Pi}_{\rm KK}^{\rm PT}(q_E^2;m_{\vec n}^2) =
  \frac{c_1}{(4\pi)^2}(q_E^2 R^2) \frac{(\Lambda R)^{\delta-2}}{\delta-2}
   \cdot 2r \cdot \left(\frac{\pi}{2r}\right)^{\delta/2} 
   + {\cal O}(q_E^4) , \label{pt-q}
\end{equation}
where $(\Lambda R)^{\delta-2}/(\delta-2)$ is replaced by 
$\ln \Lambda R$ for $\delta=2$. 

We regard the bare coupling $g_0$ as a quantity defined at 
the cutoff $\Lambda$:
\begin{equation}
  g_0 \to g_{\rm PT}(\Lambda) .
\end{equation}
Since the effective coupling $g_{\rm eff}(q_E^2)$ should not 
depend on the cutoff $\Lambda$ at the zero momentum $q_E^2 \to 0$,
we can deduce the ``$\beta$-function'' for $g_{\rm PT}(\Lambda)$:
\begin{equation}
  (4\pi)^2 \frac{\partial g_{\rm PT}}{\partial \ln \Lambda}
 = (b^{\rm SM}+b^{\rm KK}_{\rm PT}(\Lambda))g_{\rm PT}^3 
\end{equation}
with the ``RGE coefficient'' $b^{\rm KK}_{\rm PT}$,
\begin{equation}
 b^{\rm KK}_{\rm PT}(\Lambda) \equiv 
 -\frac{11}{3} \, C_A \, K_g(r \Lambda^{-2})
 +\frac{\delta}{6} \, C_A \, K_{gs}(r \Lambda^{-2})
 +\frac{4 T_R}{3} \, n_f \, K_f(r \Lambda^{-2})
 +\frac{T_R}{3} \, n_h \, K_h(r \Lambda^{-2}) .
\end{equation}
Matching the RGE for the $\overline{\rm MS}$-coupling with 
$\mu = \Lambda$ in the limit of $\Lambda R \gg 1$,
we easily find~\cite{Dienes:1998vh}
\begin{equation}
  r = \pi X_\delta^{-2/\delta}, \quad 
  X_\delta = \frac{\pi^{\delta/2}}{\Gamma (1+\delta/2)} . \label{pt-r}
\end{equation}
We identify the effective coupling at the zero momentum
calculated in the PT-scheme to the SM one with $\mu=\Lambda=R^{-1}$:
\begin{equation}
  \frac{1}{g_{\rm PT}^2(R^{-1})} -\sum_{|\vec n|^2 > 0}\Pi_{\rm KK}^{\rm PT}
  (q_E^2=0;m_{\vec n}^2)|_{\Lambda=R^{-1}}
  +\frac{c_0^{\rm PT}}{(4\pi)^2}-\frac{b^{\rm SM}}{(4\pi)^2}\ln \frac{r}{4}
  = \frac{1}{g_{\overline{\rm MS}}^2(R^{-1})} 
  +\frac{c_0^{\overline{\rm MS}}}{(4\pi)^2} . \label{gpt-mz}
\end{equation}

We comment on the coefficients of $q_E^2 R^2 (\Lambda R)^{\delta-2}$
in the vacuum polarization function Eq.~(\ref{pt-q}). 
Plugging Eq.~(\ref{pt-q}) with $r$ determined as Eq.~(\ref{pt-r}),
we can rewrite Eq.~(\ref{pt-q}),
\begin{equation}
 \sum_{|\vec n|^2 > 0}
  \overline{\Pi}_{\rm KK}^{\rm PT}(q_E^2;m_{\vec n}^2) =
  \frac{c_1}{(4\pi)^2}(q_E^2 R^2) \frac{(\Lambda R)^{\delta-2}}{\delta-2}
   \, \frac{2 \pi^{\delta/2}}{2^{\delta/2}\Gamma(\delta/2)} \,
   \frac{2r}{\delta}
   + {\cal O}(q_E^4) . \label{pt-q2}
\end{equation}
\end{widetext}
Although the coefficients of $q_E^2 R^2 (\Lambda R)^{\delta-2}$ 
in the $\overline{\rm MS}$-scheme and the PT-scheme are accidentally 
identical in $\delta=2\; (D=6)$, they have generally the regularization 
dependence and their difference is the factor $2r/\delta$ in Eq.~(\ref{pt-q2}).

\end{document}